\documentclass{jnmp}

\usepackage{amsmath}

\JNMPnumberwithin{equation}{subsection}
\JNMPnumberwithin{footnote}{section}

\theoremstyle{definition}
\newtheorem{example}{Example}[section]

\begin{document}

\renewcommand{\evenhead}{P~G~L Leach and G~P Flessas}
\renewcommand{\oddhead}{Generalisations of the Laplace--Runge--Lenz Vector}

\thispagestyle{empty}

\FirstPageHead{10}{3}{2003}{\pageref{leach-firstpage}--\pageref{leach-lastpage}}{Review
Article}

\copyrightnote{2003}{P~G~L Leach and G~P Flessas}

\Name{Generalisations\\ of the Laplace--Runge--Lenz Vector}
\label{leach-firstpage}

\strut\hfill

\Author{P~G~L LEACH~$^{\dag^1\dag^2\dag^3}$ and G~P FLESSAS~$^{\dag^1\dag^4}$}

\Address{$^{\dag^1}$~GEODYSYC, School of Sciences, University of the Aegean, Karlovassi 83 200, Greece\\[10pt]
$^{\dag^2}$~Department of Mathematics, University of the Aegean, Karlovassi 83 200, Greece\\[10pt]
$^{\dag^3}$~Permanent address: School of Mathematical and Statistical Sciences,\\
$\phantom{^{\dag^3}}$~University of Natal, Durban 4041, Republic of South Africa\\[10pt]
$^{\dag^4}$~Department of Information and Communication Systems Engineering, \\
$\phantom{^{\dag^4}}$~University of the Aegean, Karlovassi 83 200, Greece}

\Date{Received October 17, 2002; Revised January 22, 2003;
Accepted January 27, 2003}

\begin{abstract}
\noindent
The characteristic feature of the Kepler Problem is
 the existence of the so-called Laplace--Runge--Lenz vector
 which enables a very simple discussion of the properties
of the orbit for the problem.  It is found that there are many classes of problems,
some closely related to the Kepler Problem and others somewhat remote, which
 share the possession of a conserved vector which plays a significant r\^ole in
the analysis of these problems.
\end{abstract}

\tableofcontents

\input{leach-commands}

\setcounter{footnote}{0}
\renewcommand{\thefootnote}{\arabic{section}.\arabic{footnote}}

\renewcommand{\theequation}{\arabic{section}.\arabic{subsection}.\arabic{equation}}

\section{Introduction}

\resetfootnoterule

Cosmology and its precursor, Cosmogeny, have a very long history
dating back to the earliest records of ancient civilisations.  The
practical tasks of Astronomy and its more imaginative relation,
Astrology, seemed to have prompted speculation as to the nature of
the Universe almost as an automatic concommitent.  One aspect of
these speculations related to the size and distribution of the sun
and its planets.  In classical times there were two schools of
thought about the structure of the solar system.  One placed the
Earth at the centre of the Universe with the Sun, Moon and the
then known planets revolving around it, the geocentric hypothesis,
and the other had the Earth and the planets revolving about the
Sun, the heliocentric hypothesis. The latter was proposed about
250 BC by Aristarchos of Samos.  The former was the majority
opinion of classical philosophers, endorsed by one of the more
influential opinion-formers of that period, Aristotle, and adopted
by the computer of orbits, Claudius Ptolemaos.  Aristarchos found
no support until the present era.  The dominance of Scholasticism
in the Medi\ae val Ages and its dominance by Aristotlean Physics
in the profane realm meant that the hypothesis of Aristarchos was
known to very few by the time of the Renaissance.  The
heliocentric hypothesis was revived by the Pole, Nicolas
Copernicus, in his book {\it Revolutionibus Orbium Coelestium}
published in 1543.  Evidently Copernicus knew of the work of
Aristarchos of Samos, but the reference was deleted from the first
edition, apparently by its theologically cautious publisher, the
Lutheran Andrias Osiander \cite[p.~319]{Dreyer 53}.  So
effectively was the contribution of Aristarchos neglected that we
find the chronicler of his work, Sir Thomas Heath, moved to
entitle the book {\it Aristarchus of Samos: the ancient
Copernicus} \cite{Heath} and we find the historian of Mechanics,
Dugas \cite[p.~83]{Dugas 88}, writing  `From Antiquity there had
been writers whose opinions were similar to those of Copernicus'!
In an attempt to counteract such neglect the municipality of the
town of Karlovassi on the island of Samos erected a bust of the
island's second most famous son in the main square some 200 m from
where this is written with a legend (in English; one assumes that
the Greeks would not be so ignorant) indicating that Copernicus
was an 1800 year Johnny-come-lately.

The timing of Copernicus was not perfect.  The religious turmoil which was to
shake Europe for a century and a half had commenced about a quarter of a
century before the book was published.  Not only did physicists, entirely
entangled in Aristotlean ideas, take exception to the impossible strains
placed upon the fabric of the Earth and its environment, but the theologians
-- particularly on the reformed side -- categorically rejected the theory of
the motion of the Earth because they considered it contrary to Holy Writ
\cite[p.~22]{Caspar}.

The eventual triumph of the heliocentric hypothesis
can reasonably be attributed to the work of Kepler who, painstakingly
analysing his observations and those of his predecessor as Imperial
Mathematician at Prague, Tycho Brahe, endowed the hypothesis with three
empirical laws.  In the books {\it Astronomia Nova} of 1609
\cite{Kepler 1} and {\it Harmonices Mundi} of 1619~\cite{Kepler 2} Kepler proposed
\begin{description}
\itemsep0mm

\item[]{\it Itaque plane hoc est: orbita planet\ae\ non est
circulus, sed ingrediens
ad latera utraque paulatim, iterumque ad circuli amplitudinem in perig\ae o
exiens: cujusmodi figuram itineris ovalem appellitant.  \ldots orbitam
planet\ae\ non esse circulum, sed figura ovalis.}
\cite[pp.~336--337]{Kepler 1}\footnote{Kepler's First Law:  `Therefore this is
obvious: the path of the planet is not a circle, but gradually curves inwards
on both sides and again departs to the width of the orbit at perigee: such a
path is called an oval figure.  \ldots the orbit of the planet is not a circle
but an oval figure.'}
\item[]{\it Cumque scirem, infinita esse puncta eccentrici, et distantias earum
infinitas, subiit, in plano eccentrici has distantias omnes inesse.  Nam
memineram, sic olim et Archimedem, cum circumferenti\ae\ proportionem ad
diametrum qu\ae reret, circulum in infinita triangula dissecuisse
\ldots} \cite[p.~321]{Kepler 1}\footnote{Kepler's Second Law:  `Since I was
aware that there exists an infinite number of points on the orbit and
accordingly an infinite number of distances [from the Sun] it occurred to me
that the sum of these distances is contained in the area of the orbit.  For
I remembered that in the same manner in times past, Archimedes too divided
[the area] of a circle, for which he found the circumference proportional
to the diameter, into an infinite number of triangles \ldots'}
\item[] {\it Sed res est certissima, quod proportio, qu\ae\ est inter binorum
quorumcunque planetarum tempora periodica, sit pr\ae cise sesquialtera
proportionis mediarum distantiarum, id est orbium ipsorum \ldots}
\cite[p.~279]{Kepler 2}\footnote{Kepler's Third Law:  `But it is absolutely
certain and exact that the ratio which exists between the periodic times of
any two planets is precisely the ratio of the 3/2th power of the mean
distances, \ie of the spheres themselves \ldots'}
\end{description}

Johannes Kepler had presented simple quantitative facts about the orbits of
the planets.  In August, 1684, Edmund Halley visited Issac Newton at Trinity
College, Cambridge, and informed him of Christopher Wren's challenge, to which
there was attached a sizeable monetary reward, to derive the shape of the
orbits of the planets about the sun.  Newton described his solution in the
tract {\it De Motu Corporum\,} and set about improving and expanding his ideas.
During this period he proposed his law of universal gravitation and his three
laws of motion which provided the first quantitative explanation of the motion
of visible bodies.  In 1686 these results were published in Newton's
masterpiece, {\it Philosophi\ae\ Naturalis Principia Mathematica}, which has
been claimed to mark the birth of modern science \cite[p.~3]{Gorringet}, if
one can isolate one moment in a period of evolution.

In the {\it Principia\,}
Newton attempted to verify Kepler's First Law by proving that, if
a particle moves on an elliptical orbit under the influence of a centripetal
force directed towards one focus, that force must be inversely proportional
to the square of the radius \cite[Prop.~XI, p.~56]{Newton}.  He obtained the
same results for focus-centred parabolic \cite[Prop.~XIII, p.~60]{Newton}
and hyperbolic motion
\cite[Prop.~XII, p.~57]{Newton}.  Newton proved that equal areas were swept out
by the radius vector in equal times \cite[Propp.~I--II, p.~40ff]{Newton}, which
was the content of Kepler's Second Law, and also that the square of the
periodic time of an inverse square law ellipse was proportional to the cube of
the semimajor axis length, thereby verifying Kepler's Third Law
\cite[Prop.~XV, p.~62]{Newton}.

Newton is also generally credited with having proved the converse of
Propositions XI-XIII, \ie that all possible orbits of a particle moving under
an inverse square law force are conic sections.  This is of greater
significance than the proof of the Propositions above since this result verifies
Newton's law of universal gravitation by showing that elliptical planetary
orbits are a natural consequence of the gravitational interaction between
the Sun and the planets.  The validity of this credit has been questioned
by Weinstock \cite{Weinstock 82, Weinstock 89} who argues that proof that
motion along a conic section orbit under the influence of a force directed
towards one focus implies an inverse square law force does not demonstrate
that a~particle moving under the influence of an inverse square law force
will move along a conic section orbit.  The conic section orbit, which was
assumed initially, is only a geometrical construction and not necessarily
the actual path described by the particle.

The initial approach was to determine the gravitational force law
assuming that the planets move along conic sections.  In 1710
Ermanno\footnote{Ermanno made a considerable impact on Italian
Mathematics during his brief sojourn at Padova (1707--1713) and
his place in history is based more on that.  For an account of his
work and extracts of many original letters and documents see the
work of Mazzone and Poero~\cite{Mazzone 97}.}, a student of Johann
Bernoulli, used the new techniques of the Leibnizian calculus to
find directly the orbits given an inverse square law force.
Ermanno obtained a constant of integration related to the
eccentricity of the conic section~\cite{Ermanno 1} and summarised
his results in a letter~\cite{Ermanno 2} to Johann Bernoulli who
proceeded to generalise the results to allow for an arbitrary
orientation of the orbit in the
plane~\cite{Bernoulli}\footnote{Some aspects of the Kepler Problem
in the period from Newton to Johann Bernoulli are to be found in
Speiser~\cite{Speiser 96}.}.  The Ermanno--Bernoulli constants are
the equivalents of the conserved vector generally known as the
Laplace--Runge--Lenz vector \cite{Laplace, Runge, Lenz}.  Almost
90 years later in his {\it Trait\'e de M\'ecanique
C\'eleste}~\cite{Laplace} Laplace obtained seven first integrals
for the Kepler problem.  These were the three cartesian components
of the angular momentum, the three cartesian components of the
Laplace--Runge--Lenz vector and what was essentially the total
energy.  Since only five of these autonomous functions could be
independent, Laplace determined the two relationships between the
seven integrals.  He showed that the conservation of the angular
momentum implied that the orbit was planar and that the orbit
equation could be expressed as a conic section~\cite{Goldstein
75}.

In July 1845 Hamilton presented to the Royal Irish Academy a new constant
of the motion for the Kepler problem which he subsequently named the
`eccentricity' vector~\cite{Hamilton 45} and which is commonly called
Hamilton's vector.  In another of the ironies of science the devisor of
vector notation, Josiah Gibbs, is believed to have been the first to express
the classical Laplace--Runge--Lenz vector in terms of modern vector notation,
thereby antedating Runge's derivation by roughly twenty years~\cite{Goldstein 76}.

In 1785 Charles Coulomb discovered the inverse square law of the electrostatic
force.  In 1913 Neils Bohr used the Kepler--Coulomb potential energy for a
Rutherford model of the hydrogen atom to obtain a theoretical formula for
the wavelengths of the atomic spectrum which agreed with experimental
observations.  The Laplace--Runge--Lenz vector\footnote{Known amongst physicists
as the Lenz vector following the 1924 paper of Lenz~\cite{Lenz} or the
Runge--Lenz vector since Lenz had referred to the textbook on vector analysis
by Runge~\cite{Runge}, the contribution by Laplace has been noticed in the
title of the vector only in recent decades.  The lack of recognition of
Ermanno and Bernoulli parallels that of Aristarchos and makes one wonder
if there is some sinister influence for those who are initiators in the
field of the motions of the planets.} came to prominence with the mathematical
development of Quantum Mechanics.  In 1926 Pauli~\cite{Pauli} used the vector
to derive an expression for the energy levels of the hydrogen atom using the
new techniques of matrix mechanics.  In 1935 Fock~\cite{Fock} showed that the
hydrogen atom possessed the four-dimensional rotation group for the bound
states and the Lorentz group for positive energy states.  These results were
confirmed by Bargmann~\cite{Bargmann} in 1936 when he calculated the
commutation relations between the components of the angular momentum and the
Laplace--Runge--Lenz vectors.

The characteristic feature of the Kepler problem is the existence of the
conserved vectors which enable one to determine the orbit easily.  In fact
what is potentially a very difficult nonlinear system can be trivially
integrated~\cite{Runge,Kustaanheimo, Collinson 73,Collinson 77} to obtain
the orbit and the only complexity is determining the time evolution of
system.  These conserved vectors play the same role for the Kepler problem as
do the tensorial first integrals -- the angular momentum tensor and the
Jauch--Hill--Fradkin tensor~\cite{Hill, Fradkin} -- for that other great
paradigm of Mechanics, the isotropic harmonic oscillator.  What is missing
for the Kepler problem are the explicitly time-dependent tensor invariants
which enable one easily to specify the trajectory in the configuration space
of the oscillator~\cite{Lemmer 96}.  The Kepler problem is not the only
problem in Mechanics for which there exists conserved vectors.  Already in
1896 Poincar\'e had identified a conserved vector with properties reminiscent
of angular momentum for the classical monopole problem~\cite{Poincare 96}.

The advent of the space age brought about some generalisations of
the Kepler problem to deal with the motion of low altitude
satellites, a model was proposed by Danby in 1962~\cite{Danby 62}.
Danby was able to provide an approximate solution.  In the early
Eighties Jezewski and Mittleman \cite{Jezewski 82, Jezewski 83}
were able to demonstrate the existence of first integrals for this
problem.  Subsequently many problems which could loosely be
regarded as relaxations of the Kepler problem were shown to
possess vector first integrals which were sufficient to provide
the orbit just as the angular momentum and Laplace--Runge--Lenz
vectors provided for the Kepler problem \cite{Gorringe 86,
Gorringe 87, Gorringe 88a,Gorringe 88b,Gorringe 89,Gorringe 91,
Gorringe 93a,Gorringe 93b,Leach 85,Leach 87a,Leach 87b, Leach 88,
Leach 90}.

In 1994 Krause \cite{Krause 94} used the Kepler problem as the vehicle for
the introduction of his concept of the complete symmetry group of a
differential equation or system of differential equations.  Essentially the
complete symmetry group of a differential equation is the group of the algebra
of the Lie symmetries required to specify completely the differential equation.
In the case of the Kepler problem the Lie symmetries required included nonlocal
symmetries although it is not the case for every differential
equation~\cite{Andriopoulos 00}.  In the subsequent investigations of the complete
symmetry groups of the Kepler and related problems by the technique of
reduction of order~\cite{Nucci 96b} by Nucci~\etal\ \cite{Nucci 2000} the
underlying symmetry was found to be the same for a number of these problems
and the critical role of the Ermanno--Bernoulli constants clearly delineated.

It is the purpose of this review to bring the details of these divers
problems together and to highlight the essential role played by certain
requirements on the equations of motion.  Our work may be divided into
three parts with the nature of the division being dependent upon the angular
momentum.  In the case of the Kepler problem the angular momentum is
conserved.  In Danby's model for the motion of a low altitude satellite the
magnitude of the angular momentum is no longer conserved.  Poincar\'e's comments
upon the experiment of Birkland led to the introduction of a generalisation
of the angular momentum as the conserved vector.  The magnitude of the angular
momentum was constant but not its direction.  Consequently we discuss the
three classes of problem, that of the conservation of angular momentum,
that of the conservation of the direction of angular momentum and that of
the conservation of the magnitude of angular momentum.  Finally we look to the
symmetry properties of these problems to reconcile the results.  In this we
must go beyond the point symmetries initially introduced by Lie~\cite{Lie:cg,
Lie:dgl} to the nonlocal symmetries introduced by Krause~\cite{Krause 94} to
construct the complete symmetry group of the Kepler problem and extended in
both number and scope by Nucci~\etal\ \cite{Nucci 2000}.

In the various generalisations of the
Kepler Problem which we treat in this paper we generally start with
an equation of motion which has a certain structure consistent
with the angular momentum model under consideration.  These equations
of motion contain arbitrary functions of, usually, unspecified variable dependence.
Part of the task of the analysis is to identify the nature of the variable dependence
in these functions which is possible for a conserved vector having the nature of the
Laplace--Runge--Lenz vector to exist.  Consequently the reader should understand
that the arguments of these arbitrary functions become evident during the process
of analysis.  Usually the function itself remains of arbitrary form.  Specific functions
are used by way of example.

The position taken here in respect to the identification of these vectors
as being generali\-sa\-tions of the Laplace--Runge--Lenz vector is based on
the notion of function rather than form.  The Laplace--Runge--Lenz
vector of the Kepler Problem has a specific functional structure.  In particular it
is quadratic in the components of the velocity.  Some of the conserved vectors found
here are very evidently not of that structure.  However, they serve the same function
for their equations of motion as the Laplace--Runge--Lenz
vector does for the equation of motion of the Kepler Problem.
All of these conserved vectors provide a simple route to the determination
of the equation of the orbit, \ie they perform the same function.
To argue that for a conserved vector to be associated with the
notion of the Laplace--Runge--Lenz vector means that the vector should
be quadratic in the components of the velocity is the equivalent of using
a sample of one to specify the properties of a~population, a practice
which is clearly statistically incorrect.

This preferential option for function over form has been given renewed
strength by the results of Nucci \etal\ \cite{Nucci 2000} discussed in Chapter~5.
There the various models treated in earlier chapters are seen to be reducible
to the same linear system so that the different types of conserved vectors
can in fact be regarded as different manifestations of the same object.

\section[Motions with conservation of the angular momentum
vector]{Motions with conservation\\ of the angular momentum vector}

\subsection[The central force problem $\ddot{\bfr} +f\bfr = 0$]{The central force problem $\boldsymbol{\ddot{\bfr} +f\bfr = 0}$}

\subsubsection[The conservation of the angular momentum vector, the Kepler\\ problem
and Kepler's three laws]{The conservation of the angular momentum vector,\\ the Kepler problem
and Kepler's three laws}

The techniques used for this and other problems to obtain conserved vectors and
scalars and equations describing the motion are natural extensions of the use
of scalar and vector products in the analysis of the reduced equation for the
Kepler problem (\cf the works of Runge \cite{Runge}, Bleuler and Kustaanheimo
\cite{ Kustaanheimo}, Collas \cite{Collas}, Collinson \cite{Collinson 73,
Collinson 77} and Pollard \cite{Pollard 76}), \viz
\begin{equation}
\ddot{\bfr} +\frac{\mu\bfr}{r^3} = 0,\label{1.1.0}
\end{equation}
in which the motion of the centre of mass of the system of two particles has
already been removed.  Such other equations as we shall study are to be
regarded as generalisations of this reduced equation,~\re{1.1.0}.  The
vector product of $\bfr $ with~\re{1.1.0} immediately gives the conservation
of the angular momentum vector
\begin{equation}
\bfL := \bfr\times\dot{\bfr},\label{1.1.1}
\end{equation}
from which we factor the mass constant.
 Since $\bfL $ is perpendicular to both $\bfr $ and $\dot{\bfr} $
for all time, the constancy of $\bfL $ implies that the orbit is confined to a plane.

In the spirit of using vectorial methods to obtain the Laplace--Runge--Lenz
and Hamilton's vectors there are two possible routes.  For one
of them we take the vector product of~\re{1.1.0} with $\bfL $ and the relationship
\begin{equation}
{\bfr}\times\bfL = -r^3\dot{\hat{\bfr}},\label{1.1.2}
\end{equation}
where $\hat{\bfr}$ denotes the unit vector in the direction of $\bfr$,
to obtain
\begin{align}
0 & =  \ddot{\bfr}\times\bfL -\mu\dot{\hat{\bfr}}\nonumber\\
\Rightarrow \bfJ & =  \dot{\bfr}\times\bfL -\mu\hat{\bfr},\label{1.1.3}
\end{align}
where $\bfJ $ is the Laplace--Runge--Lenz vector.  Hamilton's vector is obtained as
\begin{align}
\bfK & = \frac{1}{L^2}\(\bfL\times\bfJ\)\nonumber\\
& = \dot{\bfr} +\frac{\mu r^2\dot{\hat{\bfr}}}{L^2},\label{1.1.4}
\end{align}
where $\hat{\bfr}\times\bfL$ is $r^2\dot{\hat{\bfr}}$ and
$L $ is the constant magnitude of $\bfL $.  For the second route
we make use of the fact that the orbit is in a plane to specify plane
polar coordinates, $(r,\theta) $, in that plane so that the three unit
vectors, $\hat{\bfr} $, $\bht $ and $\hat{\bfL} $ have the properties that
\begin{align}
& \hat{\bfr}\times\bht  = \hat{\bfL}, &&\bht\times\hat{\bfL} =\hat{\bfr},& &
\hat{\bfL}\times\hat{\bfr} =\bht,&&\nonumber\\
& \dm{\hat{\bfr}}{t} =\dot{\theta}\bht, &&\dm{\bht}{t} = -\dot{\theta}\hat{\bfr}, && L
=r^2\dot{\theta}.&&
\label{1.1.5}
\end{align}
Then \re{1.1.0} can be written as
\begin{align}
0 & = \ddot{\bfr} +\frac{\mu}{L}\dot{\theta}\hat{\bfr}\nonumber\\
\Rightarrow \bfK & = \dot{\bfr} -\frac{\mu}{L}\bht.\label{1.1.6}
\end{align}
The two routes are reconciled when we express \re{1.1.4} in terms of the plane
polar coordinates, which the constancy of the angular momentum vector enables us to do.

From the integration of the scalar product of \re{1.1.0} with $\dot{\bfr} $
we obtain the conserved scalar
\begin{equation}
E: = \half\dot{\bfr}\cdot \dot{\bfr} - \frac{\mu}{r}\label{1.1.7}
\end{equation}
which is the mechanical energy of the system scaled
by the mass.  From the scalar products of \re{1.1.6} and \re{1.1.3} with
themselves we obtain the equivalent relationships
\begin{align}
K^2 & =  2E+\frac{\mu^2}{L^2}\label{1.1.8}\\
J^2 & =  2L^2E+\mu^2,\label{1.1.9}
\end{align}
which give two of the constraints necessary for the ten first integrals, $\bfJ$,
$\bfK$, $\bfL$ and $E$.  The other constraint is the mutual orthogonality of
the three conserved vectors.

Finally, if we choose the conserved vector, $\bfJ $,
as the reference line for the measurement of the polar angle, $\theta $,
the scalar product of $\bfr $ and $\bfJ $ gives
\begin{align}
\bfr\cdot \bfJ =rJ\cos\theta & = \bfr\cdot (\dot{\bfr}\times\bfL) -\mu \bfr\cdot \hat{\bfr}\nonumber\\
\Rightarrow r & = \frac{L^2}{\mu +J\cos\theta}\label{1.1.10}
\end{align}
which is the polar equation for a conic section with the origin at one focus.
Thus we have the orbit for the motion governed by the equation \re{1.1.0}.
The time evolution of the system may be obtained, for example, by means
of the constancy of the magnitude of the angular momentum for then
we have that the relationship
$L =r^2\dot{\theta} $ can be combined with \re{1.1.10} to give
\begin{equation}
t-t_0 =\int_{\theta_0}^{\theta}\frac{L^3\d\theta}{\(\mu +J\cos\theta\)^2}\label{1.1.11}
\end{equation}
which is moderately simple to evaluate.  In the constancy of the vector
of angular momentum and
the specific equation for the orbit we have two of Kepler's three laws.  For
a~closed orbit the third law follows from the evaluation of \re{1.1.11}
over an orbit to give the period, $T $, as (\cite[p.~148, 2.553.3 and 2.554.3]{GAR})
\begin{equation}
T = \frac{2\pi\mu L^3}{\(\mu^2-J^2\)^{3/2}} =\frac{2\pi\mu}{(- 2E)^{3/2}}\label{1.1.12}
\end{equation}
and the use of \re{1.1.9} and \re{1.1.10} to relate
the semimajor axis length, $R $, to the energy, $E $, according to
\begin{equation}
2ER +\mu = 0\quad\Leftrightarrow\quad R =\frac{\mu}{- 2E}.\label{1.1.13}
\end{equation}
Consequently we obtain
\begin{equation}
\frac{T^2}{R^3} =\frac{2\pi}{\mu}\label{1.1.14}
\end{equation}
in a way somewhat different compared with the manner in
which Kepler obtain the result almost 400 years ago.

It is unfortunate that not all can be calculated for the Kepler problem by
these elementary methods.  The evolution of the motion in time is not so
simple to calculate\footnote{A difficulty recognised very early in the study of
the Kepler Problem through the solution of Kepler's equation, \cf
\cite{Dutka 95} and \cite{Colwell 92}, for which see below.}.  In~\re{1.1.11}
we have the
equation to be integrated to obtain the evolution of the polar
angle, $\theta $, in time.  If we differentiate the orbit equation,
\re{1.1.10}, with respect to time, we obtain
\begin{equation}
\d t =\frac{L\d r}{J\sin\theta}.\label{1.5.16}
\end{equation}
We may eliminate $\sin\theta $ by using the orbit equation, \re{1.1.11}, to write
\begin{equation}
\sin\theta =\frac{1}{Jr}\((J^2-\mu^2)r^2+ 2L^2\mu r-L^4\)^{\half}\label{1.5.17}
\end{equation}
so that \re{1.5.16} becomes
\begin{equation}
\d t =\frac{r\d r}{\((J^2-\mu^2)r^2/L^2+ 2\mu r-L^2\)^{\half}}.\label{1.5.18}
\end{equation}
Naturally, if we use \re{1.1.9}, \re{1.5.18} becomes the same
equation as we would obtain from the energy expression after
the substitution $\dot{\theta} =L/r^2 $.  When we make this
substitution and perform the integration (\cite[2.264.2 and 2.261.3]{GAR}), we obtain
\begin{equation}
t-t_0 =\frac{\left(2Er^2+ 2\mu r-L^2\right)^{\half}}{2E} +\frac{\mu}{2E (- 2E)^{\half}}
\arcsin\left.
\(\frac{2Er +\mu}{(2EL^2+\mu^2)^{\half}}\)\right|_{r_0}^r.\label{1.5.45}
\end{equation}
Although it is a simple matter to recover Kepler's Third Law from this,
the inversion to obtain an explicit function, $r (t) $, is not globally possible.
The integration of \re{1.1.11} gives an equally unpleasant expression
\cite[2.554.3 and 2.553.3]{GAR}  for the relationship between $\theta $ and $t $, \viz
\begin{equation}
t-t_0 = \frac{L}{2E}\(\frac{J\sin\theta}{\mu +J\cos\theta} -\frac{2\mu}{L (- 2E)^{\half}}
\arctan\frac{L (- 2E)^{\half}\tan\half\theta}{\mu +J}\).\label{1.5.50}
\end{equation}
The same comment applies.

There is another approach which can be taken and which dates back
for several centuries.  The standard equation for a conic section in polar coordinates is
\begin{equation}
r = \frac{1}{c\(1+e\cos\theta\)},\label{1.5.51}
\end{equation}
where $c $ and the eccentricity, $e $, are constants.  It is evident that
\begin{equation}
e =\frac{J}{\mu}\label{1.5.52}
\end{equation}
and so we may write
\begin{equation}
L^2 =\mu R \left(1-e^2\right).\label{1.5.53}
\end{equation}
consequently \re{1.5.18} can be rewritten as
\begin{equation}
\d t =\frac{r\d r}{\(-\mu r^2/R+ 2\mu r-\mu R (1-e^2)\)^{\half}}.\label{1.5.54}
\end{equation}
The introduction of another angular variable, $\psi $, which is known as the
eccentric anomaly (in contrast to $\theta $ which has been known as
the true anomaly since medi\ae val times \cite[p.~99]{Goldstein 80}), through the substitution
\begin{equation}
r = R (1- e\cos\psi)\label{1.5.55}
\end{equation}
simplifies \re{1.5.54} to
\begin{equation}
\d t = \(\frac{R^3}{\mu}\)^{\half} (1- e\cos\psi)\d \psi\label{1.5.56}
\end{equation}
which is trivially integrated to give
\begin{equation}
t = \(\frac{R^3}{\mu}\)^{\half} (\psi - e\sin\psi), \label{1.5.57}
\end{equation}
where the lower limits of integration have been taken to be zero.
When \re{1.5.57} is evaluated at $2\pi $, Kepler's Third Law is recovered.
Using this we may write \re{1.5.57} as
\begin{equation}
\omega t =\frac{2\pi}{T}t =\psi - e\sin\psi,\label{1.5.58}
\end{equation}
which is known as Kepler's equation.  The quantity $\omega t $ varies
between $0 $ and~$2\pi $ as do~$\psi $ and~$\theta $
and is known as the mean anomaly.  This equation is useful in the process of
calculating~$r $ and~$\theta $ at a given time.  It is solved to give the value
of~$\psi $ at that time.  From \re{1.5.55} one obtains~$r $.
 By a manipulation of the orbit equation and \re{1.5.55} one obtains
\begin{equation}
\tan\half\theta =\(\frac{1+e}{1-e}\)^{\half}\tan\half\psi.\label{1.5.62}
\end{equation}
Before the days of computers the solution of Kepler's equation
attracted much attention for accurate astronomical calculations.

Many treatments dealing with the Laplace--Runge--Lenz vector have been
criticised as being too {\it ad hoc}
\cite{Kaplan 86} in their derivations.  Apart from the elegant simplicity of the
above discussion, the same could be said of the method used here.  However,
in the sequel we see that the same techniques {\it mutatis mutandis\,} may be
applied to a wide variety of problems and so, by this generality of application,
are free from the taint of being too
{\it ad hoc}.  Certainly more sophisticated methods can be used to obtain the same results.
Here we prefer to establish the classes of systems by elementary methods and then
use the more advanced techniques to analyse the natures of these different
classes of problems.

\subsubsection[The model equation $\ddot{\bfr} +f\bfr= 0$]{The model equation $\boldsymbol{\ddot{\bfr} +f\bfr= 0}$}

The central force equation
\begin{equation}
\ddot{\bfr} +f\bfr = 0,\label{1.1.15}
\end{equation}
where the variables in $f$ are specified below,
can formally be integrated to give a conserved vector of Hamilton's type, \viz
\begin{equation}
\bfK =\dot{\bfr} +\int f\bfr\d t.\label{1.1.16}
\end{equation}
The formality is removed if the integration in \re{1.1.16}
can be rendered meaningful without the solution of the original
equation~\re{1.1.15} as an explicit function of time.
The conservation of the angular momentum leads to a conserved vector
of Laplace--Runge--Lenz type and immediately
we can make sense of \re{1.1.16}.  More usefully
 for the moment we can make use of the planar nature of the
motion to express everything in terms of plane polar coordinates.
For the moment we measure the polar angle, $\theta $, from the cartesian
direction,~$\bhi $.  Then we can use the relations
\begin{equation}
\hat{\bfr} =\bhi\cos\theta +\bhj\sin\theta,\qquad
\bht = -\bhi\sin\theta +\bhj\cos\theta\label{1.1.17}
\end{equation}
between the variable unit vectors in the polar coordinates and the constant
unit vectors in cartesian coordinates.

In central force problems the conservation of angular momentum,
hence the confinement of the motion to a plane, enables one to use
 the relationship $\dot{\theta} =L/r^2 $ to pass from an integration
with respect to time to an integration with respect to polar angle.  This
technique is well-established \cite[p.~80]{Whittaker 44},
\cite[Chap.~3]{Goldstein 80} and \cite[Chap.~3]{Symon}.  To reduce
the formal integral in \re{1.1.16} to a quadrature we make the {\it Ansatz}
\begin{equation}
fr =v (\theta)\dot{\theta},\label{1.1.18}
\end{equation}
where $v (\theta) $ is an arbitrary function of its argument.  (The only
mathematical requirement on $v $ is that it be Riemann integrable.) With the
substitution \re{1.1.18} we have
\begin{equation}
\int f\bfr\d t = \int v(\theta)\hat{\bfr}\d\theta. \label{1.1.19a}
\end{equation}
Using the cartesian decomposition of the unit vectors in plane polar
coordinates in terms of the cartesian unit vectors and their inverses, {\it
videlicet}
\begin{gather}
\hat{\bfr} = \bhi\cos\theta+\bhj\sin\theta, \qquad
\bht = -\bhi\sin\theta+\bhj\cos\theta,\nonumber\\
\bhi = \hat{\bfr}\cos\theta-\bht\sin\theta,\qquad
\bhj = \hat{\bfr}\sin\theta+\bht\cos\theta
\label{1.1.18b}
\end{gather}
we may write \re{1.1.19a} as
\begin{gather}
\int f\bfr\d t = \bhi\int \cos\theta v(\theta)\d\theta
+  \bhj\int \sin\theta v(\theta)\d\theta\nonumber\\
\phantom{\int f\bfr\d t}{}= \hat{\bfr}\lb\cos\theta\int_{\theta_0}^{\theta}\cos\eta v(\eta)\d\eta
+ \sin\theta\int_{\theta_0}^{\theta}\sin\eta v(\eta)\d\eta\rb \nonumber\\
\phantom{\int f\bfr\d t=}{}+ \bht\lb-\sin\theta\int_{\theta_0}^{\theta}\cos\eta v(\eta)\d\eta
+ \cos\theta\int_{\theta_0}^{\theta}\sin\eta v(\eta)\d\eta\rb   \nonumber\\
\phantom{\int f\bfr\d t}{}= \hat{\bfr}z'(\theta) -\bht z (\theta),\label{1.1.19}
\end{gather}
where
\begin{gather}
z (\theta)  =  \int_{\theta_0}^{\theta}v (\eta)\sin (\theta -\eta)\d \eta\nonumber\\
z' (\theta)  =  \int_{\theta_0}^{\theta}v (\eta)\cos
(\theta -\eta)\d \eta,\label{1.1.20}
\end{gather}
and consequently we may write $z (\theta) $ in terms of the differential equation
\begin{equation}
z''(\theta) + z (\theta) = v (\theta)\label{1.1.21}
\end{equation}
subject to the initial conditions $z (\theta_0) = 0 $
and $z'(\theta_0) = 0 $.  Hence we have the two conserved vectors
\begin{gather}
\bfK  =\dot{\bfr} +z'(\theta)\hat{\bfr} -z (\theta)\bht\label{1.1.22}\\
\bfJ  = \dot{\bfr}\times\hat{\bfL} -z (\theta)\hat{\bfr} -z'(\theta)\bht,\label{1.1.23}
\end{gather}
generalisations of Hamilton's vector and the Laplace--Runge--Lenz vector
respectively.

Now that we have the Ermanno--Bernoulli constants in terms of the vector first
integ\-ral,~$\bfJ $, the equation of the orbit in the plane is obtained by taking the
scalar product of~$\bfr $ with $\bfJ $.  Unlike the case of the Kepler problem treated
in the previous subsection here we have not assumed that $\theta $ is measured from
$\bfJ $ but from a standard cartesian direction denoted by $\bhi$.  We
take the angle between $\bfJ $ and $\bhi$ to be $\theta_0 $ so that the equation
of the orbit in the plane is given by
\begin{equation}
r (\theta) =\frac{L}{z (\theta) +J\cos (\theta -\theta_0)}.\label{1.2.9}
\end{equation}

The evolution of the motion in time can be obtained by using the conservation
of angular momentum to give $t $ as a function of $\theta $ or by using the
equation of the orbit, \re{1.2.9}, to give~$t $ as a function of $r $.  In
either case the ease, or more likely lack of ease, in evaluating these
functions will depend upon the particular function $v (\theta) $, which one
assumes is dictated by physical considerations.  Supposing that the quadratures
can be performed we then have the problem of inversion to find the dependent
variables~$r $ and/or~$\theta $ as explicit functions of time.  Differentiating \re{1.2.9}
with respect to time we obtain
\begin{equation}
\dot{r} = -z'+J\sin (\theta -\theta_0).  \label{1.3.1}
\end{equation}
We can evaluate $z'(\theta) $ from \re{1.1.20} and, provided
all terms involving $\theta $ can be expressed in terms of~$r $,
the relationship between~$t $ and~$r $ is given by
\begin{equation}
t =\int\frac{\d r}{\(-z'(\theta (r)) +J\sin (\theta (r) -\theta_0)\)}.\label{1.3.3}
\end{equation}
The angular momentum route, the more familiar one, leads to
\begin{equation}
t =\int\frac{L\d\theta}{\(z (\theta) +J\cos (\theta -\theta_0)\)^2}.\label{1.3.7}
\end{equation}

Even though it may not be possible to evaluate these integrals in closed
form or, if this be possible, to invert the results to give the
dependent variables in terms of the independent variable~$t $,
one does know that the results of numerical integration
can be trusted since the reduction of the problem to
explicit quadratures removes the possibility of chaos.

Since the angular momentum is conserved,
the areal velocity is a constant.  In fact the area swept out by the radial
vector~$\bfr $ in a time~$\d t $ is given by the element of area
\begin{equation}
\d A =\half r (r\d\theta).\label{1.3.8}
\end{equation}
Consequently the areal velocity is given by
\begin{equation}
\dm{A}{t} =\half L\label{1.3.9}
\end{equation}
and, since $L $ is a constant
for the class of problems being considered here, \re{1.3.9} is integrated to give
\begin{equation}
A =\half Lt.\label{1.3.10}
\end{equation}
In the case that the orbit for a given choice of $v (\theta) $
has nicely defined geometric properties there may be some
interest in relating \re{1.3.10} to these geometric properties,
for example the semimajor axis length in the case of an elliptical orbit and the
period which would arise from the area, $A $, in \re{1.3.10}
assuming a multiple of the area enclosed by the orbit.  In general
one would not expect to be able to make a satisfactory relationship
between the geometrical features of the orbit and the time.

We consider some examples.
\begin{example}
The equation of motion
\begin{equation}
\ddot{\bfr} +\frac{a\(\theta -\theta_0\) +b}{r^3}\bfr = 0,\label{1.4.1}
\end{equation}
where $a$, $b$ and $\theta_0$ are constants,
represents a force of the Kepler type with an additional inverse
square law term which depends upon the polar angle~$\theta $.
The quadrature in \re{1.1.20} may be performed to give
\begin{gather}
z (\theta)  =  \frac{a}{L}\(\theta -\theta_0-\sin (\theta -\theta_0)\)
+\frac{b}{L}\(1-\cos (\theta -\theta_0)\),\label{1.4.3}\\
z'(\theta)  = \frac{a}{L}\(1-\cos (\theta -\theta_0)\) +\frac{b}{L}\sin (\theta -\theta_0).\label{1.4 24}
\end{gather}
The expressions for the conserved vectors and the orbit equation followed immediately.
They are rather complicated and are not reproduced here.  For increasing~$\theta $
the attraction to the centre increases and the orbit spirals inwards
in a counterclockwise direction towards the origin.
\end{example}
\begin{example}
By way of contrast the equation of motion
\begin{equation}
\ddot{\bfr} +\frac{a\e^{- (\theta -\theta_0)} +b}{r^3}\bfr = 0,\label{1.4.5}
\end{equation}
for which, again, $a$, $b$ and $\theta_0$ are constants,
contains a force of Kepler type with an additional inverse
square law term which decays exponentially with increasing polar angle~$\theta $.
In this case we have
\begin{gather}
z (\theta)  = \frac{a}{2L}\(\e^{- (\theta -\theta_0)} +\sin (\theta -\theta_0) -\cos (\theta -\theta_0)\)
+\frac{b}{L}\(1-\cos (\theta -\theta_0)\),\label{1.4.7}\\
z'(\theta) = \frac{a}{2L}\(-\e^{-(\theta -\theta_0)} +
\sin (\theta -\theta_0) +\cos (\theta -\theta_0)\)
+\frac{b}{L}\sin (\theta -\theta_0).\label{1.4.8}
\end{gather}
The orbit contracts quickly to what appears to be a limit cycle as $\theta $ increases.
\end{example}
\begin{example}
For the equation of motion
\begin{equation}
\ddot{\bfr} +\frac{a\sin 3 (\theta -\theta_0) +b}{r^3}\bfr = 0,\label{1.4.9}
\end{equation}
which couples a force of Kepler type with an additional inverse square law
term depending
periodically on the polar angle $\theta $, we obtain
\begin{gather}
z (\theta)  = \frac{a}{L}\(-\oei\sin 3 (\theta -\theta_0) +
\tei\sin (\theta -\theta_0)\)
+\frac{b}{L}\(1-\cos (\theta -\theta_0)\),\nonumber\\
z'(\theta) = \frac{a}{L}\(-\tei\cos 3 (\theta -\theta_0) +\tei\cos (\theta -\theta_0)\)
+\frac{b}{L}\sin (\theta -\theta_0).\label{1.4.12}
\end{gather}
The periodicity of the central force is reflected in the
orbit which has three lobes spaced at intervals of $2\pi/3 $ in $\theta $.
\end{example}
\begin{example}
In the present context the equation for the Kepler problem, \re{1.1.0}, represents the choice
\begin{equation}
v (\theta) =\frac{\mu}{L}\label{1.5.2}
\end{equation}
for which
\begin{gather}
z (\theta)  = \frac{\mu}{L}\(1-\cos (\theta -\theta_0)\),\label{1.5.3}\\
z'(\theta) = \frac{\mu}{L}\sin (\theta -\theta_0).\label{1.5.4}
\end{gather}
The conserved vectors are
\begin{gather}
\bfK  =  \dot{\bfr} -\frac{\mu}{L}\(\bht -\bht_0\),\label{1.5.5}\\
\bfJ  =  \dot{\bfr}\times\hat{\bfL} -\frac{\mu}{L}\(\hat{\bfr} -\hat{\bfr}_0\).\label{1.5.6}
\end{gather}
The expressions given here differ from the standard expressions by a factor of
$L $ in $\bfJ $ and the lack of identification of the reference line for
$\theta $ with $\bfJ $.  When he obtained his so-called `eccentricity vector',
Hamilton~\cite{Hamilton 45} immediately recognised that the velocity hodograph
for the Kepler problem was a circle with its centre given by Hamilton's vector
(putting $\bht_0 = 0 $) and radius of length $\mu/L $.  The velocity hodograph
is generally attributed to Hamilton, but his biographer gives the credit to
M\"obius
\cite[p.~333]{Hankins}\footnote{In the nineteenth century the velocity
hodograph attracted a considerable amount of attention for which see the works
of Tait~\cite{Tait}, Thompson (later Lord Kelvin) and Tait
\cite{Kelvin 94,Kelvin 23}, Kelland and Tait \cite{ Kelland}, Maxwell
\cite{Maxwell} and Routh \cite{Routh}.  The notation for the conserved
vectors makes the equation of the circle obvious.  The first known
representation of the Laplace--Runge--Lenz vector in modern vector notation
is found in the text of Gibbs and Wilson~\cite{Gibbs}.  The transparency
of the present representation makes for a dramatic contrast for the original
representation to be found in the papers of Ermanno and Bernoulli
\cite{Ermanno 1, Ermanno 2,Bernoulli}.  Early in the twentieth century
Child~\cite{Child} made a complete graphical study of the velocity hodograph
for the three possible orbits of the Kepler problem.  There appear to be some
inconsistencies in the depictions which are possibly due to roundoff errors
in the numerical calculations which may well have been done on the back of
an envelope.  More recently the velocity hodograph has been discussed by
Goldstein~\cite{Goldstein 76, Goldstein 80} and Stickforth~\cite{Stickforth}.
Possibly the most detailed discussion of the velocity hodograph
as applied to the Kepler and related problems is to be found in the thesis
of Gorringe~\cite{Gorringet}.  Gorringe discusses also the acceleration hodograph.}.
\end{example}
\begin{example}
In the examples above $\dot{\theta}$ does not appear in the equation of motion.
In evaluating the function $z(\theta)$ one must use the relationship
$L=r^2\dot{\theta}$ to introduce the required $\dot{\theta}$ for the quadrature.
Here we present an elementary example in which $f$ is explicitly dependent
upon $\dot{\theta}$ in the equation of motion.  The equation of motion which we use is
\begin{equation}
\ddot{\bfr} + \frac{a\dot{\theta}}{r}\bfr = 0.
\end{equation}
We find that
\begin{equation}
z(\theta) = a\left(1 - \cos(\theta-\theta_0)\right)
\end{equation}
and that the equation of the orbit is
\begin{equation}
r = \frac{L}{a\left(1-\cos(\theta-\theta_0)\right)+ J\cos(\theta-\phi_0)},
\end{equation}
where $\phi_0$ is the angle the reference line makes with $\bfJ$.
The orbit is an ellipse, a~parabola or an hyperbola depending upon
the relative values of the constants.
\end{example}

\subsection[Algebraic properties of the first integrals
of the Kepler problem]{Algebraic properties of the first integrals of the Kepler problem}

The Hamiltonian of the Kepler problem described by equation \re{1.1.0} is
\begin{equation}
H = \half\bfp\cdot \bfp -\frac{\mu}{r},\label{1.5.7}
\end{equation}
where the canonical momentum is related to the velocity
by $\bfp =\dot{\bfr} $, and is simply the mechanical energy which is conserved.
In terms of canonical coordinates the conserved vectors are
\begin{gather}
\bfL  = \bfr\times\bfp,\label{1.5.8}\\
\bfK  = \bfp -\frac{\mu}{L}\bht,\label{1.5.9}\\
\bfJ  =\bfp\times\bfL -\mu\hat{\bfr}\label{1.5.10}
\end{gather}
and the components of all three vectors have zero Poisson Bracket
with the Hamiltonian.  Consequently, in quantum mechanical terms,
the physical quantities represented by the corresponding quantum
mechanical operators are simultaneously observable with the Hamiltonian.
Since the three vectors are not independent, a knowledge of two means
a~knowledge of three.  Normally the two vectors considered are the
angular momentum and the Laplace--Runge--Lenz\footnote{There
appears to be no reason why Hamilton's vector should not be used in place
of the Laplace--Runge--Lenz vector.
The use of the latter is possibly an accident of history.
We recall that in the early twentieth century the Ermanno--Bernoulli
constants had been `popularised' by the German writers Runge and Lenz.
A large contribution to the early development of Quantum Mechanics had
a Teutonic origin and it is possible that this recent work was more
familiar than the older work of the quaternion-fancying Irishman, Hamilton,
amongst German speakers.  The Hamiltonian formulation of Quantum Mechanics
by Dirac~\cite{Dirac} was probably too late to change the situation.}.
The Laplace--Runge--Lenz vector was used to explain the degeneracy
of the spectrum of the hydrogen atom and is described by physicists
as a~dynami\-cal symmetry in contrast to the obvious geometric
symmetry of the angular momentum.  The angular momentum
vector may be a more frequently occurring conserved quantity than
the Laplace--Runge--Lenz vector, but there is no reason to distinguish between the two.
Both vectors are a consequence of the equation of motion.

If one writes each vector in terms of cartesian components, \ie
 as $\bfA =A_i\hat{\bfe}_i $, the Poisson Bracket relationships are
\begin{align}
& \lb L_i,H\rb = 0, && \lb J_i,H\rb = 0,&&&\nonumber\\
& \lb L_i,L_j\rb = \varepsilon_{ijk}L_k, && \lb J_i,L_j\rb = \varepsilon_{ijk}J_k,&&& \nonumber\\
& &&\lb J_i,J_j\rb = - 2E\varepsilon_{ijk}L_k,&&&
\label{1.5.11}
\end{align}
where $\varepsilon_{ijk} $ is Kronecker's epsilon and $E $ is the value of the
Hamiltonian.  The three components of the angular momentum vector form the
compact rotation algebra $so (3) $.  The algebra of the two vectors is $so (4)
$, the representation of the four-dimensional compact rotation group,
for negative energies, $so (3,1) $, the representation of the four-dimensional noncompact
rotation group, \ie the Lorentz group,
for positive energies and $e (3) $, the representation of the
Euclidean group for zero energy.  A better representation
of the algebra for nonzero energy is given by defining
\begin{equation}
{\cal{J}}_{i\pm} = \frac{J_i}{|2E|^{1/2}}\label{1.5.12}
\end{equation}
for negative and positive energies respectively.  Then \re{1.5.11} is replaced by
\begin{align}
& \lb L_i,H\rb = 0, && \lb {\cal{ J}}_i,H\rb = 0,&&&\nonumber\\
& \lb L_i,L_j\rb = \varepsilon_{ijk}L_k,& & \lb {\cal{ J}}_i,L_j\rb =
\varepsilon_{ijk}{\cal{J}}_k,&&&\nonumber\\
&&&\lb {\cal{J}}_i,{\cal{J}}_j\rb = \pm\varepsilon_{ijk}L_k,&&&
\label{1.5.13}
\end{align}
for negative and positive energies respectively, which has an \ae sthetic
appeal greater than \re{1.5.11}.

\subsection[Extension of the model equation to nonautonomous
systems]{Extension of the model equation to nonautonomous systems}

In \S~2.2 the model equation \re{1.1.10} was tacitly assumed to be autonomous.
 However, Katzin and Levine~\cite{Katzin 83}, using a direct method,
and Leach~\cite{Leach 85}, using a variant of Noether's
theorem~\cite{Noether}, showed that the time-dependent central
force problem with the equation of motion
\begin{equation}
\ddot{\bfr} = F (r,t)\hat{\bfr}\label{1.10.1}
\end{equation}
possesses a time-dependent vector first integral of the form
\begin{equation}
\bfJ = U (r,t) (\bfL\times\dot{\bfr}) +Z (r,t) (\bfL\times\bfr) +W (r,t)\bfr,\label{1.10.2}
\end{equation}
provided
\begin{align}
& U =u (t)\neq 0, && Z = -\dot{u} (t),&&&\nonumber\\
& W = \dsp{\frac{\mu}{r}}, &&F = \(\dsp{\frac{\ddot{u}r}{u}}\) -
\(\dsp{\frac{\mu}{ur^2}}\).&&&
\label{1.10.3}
\end{align}
This system includes both a time-dependent Kepler problem ($\mu\neq 0 $ and
$u = (\alpha t+\beta)/\lambda $ with $\alpha $, $\beta $ and $\lambda $ being
constant -- this was the subject of an earlier study by Katzin and
 Levine~\cite{Katzin 82}) -- and a particular case of the time-dependent harmonic oscillator when
$\mu = 0 $\footnote{The time-dependent harmonic oscillator was solved in one
dimension by Lewis~\cite{ Lewis 67, Lewis 68} using Kruskal's asymptotic
method~\cite{Kruskal 62}.  Lewis and Riesenfeld~\cite{Lewis 69}
used the time-dependent harmonic oscillator to construct
an invariant for a charged particle moving in a time-dependent electromagnetic field.
Further techniques and results have been given by Lewis and
Leach~\cite{Lewis 82}, Moreira \cite{Moreira 83, Moreira 85}
and G\"unther and Leach~\cite{Leach 77}.}.  Previous techniques used
to solve this problem have been algebraically rather complex.  The techniques used here
have been reported in Gorringe and
 Leach~\cite{Gorringe 86, Gorringe 88a} and Leach and Gorringe~\cite{Leach 87b}.

We rewrite our model equation to include the possible dependence on time explicitly as
\begin{equation}
\ddot{\bfr} +f (r,\theta,t)\bfr = 0\label{1.10.4}
\end{equation}
of which the system above is a special case.  The conservation
of the vector of angular momentum, $\bfL
: =\bfr\times\dot{\bfr} $, follows from the vector product of $\bfr $ with \re{1.10.4}.  We seek
a~Hamilton-like vector for \re{1.10.4} through the use of an integrating factor, $g $.  Writing
\begin{equation}
g\ddot{\bfr} = (g\dot{\bfr} -\dot{g}\bfr)^{\dsp{{\bf .}}}
+\ddot{g}\bfr\label{1.10.6}
\end{equation}
and multiplying \re{1.10.4} by $g $ we obtain
\begin{equation}
(g\dot{\bfr} -\dot{g}\bfr)^{\dsp{{\bf .}}}  + (\ddot{g} +gf)\bfr = 0.\label{1.10.7}
\end{equation}
A second conserved vector exists if the second term
of \re{1.10.7} can be written as a total time derivative.
The presence of $\hat{\bfr} $ which is a function of $\theta $ signals the requirement that
\begin{equation}
(\ddot{g} +gf)r =v (\theta)\dot{\theta},\label{1.10.8}
\end{equation}
for some function $v (\theta) $.  The explicit dependence
upon $\dot{\theta} $ may be removed by using $L =
r^2\dot{\theta} $ and we then solve \re{1.10.8} to obtain
\begin{equation}
f =\frac{Lv (\theta)}{gr^3} -\frac{\ddot{g}}{g}.\label{1.10.10}
\end{equation}
Since $f $ was assumed a function of $r $, $\theta $
and $t $, $g =g (t) $.  The equation \re{1.10.4} is now
\begin{equation}
\ddot{\bfr} +\frac{Lv (\theta)}{g (t)r^3}\bfr -\frac{\ddot{g} (t)}{g (t)}\bfr = 0,\label{1.10.11}
\end{equation}
where $L $ is treated as a constant in the equation
of motion, although it is expressed as $r^2\dot{\theta} $ during calculations.

Direct integration of the equation of motion \re{1.10.11} yields the Hamilton-like vector
\begin{equation}
\bfK = g\dot{\bfr} -\dot{g}\bfr +z'(\theta)\hat{\bfr} -z (\theta)\bht,\label{1.10.12}
\end{equation}
where $z (\theta) $ and $z'(\theta) $ are given by \re{1.1.19} and \re{1.1.20}.
 The analogue of the
Laplace--Runge--Lenz vector is obtained by taking
the vector product of $\bfK $ with $\hat{\bfL} $ and is
\begin{equation}
\bfJ = (g\dot{\bfr} -\dot{g}\bfr)\times\hat{\bfL} -z (\theta)\hat{\bfr} -z'(\theta)\bht.\label{1.10.13}
\end{equation}
The presence of $\bfr $ in the Hamilton-like vector $\bfK $
precludes the possibility to express the velocity hodograph
in a simple form as was the case for the Kepler problem.
However, this can be partially rectified by a
time-dependent linear transformation of Kummer--Liouville
type~\cite{Kummer, Liouville} (in Hamiltonian Mechanics it is a time-dependent
linear generalised canonical transformation~\cite{Leach 77, Pillay 98}).

The scalar product of $\bfJ $ with $\bfr $ gives
\begin{equation}
r (\theta) =\frac{Lg (t)}{z (\theta) +J\cos (\theta -\theta_0)}.\label{1.11.1}
\end{equation}
Since \re{1.11.1} contains an explicit function of time, it is not
the required orbit equation.  We eliminate
the function of time to obtain the orbit equation.  From the expression
for the constant magnitude of the angular momentum we have
\begin{equation}
L =r^2\dm{\theta}{t} =\frac{L^2g^2 (t)}{\(z (\theta) +J\cos (\theta -\theta_0)\)^2}\dm{\theta}{t}
\label{1.11.2}
\end{equation}
which can be rearranged to give
\begin{equation}
\int_0^t\frac{\d t}{g^2 (t)} = \int_{\theta_0}^{\theta}\frac{L\d\theta}{\(z (\theta) +J\cos (\theta -\theta_0)
\)^2}.\label{1.11.3}
\end{equation}
Provided the integrands are continuous and nonzero we can, according to the Implicit Function
 Theorem~\cite[p.~165]{Brand}, write the result of the integration as
\begin{equation}
t =N (\theta)\label{1.11.04}
\end{equation}
to obtain the orbit equation
\begin{equation}
r (\theta) =\frac{Lg\circ N (\theta)}{z (\theta) +J\cos (\theta -\theta_0)},\label{1.11.5}
\end{equation}
where $g\circ N $ denotes the composition of the functions $g $ and $N $.

We consider some examples.

\begin{example} $v (\theta) = 0 $. The equation of motion, \re{1.10.11}, is
\begin{equation}
\ddot{\bfr} =\frac{\ddot{g}}{g}\bfr\label{1.12.1}
\end{equation}
with the conserved vectors
\begin{gather}
\bfK  = g\dot{\bfr} -\dot{g}\bfr,\nonumber\\
\bfJ  =  (g\dot{\bfr} -\dot{g}\bfr)\times\hat{\bfL}\label{1.12.2}
\end{gather}
and an energylike invariant
\begin{gather}
I  = \half\bfK\cdot \bfK\nonumber\\
\phantom{I}{} = \half (g\dot{\bfr} -\dot{g}\bfr)^2.\label{1.12.3}
\end{gather}
The Hamiltonian for \re{1.12.1} is
\begin{equation}
H =\half\(\bfp\cdot \bfp -\frac{\ddot{g}}{g}r^2\).\label{1.12.4}
\end{equation}
If the integral on the left side of \re{1.11.3}
is denoted by $M (t) $, the orbit equation is
\begin{equation}
r (\theta) =\frac{Lg\circ M^{- 1}
\circ (L\tan (\theta -\theta_0)/J^2)}{J\cos (\theta -\theta_0)}.\label{1.12.7}
\end{equation}
\end{example}
\begin{example} $v (\theta) =\mu/L $.
The equation of motion is
\begin{equation}
\ddot{\bfr} =\frac{\ddot{g}}{g}\bfr -\frac{\mu}{gr^3}\bfr\label{1.12.8}
\end{equation}
and the conserved vectors are
\begin{gather}
\bfK  =  g\dot{\bfr} -\dot{g}\bfr -\frac{\mu}{L} (\bht -\bht_0),\nonumber\\
\bfJ = (g\dot{\bfr} -\dot{g}\bfr)\times\hat{\bfL} -\frac{\mu}{L} (\hat{\bfr} -\hat{\bfr}_0).\label{1.12.9}
\end{gather}
The energylike invariant and Hamiltonian are
\begin{gather}
I  =\half (g\dot{\bfr} -\dot{g}\bfr)^2-\frac{\mu g}{r}\qquad\mbox{\rm and}\nonumber\\
H  = \half\(\bfp\cdot \bfp -\frac{\ddot{g}}{g}r^2\) -\frac{\mu}{gr}\label{1.12.11}
\end{gather}
respectively.  The orbit equation is
\begin{equation}
r (\theta) =\frac{L^2 g\circ N (\theta)}{\mu + (JL -\mu)\cos (\theta -\theta_0)}.\label{1.12.12}
\end{equation}
Leach \cite{ Leach 85} gave a simple example of an orbit in the case that
\begin{equation}
g (t) = (a +b\cos t)^{\half}\label{1.12.13}
\end{equation}
which resulted in an orbit
not greatly different from the standard elliptical orbit except for its elongation.  The choice
\begin{equation}
g (t) = \left(a +b\cos\half t\right)^{\half}\label{1.12.14}
\end{equation}
provides more complex orbits with
the degree of complexity depending upon the values of the parameters.
In the case that $a^2>b^2 $ and $\mu^2> (JL -\mu)^2 $ the equation for the orbit is
 (\cite[2.553.3 and 2.554.3]{GAR})
\begin{gather}
r (\theta)  =  a^{\half}\lb 1+k\cos\(2\arctan\(\(\frac{1+k}{1-k}\)^{\half}
\right.\right.\right.\nonumber\\
\phantom{r (\theta)  =}{}\left.\left.\left.\times\tan\(\frac{a (1-k^2)^{\half}L^3}{
4\mu^2 (1-l^2)}\(\frac{-l\sin (\theta -\theta_0)}{1+l\cos (\theta -\theta_0)}
\right.\right.\right.\right.\right.\nonumber\\
\phantom{r (\theta)  =}{} \left.\left.\left.\left.\left.+\frac{2}{(1-l^2)^{\half}}
\arctan\(\(\frac{1-l}{1+l}\)^{\half}
\tan\(\half\(\theta -\theta_0\)\)\)\)\)\)\)\rb^{\half},
\label{1.12.50}
\end{gather}
where $b =ka $ and $J =\mu (1+l)/L $.
The complexity of the orbits depends upon the choice of parameters.
 In general the orbits are not closed, but for suitable values of $L $ they can be made closed.

As the mass has been taken as unity, the momentum $\bfp $ is just $\dot{\bfr}$.
For both of these examples the Poisson Bracket relations are
\begin{align}
& \lb L_i,L_j\rb_{PB} =\varepsilon_{ijk}L_k,&&
 \lb J_i,J_j\rb_{PB} = (- 2I)\varepsilon_{ijk}L_k,&&&\nonumber\\
& \lb J_i, L_j\rb_{PB} =\varepsilon_{ijk}J_k, && \lb L_i,H\rb_{PB} = 0,&&&\nonumber\\
& \lb L_i,I\rb_{PB} = 0, && \lb J_i,I\rb_{PB} = 0,&&&
\label{1.12.16}
\end{align}
where the Hamiltonian is given by \re{1.12.4} for the first example and
\re{1.12.11} for the second example.  The Hamiltonian has the symmetry
group $SO (3) $, the rotation group in three dimensions.  The rotation
group in four dimensions $SO (4) $ of the time-independent Kepler problem
has been transferred to the energylike invariant,~$I $.  A~similar transferral
occurs in the case of the $n $-dimensional time-dependent oscillator~\cite{Gunther}.
\end{example}

\begin{example}
Another system for which an energylike integral can be found is described by
the equation of motion
\begin{equation}
\ddot{\bfr} =\frac{\ddot{g}}{g}\bfr -\frac{\mu}{g^3r}\bfr.\label{1.12.17}
\end{equation}
An integrating factor for \re{1.12.17} is $g (g\dot{\bfr} -\dot{g}\bfr) $
as a scalar product.  The invariant obtained is
\begin{equation}
I =\half (g\dot{\bfr} -\dot{g}\bfr)^2+{\mu r}{g}\label{1.12.18}
\end{equation}
and the Hamiltonian is
\begin{equation}
H =\half\bfp\cdot \bfp -\half\frac{\ddot{g}}{g}r^2+\frac{\mu r}{g^3}.\label{1.12.19}
\end{equation}
There does not seem to be a simple way to construct either a Hamilton's or
Laplace--Runge--Lenz vector for \re{1.12.17} by the methods which
have been considered here.  If one follows the principles of Fradkin and
others, this should be possible.  However, the conserved vectors would be
piecewise continuous at best and would not possess group properties but only
algebraic properties.  Although it is convenient to work with algebras, one
must agree with Bacry~\cite{Bacry 66, Bacry 91} that the important object,
as far as Physics is concerned, is the group.
\end{example}

\section[Conservation of the direction of angular momentum only]{Conservation
of the direction of angular momentum only}

\subsection[Vector conservation laws for the equation
 of motion $\ddot{\bfr}
+g\hat{\bfr} +h\bht = 0$]{Vector conservation laws for the equation\\
 of motion $\boldsymbol{\ddot{\bfr}
+g\hat{\bfr} +h\bht = 0}$}

In the previous chapter we considered equations of motion for which the vector
of angular momentum is conserved in both magnitude and direction.  We saw that
it was possible to obtain generalisations of Hamilton's vector and the
Laplace--Runge--Lenz vector.  In this chapter we examine the extent
to which we may still make reasonable progress with a relaxation of
one of the requirements previously imposed upon the angular momentum.
The relaxation is that we require only the constancy of the direction
of angular momentum and not of its magnitude.  The motion is still in
a plane and, by selecting the origin to be in that plane, we may still use
plane polar coordinates, $(r,\theta) $, to describe the orbit.
Consequently, for the content of this chapter at least, we do not have
to move out of two dimensions.  A particular advantage is that we preserve the relationship
\begin{equation}
\hat{\bfL} =\hat{\bfr}\times\bht\label{2.1.1}
\end{equation}
between the direction of the angular momentum and the two unit vectors of the
coordinate system.  Since $\dot{\bfL} =\dot{L}\hat{\bfL} $ and so
$\dot{\bfL}\times\bfL = 0 $, $\dot{\bfL} $ and $\bfL $ are collinear and we
may write
\begin{equation}
\dot{\bfL} +h_1\bfL = 0,\label{2.1.3}
\end{equation}
 where $h_1 $ is some arbitrary
function\footnote{That (\ref{2.1.3}) in turn
implies the constancy of $\hat{\bfL}$ follows
from writing $\dot{\bfL} = \dot{L}\hat{\bfL}+L\dot{\hat{\bfL}}$
in (\ref{2.1.3}) and taking the scalar product with $\dot{\hat{\bfL}}$
and the vector product with $\bfL$ in turn.}.  Making
use of \re{2.1.1} we may rewrite \re{2.1.3} as
\begin{equation}
\bfr\times\ddot{\bfr} +g_1\bfr\times{\bfr} +\frac{h_1L}{r}\bfr\times\bht = 0,\label{2.1.4}
\end{equation}
where $g_1 $ is also an arbitrary function.  Consequently the equation of motion
\begin{equation}
\ddot{\bfr} +g\hat{\bfr} +h\bht = 0,\label{2.1.5}
\end{equation}
where $g =g_1r $ and $h =h_1L/r $, describes motion in a plane
with~$\hat{\bfL} $ conserved.  Equation \re{2.1.5} is not the most
general equation of motion in the plane, but belongs to a class
of problems for which the techniques to be discussed below can be
applied.  For example we could introduce additional unrestricted
functions by adding terms such as $e\bfr\times\dot{\bfr} $ and
in a later section of this chapter \re{2.1.3} is rewritten in terms
of a different set of vector products and shown to describe a model
for the motion of low altitude artificial satellites in the atmosphere.

We employ a direct method, attributed to J~Bertrand\footnote{See Whittaker
\cite[p.~332, \S~152]{Whittaker 44}.},
to find a vector of the class of Laplace--Runge--Lenz
for the equation of motion \re{2.1.5}.  The general principle of a direct method is to impose some
{\it Ansatz} on the structure of the first integral or the equation of motion so that the
partial differential equation resulting from the imposition that the total time derivative
of the first integral be zero can be separated into subequations.
In this instance we assume that the functions $g $ and $h $
depend only on $r $ and $\theta $.  From the vector product of $\bfr $ with \re{2.1.5} we obtain
\begin{equation}
\dot{L} = -hr.\label{2.1.6}
\end{equation}
The vector product of \re{2.1.5} with $\bfL $ gives
\begin{equation}
\ddot{\bfr}\times\bfL -gL\bht +hL\hat{\bfr} = 0\label{2.1.7}
\end{equation}
which can be rewritten as
\begin{equation}
\frac{\d }{\d t} (\dot{\bfr}\times\bfL) -\dot{\bfr}\times\dot{\bfL} -gL\bht +hL\hat{\bfr} = 0.\label{2.1.8}
\end{equation}
A Laplace--Runge--Lenz-type vector exists of the form
\begin{equation}
\bfJ =\dot{\bfr}\times\bfL -u (r,\theta)\hat{\bfr} -v (r,\theta)\bht\label{2.1.9}
\end{equation}
provided
\begin{equation}
\frac{\d}{\d t}\(-u\hat{\bfr} -v\bht\) = hr\dot{\bfr}\times\hat{\bfL}
-gL\bht +hL\hat{\bfr}.\label{2.1.10}
\end{equation}
Equating the coefficients
of $\hat{\bfr} $ and $\bht $ separately to zero and using $L =r^2\dot{\theta} $ we have
\begin{gather}
\hat{\bfr}:\quad \dot{u} -v\dot{\theta} = - 2r^2h\dot{\theta},\nonumber\\
\bht: \quad u\dot{\theta} +\dot{v} =r\dot{r}h+gr^2\dot{\theta}.
\label{2.1.11}
\end{gather}
We now use the assumed functional dependence (\ref{2.1.9}) to obtain the set of four
partial differential equations as a sufficient condition
\begin{align}
&u_r = 0, &&v_r =hr,&&&\nonumber\\
& u_{\theta} -v = - 2hr^2, && v_{\theta} +u =gr^2,&&&
\label{2.1.12}
\end{align}
where the subscripts $_r$ and $_{\theta}$ represent partial differentiation
with respect to $r$ and $\theta$ respectively.
We solve \re{2.1.12} easily to obtain
\begin{align}
& u =U (\theta),&&v =U'(\theta) + 2r^{\half}V (\theta),&&&\nonumber\\
& h =r^{-\tha}V (\theta),&&g =r^{-2}\lb U''(\theta) +U (\theta) + 2r^{\half}V'(\theta)\rb,&&&
\label{2.1.13}
\end{align}
where $U (\theta) $ and $V (\theta) $ are arbitrary
functions of their argument.  The system with equation of motion
\begin{equation}
\ddot{\bfr} +\lb\frac{U''(\theta) +U (\theta)}{r^2} +\frac{2V'(\theta)}{r^{\tha}}\rb\hat{\bfr}
+\frac{V (\theta)}{r^{\tha}}\bht= 0\label{2.1.14}
\end{equation}
has the conserved vector
\begin{equation}
\bfJ =\dot{\bfr}\times\bfL -U (\theta)\hat{\bfr} -\lb U'(\theta) + 2r^{\half}V (\theta)\rb\bht.\label{2.1.15}
\end{equation}

We can construct an analogue of Hamilton's vector from this vector by
taking the vector product of $\hat{\bfL} $ with $\bfJ$.  It is
\begin{gather}
\bfK  = \hat{\bfL}\times\bfJ\nonumber\\
\phantom{\bfK}{} = L\dot{\bfr} -U (\theta)\bht +\lb U'(\theta) + 2r^{\half}V (\theta)\rb\hat{\bfr}.\label{2.1.16}
\end{gather}
Since $L $ is no longer conserved, we can no
longer rescale either \re{2.1.15} or \re{2.1.16} by $L $.

\subsection{The orbit equation and motion in time}

We construct the orbit equation from the conserved vector, $\bfJ $.
If we let $\theta_0 $ be the fixed angle between $\bfJ $ and the cartesian
unit vector $\bhi $ and take the scalar product of $\bfJ $ with
$\bfr $, we find after the usual rearrangement that
\begin{equation}
r =\frac{L^2}{U(\theta)+J\cos (\theta -\theta_0)}.\label{2.2.3}
\end{equation}
This is not the orbit equation since $L $ is not a constant.  To express
$L $ in terms of $\theta $ we write
\begin{equation}
\dot{L} =\dm{L}{\theta}\dot{\theta} =\dm{L}{\theta}\frac{L}{r^2},\label{2.2.4}
\end{equation}
use \re{2.1.6} and the third of \re{2.1.13} to obtain
\begin{equation}
L\dm{L}{\theta} = -r^{\tha}V (\theta).\label{2.2.5}
\end{equation}
We eliminate $r $ from \re{2.2.3} and \re{2.2.5} to express $L $ in terms of
the quadrature
\begin{equation}
\frac{1}{L} =\frac{1}{L_0} +\int_{\theta_0}^{\theta}
\frac{V (\eta)\d\eta}{\lb U (\eta) +J\cos (\eta -\theta_0)\rb^{\tha}}.\label{2.2.7}
\end{equation}
The orbit equation is
\begin{equation}
r (\theta) =\frac{1}{U (\theta) +J\cos (\theta -\theta_0)}
\times\lb\frac{1}{L_0} +\int_{\theta_0}^{\theta}
\frac{V (\eta)\d\eta}{\lb U (\eta) +J\cos (\eta -\theta_0)\rb^{\tha}}\rb^{- 2}.
\label{2.2.8}
\end{equation}
The orbit equation for the standard Kepler problem is recovered when we set
$U (\theta) =U_0 $ and $V (\theta) = 0 $.

The explicit determination of
$r (t) $ is not possible unless one can express $\theta $ in terms of~$r $.
If we differentiate \re{2.2.3}, use \re{2.1.6} to replace $\dot{L} $ and
again \re{2.2.3}, we obtain
\begin{equation}
\dot{r} = -\frac{2hr^2}{L} -\frac{1}{L}\(U'(\theta)-J\sin (\theta -\theta_0)\).
\label{2.3.1}
\end{equation}
The evident impossibility of expressing the right-hand side of \re{2.3.1} in
terms of~$r $ for a~general function of $\theta $ means that the radial motion
can only be evaluated numerically.

The function $\theta (t) $ can be calculated from \re{2.2.7} and the scalar
equation for the angular momentum
\begin{equation}
\dot{\theta} =\frac{\(U (\theta) +J\cos(\theta-\theta_0)\)^2}{L^3}.\label{2.3.2}
\end{equation}
Naturally it may not be possible to invert the results of this calculation even if the quadrature can be performed to give $\theta $ in terms of $t $.

The calculation of the areal velocity also poses serious practical problems.
If the angular motion in time can be solved using \re{2.3.2} so that
\begin{equation}
t =f_1 (\theta)\label{2.3.3}
\end{equation}
and this can be inverted to give
\begin{equation}
\theta =f_2 (t),\label{2.3.4}
\end{equation}
we can use
\begin{equation}
\dm{A}{t} =\half L\(f_2 (t)\),\label{2.3.5}
\end{equation}
where $L $ is calculated from \re{2.2.7}, and the areal velocity can in
principle be determined.  One would expect to take the numerical route normally.

We note that the existence of the conserved vectors to enable us
to perform the formal calculation of the orbit in the same, or a very
similar, manner as for the Kepler problem by no means guarantees
that we can perform in closed form the necessary quadratures.
However, we should bear in mind that motions in the plane can be chaotic.
That we are able to reduce these problems to quadratures indicates
that they do not belong to the chaotic class.

\begin{example}
The orbit equation \re{2.2.8} can yield a wide variety of orbits for different
choices of the functions $U (\theta) $ and $V (\theta) $, not to mention
for values of the parameters contained in particular choices of the functions.
 If $V\equiv 0 $ (indicating the conservation of the angular momentum vector
and so not strictly belonging to this Chapter, but still of some interest for
purposes of comparison) and $U $ is an increasing function of~$\theta $
and $U (0)>J> 0 $, the orbit spirals inwards.  If still $V\equiv 0 $,
but now $U $ is a decreasing function of $\theta $ with $U (0) >J > 0 $,
the orbit spirals outwards.  For $V\equiv 0 $ and $U $ periodic
with a~period incommensurate with $2\pi $
the orbit does not close.  For suitable initial conditions
the orbit closes with period $2\pi $ if $U $ has the
period $2\pi/n $, $n\in N $.
If $U $ has a period $2n\pi $, $n\in N $, the orbit
circles the origin $n $ times before closing, again for suitable
values
of the initial conditions.  A simple model to illustrate these possibilities is
\begin{equation}
U (\theta) = a\sin k (\theta -\theta_0) +b\label{2.4.1}
\end{equation}
for which the corresponding equation of motion, \re{2.1.14}, is
\begin{equation}
\ddot{\bfr} +\(\frac{a (1-k^2)\sin k (\theta -\theta_0) +b}{r^2}\)\hat{\bfr} = 0\label{2.4.2}
\end{equation}
and for which the conserved vectors are
\begin{gather}
\bfK  = L\dot{\bfr} -\(a\sin k (\theta -\theta_0) +b\){\bht} +
ak\cos k (\theta -\theta_0)\hat{\bfr},\label{2.4.3}\\
\bfJ  = \dot{\bfr}\times\bfL -\(a\sin k (\theta -\theta_0) +b\)\hat{\bfr} -
ak\cos k (\theta -\theta_0)\bht.
\label{2.4.4}
\end{gather}
\end{example}

\begin{example}
In the case that $U $ is constant and $V $ is periodic with period $2n\pi $, $n\in N/{1} $,
the orbit circles the origin $n $ times before closing in the case that the energy
permits bound orbits.  For
\begin{equation}
U =a,\qquad V =b\sin k (\theta -\theta_0)\label{2.4.5}
\end{equation}
the equation of motion is
\begin{equation}
\ddot{\bfr} +\(\frac{a}{r^2} +\frac{2bk\cos k (\theta -\theta_0)}{r^{\tha}}\)\hat{\bfr}
+\frac{b\sin k (\theta -\theta_0)}{r^{\tha}}\bht = 0\label{2.4.6}
\end{equation}
and the conserved vectors are
\begin{gather}
\bfK  = L\dot{\bfr} +\(2r^{\half}b\sin k (\theta -\theta_0)\)\hat{\bfr} - a\bht, \label{2.4.7}\\
\bfJ  = \dot{\bfr}\times\bfL -\(2r^{\half}b\sin k (\theta -\theta_0)\)
\bht -a\hat{\bfr}.\label{2.4.8}
\end{gather}
This example is true to the genre in that the angular momentum is constant
only in direction and not in magnitude.
\end{example}

Some illustrative orbits are given in Gorringe and Leach~\cite{Gorringe 87}.

\subsection{The related Hamiltonian system}

We now consider the most general autonomous Hamiltonian
which possesses a Laplace--Runge--Lenz
vector and belongs to the class of problems specified by the equation of motion, \re{2.1.14}.
We assume an Hamiltonian of the form
\begin{equation}
H =\half \bfp \cdot \bfp +W (r,\theta)\label{2.5.1}
\end{equation}
for which the Newtonian equation of motion can be calculated from
\begin{equation}
\ddot{\bfr} = -\nabla W (r,\theta) =\bfF (r,\theta).\label{2.5.2}
\end{equation}
In plane polar coordinates
\begin{equation}
\nabla W (r,\theta) =\pa{W}{r}\hat{\bfr} +\frac{1}{r}\pa{W}{\theta}\bht\label{2.5.3}
\end{equation}
and by comparison with \re{2.1.13} we have
\begin{gather}
\pa{W}{r}  =  r^{-2}\lb U''(\theta) +U (\theta)\rb + 2r^{-\tha}V'(\theta),\label{2.5.6}\\
\pa{W}{\theta}  = r^{-\half}V (\theta).\label{2.5.7}
\end{gather}

A consistent solution of \re{2.5.6} and \re{2.5.7} places considerable
constraints on the functions $U (\theta) $ and $V (\theta) $ so that we
obtain the Hamiltonian
\begin{equation}
H =\half\(p_r^2+\frac{p_{\theta}^2}{r^2}\) -\frac{\mu}{r} -
\alpha r^{-\half}\cos\(\half (\theta -\beta)\)\label{2.5.12}
\end{equation}
with the Newtonian equation of motion and conserved vectors given by
\begin{gather}
\ddot{\bfr} +\(\frac{\mu}{r^2} +
\frac{\alpha\cos\half (\theta -\beta)}{2r^{\tha}}\)\hat{\bfr}
+\frac{\alpha\sin\half (\theta -\beta)}{2r^{\tha}}\bht = 0,\label{2.5.14}\\
\bfK  = L\dot{\bfr} + \alpha r^{\half}\sin\half (\theta -\theta)\hat{\bfr} -\mu\bht,\label{2.5.15}\\
\bfJ  = \dot{\bfr}\times\bfL -\mu\hat{\bfr} -\alpha r^{\half}\sin\half (\theta -\beta)\bht.\label{2.5.16}
\end{gather}
We note that only the constant solution for $U (\theta) $ persists.
The $\theta$-dependent part of $U$ plays no role in the Hamiltonian
or the equation of motion.  Its sole contribution is to add a constant vector to $\bfJ $
and so it may be ignored.

\subsection[The Lie algebra of the Poisson Brackets of the
first integrals]{The Lie algebra of the Poisson Brackets of the first integrals}

We note a considerable diminution in the generality of the results
from the Newtonian case to the Hamiltonian case just simply
because of the requirement that the force be the gradient of a potential.
Nevertheless it is interesting that even that requirement does not impose
the condition that the force be single-valued.  The two nonzero cartesian
components of \re{2.5.16} are
\begin{gather}
J_1  = p_rp_{\theta}\sin\theta +\frac{p_{\theta}^2}{r}\cos\theta
-\mu\cos\theta +\alpha r^{\half}\sin\half (\theta -\beta)\sin\theta,
\label{2.6.1}\\
J_2  =  -p_rp_{\theta}\cos\theta +\frac{p_{\theta}^2}{r}\sin\theta
-\mu\sin\theta -\alpha r^{\half}\sin\half (\theta -\beta)\cos\theta,\label{2.6.2}
\end{gather}
where for convenience we have written the cartesian representation in terms
of the usual plane polar coordinates.
An alternative derivation of \re{2.6.1} and \re{2.6.2} would be to
use the structure for $H $ \re{2.5.1} and $\bfJ $ \re{2.1.9} and
impose the requirement that the Poisson Bracket of $H $ with
each of $J_1 $ and $J_2 $ be zero.  The result is the same
although the machinery used is quite different.  A detailed treatment may
be found in Gorringe and Leach \cite{Gorringe 93b}.  An
alternate representation of the two constants is to write
them as
\begin{gather}
J_{\pm}  = J_1\pm iJ_2\nonumber\\
\phantom{J_{\pm}} {}= \e^{\pm i\theta}\lb\frac{p_{\theta}^2}{r}
-\mu \mp i\(p_rp_{\theta} +\alpha r^{\half}\sin\half (\theta -\beta)
\)\rb\label{2.6.2a}
\end{gather}
which is the type of combination of some use, as we see below, in the
study of the complete symmetry groups for these problems.

Whichever form of the Ermanno--Bernoulli constants we use, the two first
integrals can be used to generate a third integral using Poisson's theorem
(\cf \cite [p.~320, \S~145]{Whittaker 44} and from the Poisson Bracket we obtain
\begin{equation}
I = 2p_{\theta}H+p_r\alpha r^{\half}\sin\half (\theta -\beta)
+p_{\theta}\alpha r^{-\half}\cos\half (\theta -\beta).
\label{2.6.4}
\end{equation}
Since they are first integrals, all have zero Poisson Bracket
with the Hamiltonian and so constitute the elements of an invariance
algebra of the Hamiltonian.  The Poisson Bracket relations between the three functions are
\begin{gather}
\lb J_1,J_2\rb_{PB}  =  -I,\nonumber\\
\lb J_1,I\rb_{PB}  =  - 2HJ_2-\half\alpha^2\sin\beta, \label{2.6.5}\\
\lb J_2,I\rb_{PB}  =  2HJ_1+\half\alpha^2\cos\beta.\nonumber
\end{gather}

The Poisson Bracket relations, \re{2.6.5}, are neither very tidy nor very
attractive.  We make some adjustment to the definitions of the first integrals
by writing
\begin{gather}
A_{\pm}  =  |2H|^{-\half}\(J_1+\frac{\alpha^2\cos\beta}{4H}\),\nonumber\\
B_{\pm}  =  |2H|^{-\half}\(J_2+\frac{\alpha^2\sin\beta}{4H}\),\nonumber\\
C_{\pm}  = \frac{I}{2H}\label{2.6.6}
\end{gather}
in which the positive sign of the $\pm $ refers to positive energies and the
negative sign to negative energies.  Now the Poisson Bracket relations are
\begin{gather}
\lb A_{\pm},B_{\pm}\rb_{PB}  = \mp C_{\pm},\nonumber\\
\lb B_{\pm},C_{\pm}\rb_{PB}  =  A_{\pm},\nonumber\\
\lb C_{\pm},A_{\pm}\rb_{PB}  =  B_{\pm}\label{2.6.7}
\end{gather}
which in the case of negative energy is immediately recognisable as the Lie
algebra $so (3) $ and in the case of negative energy the noncompact algebra
$so (2,1) $.  It is remarkable how a little refinement can make things
look so attractive!

In the planar version of the Kepler problem the algebra for negative and
positive ener\-gies is the same as we have found for this problem.  Consequently
we should not be surprised if there be a relationship between these conserved
quantities as exists for the Kepler problem.  After considerable manipulation
one finds that the relationship is
\begin{equation}
\frac{1}{2H}\left\{\(J_1+\frac{\alpha^2\cos\beta}{4H}\)^2+\(J_2+\frac{\alpha^2\sin\beta}{4H}\)^2
-\(\mu -\frac{\alpha^2}{4H}\)^2\right\} =\(\frac{I}{2H}\)^2\label{2.6.10}
\end{equation}
which is also somewhat lacking in \ae sthetic appeal.  By defining
\begin{equation}
\mu_{\pm} = |2H|^{-\half}\(\mu -\frac{\alpha^2}{4H}\)\label{2.6.11}
\end{equation}
for positive and negative energies respectively we may write \re{2.6.10} as
\begin{equation}
\mu_{\pm}^2 =A_{\pm}^2+B_{\pm}^2 \mp C_{\pm}^2\label{2.6.12}
\end{equation}
for positive and negative energies respectively.  For negative energies
in the space of first integrals equation \re{2.6.12} represents
a sphere which is naturally associated with $so (3) $
symmetry and for positive energies an hyperboloid of one sheet,
the natural geometric object associated with $so (2,1) $ symmetry.

When $H $ takes the particular value of zero, the analysis
commencing at \re{2.6.4} no longer holds true.  The third integral is now
\begin{equation}
I =p_r\alpha r^{\half}\sin\half (\theta -\beta)
+p_{\theta}\alpha r^{-\half}\cos\half (\theta -\beta)\label{2.6.14}
\end{equation}
which is no longer a true first integral but, rather, a configurational
invariant as discussed by Hall \cite{Hall 82} and Sarlet \etal
\cite{Sarlet 85} since it is invariant only for the particular value, $H=0 $.
 If we introduce the unit vector $\bht $ defined by
\begin{equation}
\bht =\bhj_1\cos\beta +\bhj_2\sin\beta,\label{2.6.15}
\end{equation}
where $\bhj_1 $ and $\bhj_2 $ respectively are the
unit vectors along which $J_1 $ and $J_2 $ lie, we see that
\begin{equation}
\bfJ\cdot \bht =\frac{I^2}{\alpha^2} -\mu\label{2.6.16}
\end{equation}
or, equivalently, that
\begin{equation}
J_1\cos\beta +J_2\sin\beta =\frac{I^2}{\alpha^2} -\mu.\label{2.6.17}
\end{equation}
Equation \re{2.6.17} represents the natural geometric object associated with
the case $H = 0 $ which describes a right parabolic cylinder with
$\bht $ lying along the axis of symmetry.

A distinction can now be drawn between the standard Kepler problem,
where, for $H = 0 $, $J =\pm\mu $ from
\re{1.1.9}, and the present problem.  This is the equation
for a plane, unlike \re{2.6.17}.  This difference is also reflected
in the form of the algebra.  We define
\begin{gather}
A_0  =  -J_1\sin\beta +J_2\cos\beta,\nonumber\\
B_0  = \frac{2I}{\alpha^2},\nonumber\\
C_0  =  1.\label{2.6.18}
\end{gather}
The combination $J_1\cos\beta +J_2\sin\beta $ cannot be used as this also
is a function of $I $ from \re{2.6.17}.  The Poisson Bracket relations are
\begin{gather}
\lb A_0,B_0\rb_{PB}  = C_0,\nonumber\\
\lb B_0,C_0\rb_{PB}  =  0,\nonumber\\
\lb C_0,A_0\rb_{PB}  =  0\label{2.6.19}
\end{gather}
which represents the Weyl algebra, $W (3,1) $.  The standard Kepler problem (in
two dimensions) has the algebra $E (2) $ when $H = 0 $.

For $H\neq 0 $ the Hamiltonian can be expressed as a function of the
three first integrals, $J_1 $, $J_2 $ and $I $, as
\begin{equation}
H =\frac{1}{2\(J^2-\mu^2\)}\lb I^2-\alpha^2 (J_1\cos\beta +J_2\sin\beta +\mu)\rb.\label{2.6.20}
\end{equation}

An alternate derivation of the integrals for the Hamiltonian was proposed by
 Sen~\cite{Sen 87} as an extension but in the spirit of the work done on
first integrals polynomial in the velocities for a variety of Hamiltonian
systems by, amongst others, Gasc\'on \etal \cite{Gascon 82},
Grammaticos~\etal \cite{Grammaticos 84},
Thompson~\cite{Thompson 84}, Hietarinta~\cite{Hietarinta 85},
Leach~\cite{Leach 86} and, in the case of
time-dependent systems, Lewis and Leach~\cite{Lewis 82}.
Sen assumes that the first integral is a polynomial in the momenta of the form
\begin{equation}
I_S =\sum_{i = 0}\sum_{j = 0}d_{ij} (x,y)p_x^ip_y^j,\qquad i+j\leq n,\label{2.7.1}
\end{equation}
where $i+j $ is either even or odd in sympathy with
$n $ since autonomous invariants for an autonomous
Hamiltonian of even degree in the momenta are either even or odd in the momenta
(see Thompson~\cite{Thompson 84}).  The extension of Sen is to make a
linear canonical transformation to a complex coordinate system
\begin{align}
& z = 2^{-\half} (x+iy), &&\bar{z} = 2^{-\half} (x-iy),&&&\nonumber\\
&p_z = 2^{-\half} (p_x-ip_y),&&p_{\bar{z}} = 2^{-\half} (p_x+ip_y).&&&
\label{2.7.2}
\end{align}

In the new coordinates the Hamiltonian and first integral have the forms
\begin{gather}
H_S = 2p_zp_{\bar{z}} +V (z,\bar{z}),\nonumber\\
I_S = (zp_z-\bar{z}p_{\bar{z}})^{n- 2}p_zp_{\bar{z}}
+\sum_{i = 0}\sum_{j = 0} e_{ij} (z,\bar{z})
p_z^ip_{\bar{z}}^j,
\label{2.7.3}
\end{gather}
where now $i+j\leq n- 2 $.  The Poisson Bracket requirement that
$\lb I_S,H_S\rb_{PB} = 0 $ results in a system of equations the solution
of which gives the Hamiltonian
\begin{equation}
H_{IS} = 2p_zp_{\bar{z}} +az^{-\half} +b\bar{z}^{-\half} +c (z\bar{z})^{-\half},\label{2.7.5}
\end{equation}
the third-order first integral
\begin{equation}
I_3 = (zp_z-\bar{z}p_{\bar{z}})p_zp_{\bar{z}} +\half\lb b\bar{z}^{-\half}
+c (z\bar{z})^{-\half}\rb zp_z
-\half\lb a{z}^{-\half} +c (z\bar{z})^{-\half}\rb \bar{z} p_{\bar{z}}\label{2.7.6}
\end{equation}
and the two quadratic first integrals
\begin{gather}
I_1  =  (zp_z-\bar{z}p_{\bar{z}})p_z+\half\lb b\bar{z}^{\half} -az^{-\half}\bar{z} -cz^{-\half}\bar{z}^{\half}\rb,\label{2.7.7}\\
I_2  =  (zp_z-\bar{z}p_{\bar{z}})p_{\bar{z}} +\half\lb bz\bar{z}^{\half} -az^{\half} + cz^{\half}\bar{z}^{-\half}\rb.\label{2.7.8}
\end{gather}

When $b =\bar{a} =a_1-ia_2 $, the identification of the results of Sen with those given above is
\begin{align}
& H_{IS} = 2H,&& - 2^{-\half} (I_1-I_2) =J_1,&&&\nonumber\\
&2iI_3 =I, && - 2^{-\half}i (I_1+I_2) =J_2.&&&
\label{2.7.10}
\end{align}
Naturally the results of the two approaches can be mutually identified
when the same matters are being discussed.

The results of Sen, derived by the direct approach of assuming a structure
for the first integral, and those obtained here by means of vectorial manipulation
of the equation of motion and then the imposition of the Hamiltonian
structure are peculiar to two dimensions.  The results of a long and
tedious calculation~\cite{Gorringe 93b} show that the only potential in three
dimensions which can be permitted subject to be requirement that the
Hamiltonian be autonomous and of the form $H =T+V $ is the familiar
Kepler potential.  Sen~\cite{Sen 87} also considered the quantum mechanical
representations of the two-dimensional Hamiltonian.  The presence of the
quadratic first integrals in the classical case implies that the quantum
mechanical version of this potential is integrable (see
Hietarinta~\cite{Hietarinta 84}).  He obtained the quantum invariants,
the identical algebra to the classical algebra and the bound state energy spectrum using the
Casimir operator for $so (3) $.  Considering
equation~\re{2.6.12} above one should not be surprised.

\subsection[Invariance under time translation and
the first integrals]{Invariance under time translation and the first integrals}

Sen \cite{Sen 87} showed that the differential equation \re{2.5.14}
possesses only the single Lie point symmetry,
$G =\upt $.  Sen interpreted this in the usual way as leading to the
conservation of energy, which it does if one applies Noether's Theorem.
Using the Lie method Leach~\cite{Leach 81} constructed the energy, the
angular momentum and the components of the Laplace--Runge--Lenz
vector for the Kepler problem from this single Lie point
symmetry\footnote{As is well known (for example, see Kaplan \cite{Kaplan 86}),
a system of $N $ degrees of freedom can have at most
only $2N- 1 $
first integrals.  The excessive number obtained by Leach followed from different
routes towards obtaining the invariants of the associated Lagrange's systems
which need to be solved to obtain the integrals.  These integrals, energy,
angular momentum and the Laplace--Runge--Lenz vector,
obviously have the symmetry, $\upt $, since they
are manifestly invariant under time translation.  The single integral per
symmetry is a feature of Noether's theorem and to obtain the integrals
apart from the energy using Noether's theorem one would require
additional symmetries.  These symmetries would have to be
generalised symmetries due to the existence of
only one Lie point symmetry.}.

A Lie symmetry, $G $, gives rise to a first integral/invariant of
a differential equation for which it is a symmetry if the two
conditions
\begin{equation}
G^{\lb 1\rb}I = 0\qquad\mbox{\rm and}\quad \dm{I}{t}_{|_{E = 0}} = 0,\label{2.9.5}
\end{equation}
where the differential equation in question is $E = 0 $ and $G^{\lb 1\rb} $ is
the first extension of $G $, are satisfied.  The first extension is required
due to the required presence of first derivatives in the first integral/invariant.
The symmetry $\upt $ is its own first extension and the first of \re{2.9.5}
gives the linear partial differential equation
\begin{equation}
\pa{I}{t} + 0\pa{I}{r} + 0\pa{I}{\theta} + 0\pa{I}{\dot{r}} + 0\pa{I}{\dot{\theta}} = 0.\label{2.9.6}
\end{equation}
The characteristics obtained from the solution of the associated Lagrange's system
\begin{equation}
\frac{\d t}{1} =\frac{\d r}{0} =\frac{\d \theta}{0} =\frac{\d\dot{r}}{0} =\frac{\d\dot{\theta}}{0}\label{2.9.7}
\end{equation}
are
\begin{align}
& u_1 =r, &&  v_1 =\dot{r}, &&&&&&&\nonumber\\
& u_2 =\theta, &&v_2 =\dot{\theta},&&&&&&&
\label{2.9.8}
\end{align}
which are individually invariant under the infinitesimal transformation generated
by $G^{\lb 1\rb} $.  The second condition of \re{2.9.5} gives the associated Lagrange's system
\begin{equation}
\frac{\d u_1}{\dot{u}_1} =\frac{\d u_2}{\dot{u}_2} =\frac{\d v_1}{\dot{v}_1}
=\frac{\d v_2}{\dot{v}_2},
\label{2.9.9}
\end{equation}
where we find from the equation of motion that
\begin{gather}
\dot{u}_1  = v_1,\nonumber\\
\dot{v}_1  = u_1v_2^2-\frac{\mu}{u_1^2} -\frac{\alpha\cos\half (u_2-\beta)}{2u_1^{\tha}},\nonumber\\
\dot{u}_2 = v_2,\nonumber\\
\dot{v}_2  =  -\frac{2v_1v_2}{u_1} -\frac{\alpha\sin\half
(u_2-\beta)}{2u_1^{\fha}}.\label{2.9.10}
\end{gather}

In the case of the Kepler problem the solution of the equation corresponding
to~\re{2.9.9} was nontrivial and the righthand sides
in~\re{2.9.10} are more complex.  The underlying idea of the solution of
the associated Lagrange's system is to take combinations of the elements
of the system to obtain equivalent elements which have the total derivative
of some function in the numerator and
zero in the denominator \cite[p.~45]{Ince 27}.  To make the process of solution
more obvious, if not precisely transparent, we indicate the combinations used.
We denote the $i $th element of \re{2.9.10} by (\ref{2.9.10}.i).  The combination
\begin{gather}
\(u_1v_2^2+\frac{\mu}{u_1^2} +
\frac{\alpha\cos\half (u_2-\beta)}{2u_1^{\tha}}\) (\ref{2.9.10}.1)
\nonumber\\
\qquad {}+\frac{\alpha\sin\half (u_2-\beta)}{2u_1^{\half}} (\ref{2.9.10}.2)
+v_1 (\ref{2.9.10}.3) +u_1^2v_2 (\ref{2.9.10}.4)\label{2.9.12}
\end{gather}
gives one of the desired elements, \viz
\begin{equation}
\frac{\d\lb\half\(v_1^2+u_1^2v_2^2\) -\dsp{\frac{\mu}{u_1}} -
\dsp{\frac{\alpha\cos\half (u_2-\beta)}{
u_1^{\half}}}\rb}{0}.\label{2.9.13}
\end{equation}
The term in crochets is a characteristic
and hence a first integral.  In the original coordinates
one sees that it is the energy, which in this case is the Hamiltonian.
The components of the Laplace--Runge--Lenz vector are obtained from the combinations
\begin{gather}
\(2u_1v_1v_2\sin u_2+ 3u_1^2v_2^2\cos u_2
+\half\alpha u_1^{-\half}\sin\half (u_2-\beta)\sin u_2\)
(\ref{2.9.10}.1)\nonumber\\
\qquad {}+ \(u_1^2v_1v_2\cos u_2-u_1^3v_2^2\sin u_2
+\mu\sin u_2+\half\alpha u_1^{\half}\cos\half (u_2-\beta)
\sin u_2\right.\nonumber\\
 \qquad {}+ \left.\alpha u_1^{\half}\sin\half (u_2-\beta)\cos u_2\) (\ref{2.9.10}.2)
+\(u_1^2v_2\sin u_2\) (\ref{2.9.10}.3)\nonumber\\
\qquad  {}+ \(u_1^2v_1\sin u_2+ 2u_1^3v_2\cos u_2\) (\ref{2.9.10}.4)\label{2.9.14}
\end{gather}
for $J_1 $ and
\begin{gather}
\(-2u_1v_1v_2\cos u_2+ 3u_1^2v_2^2\sin u_2-\half\alpha u_1^{-\half}\sin\half (u_2-\beta)\cos u_2\)
(\ref{2.9.10}.1)\nonumber\\
\qquad {} + \(u_1^2v_1v_2\sin u_2+u_1^3v_2^2\cos u_2 -\mu\cos u_2-\half\alpha u_1^{\half}\cos\half (u_2-\beta)
\cos u_2\right.\nonumber\\
\qquad {} +\left.\alpha u_1^{\half}\sin\half (u_2-\beta)\sin u_2\) (\ref{2.9.10}.2)
+\(-u_1^2v_2\cos u_2\) (\ref{2.9.10}.3)\nonumber\\
\qquad {} + \(-u_1^2v_1\cos u_2+ 2u_1^3v_2\sin u_2\) (\ref{2.9.10}.4)\label{2.9.15}
\end{gather}
for $J_2 $.  Finally $I $ comes from the combination
\begin{gather}
\(4u_1v_2H+\half u_1^{-\half}v_1\alpha\sin\half (u_2-\beta) +\tha u_1^{\half}v_2\alpha\cos\half (u_2-\beta)\) (\ref{2.9.10}.1)\nonumber\\
\qquad {} + \(\half u_1^{\half}v_1\alpha\cos\half (u_2-\beta) -\half u_1^{\tha}v_2\alpha\sin\half (u_2-\beta)\) (\ref{2.9.10}.2)
\label{2.9.16}\\
\qquad {} + \(u_1^{\half}\alpha\sin\half (u_2-\beta)\) (\ref{2.9.10}.3)
+\(2u_1^2H+u_1^{\tha}\alpha\cos\half (u_2-\beta)\) (\ref{2.9.10}.4),\nonumber
\end{gather}
where the invariance of $H $, \ie $\d H = 0 $, has been used.

One cannot plausibly propose that the solution of \re{2.9.9} is transparent.
However, it is one possible method where it is not essential, {\it a priori},
to make an {\it Ansatz} for the structure of the first integrals.  This method is
more feasible when there are several symmetries available for a complex system.
Then one can impose the requirements of several symmetries simultaneously
and so reduce the number of invariants before imposing the condition $\dot{I} = 0 $,
in this case the second member of \re{2.9.5}, which is generally the major source of
difficulty\footnote{One need not be surprised at this since mathematicians and
scientists have been looking for ways to avoid the direct solution of the equations
of motion ever since they were devised several centuries ago.}.  An example
of this procedure in practice is to be found in the paper of Cotsakis \etal \cite{Cotsakis 99}
devoted to the determination of the first integrals/invariants of the Bianchi models
in Cosmology.

\subsection{The Kepler problem with `drag'}

For an equation of motion with the direction of angular momentum conserved we had,
in \S~3.1, the equation
\begin{equation}
\dot{\bfL} +h_1\bfL = 0\label{2.10.1}
\end{equation}
which we can rewrite as
\begin{equation}
\bfr\times\ddot{\bfr} +g_1\bfr\times\bfr +h_1\bfr\times\dot{\bfr} = 0,\label{2.10.2}
\end{equation}
where $g_1 $ and $h_1 $ are arbitrary functions for which the
variable dependence is specified below when it is necessary so to do.
Consequently the equation of motion
\begin{equation}
\ddot{\bfr} +f\dot{\bfr} +g\bfr = 0,\label{2.10.3}
\end{equation}
where we have replaced $h_1 $ and $g_1 $ with $f $ and $g $ respectively,
describes motion in a plane with $\hat{\bfL} $ conserved.
This equation, \re{2.10.3}, is a generalisation to a class of a
problem of physical interest, the equation which models the motion
of satellites in low altitude orbits and which was proposed by Brouwer
and Hori \cite{Brouwer 61}.  They obtained a closed-form solution which included
first-order corrections for drag\footnote{The use of the word `drag' to describe
this type of force was introduced by Jezewski and
Mittleman \cite{Jezewski 82, Jezewski 83}.  We maintain the usage,
but slightly tongue in cheek as may be inferred from the `' in the title of this section.}
acceleration involving a quadratic velocity-dependent term to account for
atmospheric effects on satellites sufficiently close to the Earth.
Danby~\cite{Danby 62} modified this assumption and used
a resistive term which was proportional to the velocity vector
and inversely proportional to the square of the radial distance.
For small values of the constant of proportionality he obtained
a first-order perturbation solution.  Mittleman and Jezewski~\cite{Jezewski 82}
provided an exact solution to the same problem and in
a subsequent paper~\cite{Jezewski 83} they demonstrated
the existence of analogues of the angular momentum, energy and Laplace--Runge--Lenz
vector for the
Danby problem.  Their approach was to manipulate the equation of motion as had
been done by Collinson \cite{Collinson 73, Collinson 77},
Pollard \cite{Pollard 76} and also Sarlet and Bahar \cite{Sarlet 80} on
various nonlinear problems.  The paper of Mittleman and Jezewski
\cite{Jezewski 82} is quite complicated and invited attempts to provide an
elegant solution to the original problem of Danby
\cite{Leach 87a} and its generalisation as represented by \re{2.10.3} \cite{Gorringe 89}.

The vector product of $\bfr $ with \re{2.10.3} gives
\begin{equation}
\dot{L} = -fL.\label{2.10.4}
\end{equation}
This equation may be used to eliminate $f $ from \re{2.10.3} to give
\begin{equation}
\ddot{\bfr} -\frac{\dot{L}}{L}\dot{\bfr} +g\bfr = 0.\label{2.10.5}
\end{equation}
The first two terms of \re{2.10.5} possess the integrating factor $L^{- 1} $ and we have
\begin{equation}
\frac{\d}{\d t}\(\frac{\dot{\bfr}}{L}\) +\frac{g\bfr}{L} = 0.\label{2.10.6}
\end{equation}
There exists a Hamilton-like vector if we can write
\begin{equation}
\bfK =\frac{\dot{\bfr}}{L} +\bfu,\label{2.10.7}
\end{equation}
where $\bfu: =\int (g\bfr/L) \d t $, explicitly without a knowledge of $r (t) $ and $\theta (t) $.
The analogue of the Laplace--Runge--Lenz vector is
\begin{equation}
\bfJ =\bfK\times\hat{\bfL} =
\frac{\dot{\bfr}\times\hat{\bfL}}{L} +\bfu\times\hat{\bfL}.\label{2.10.8}
\end{equation}

The orbit equation is obtained by taking the scalar product of \re{2.10.8} with $\bfr $ and is
\begin{equation}
r =\frac{1}{-u_{\theta} +J\cos (\theta -\theta_0)},\label{2.11.2}
\end{equation}
where $\theta_0 $ is the angle between $\bfJ $ and the cartesian unit vector,
$\bhi$, and $u_{\theta}$ is the component of $\bfu$ in the direction
$\hat{\mbt}$ (\cf\ \re{2.11.5}).  To determine $\bfu $ we use the cartesian
representation of $\hat{\bfr} $ to write
\begin{gather}
\bfu  = \int \frac{g\bfr}{L}\d t\nonumber\\
\phantom{\bfu}{} =  \bhi\int\frac{gr}{L}\cos\theta\d t+
\bhj\int\frac{gr}{L}\sin\theta\d t\label{2.11.3}
\end{gather}
and the two integrals can be evaluated provided that
\begin{equation}
\frac{gr}{L} =v (\theta)\dot{\theta}\quad \Leftrightarrow \quad g = rv(\theta)\dot{\theta}^2.\label{2.11.4}
\end{equation}
Then we may write $\bfu $ as (\cf\ \S~2.1)
\begin{equation}
\bfu =z'(\theta)\hat{\bfr} -z (\theta)\bht,\label{2.11.5}
\end{equation}
where
\begin{equation}
z (\theta) =\int_{\theta_0}^{\theta}v (\eta)\sin (\theta -\eta)\d\eta\label{2.11.6}
\end{equation}
or, alternatively,
\begin{equation}
z''(\theta) +z (\theta) =v (\theta),\qquad z (\theta_0) = 0 \qquad
\mbox{\rm and}\qquad z'(\theta_0) = 0.\label{2.11.8}
\end{equation}
The radial and angular components of $\bfu $ are
\begin{equation}
u_r =z'(\theta),\qquad u_{\theta} = -z (\theta)\label{2.11.11}
\end{equation}
and the two vectors are
\begin{gather}
\bfK  = \frac{\dot{\bfr}}{L} +z'(\theta)\hat{\bfr} -z (\theta)\bht,\label{2.11.9}\\
\bfJ = \frac{\dot{\bfr}\times\hat{\bfL}}{L} -z (\theta)\hat{\bfr} -z'(\theta)\bht.\label{2.11.10}
\end{gather}
The orbit equation, \re{2.11.2}, is now
\begin{equation}
r =\frac{1}{z (\theta) +J\cos (\theta -\theta_0)}.\label{2.11.13}
\end{equation}

We now return to \re{2.10.4}.  This can be integrated to give
\begin{equation}
M (L,r,\theta) =h,\label{2.11.16}
\end{equation}
where $h $ is the constant of integration, provided
\begin{equation}
f =\frac{\pa{M(L,r,\theta)}{r}\dot{r} +\pa{M(L,r,\theta)}{\theta}\dot{\theta}}{L\pa{M(L,r,\theta)}{L}}.
\label{2.11.14}
\end{equation}
We may invert \re{2.11.16}, at least locally, to obtain
\begin{equation}
L =N (h,r,\theta).\label{2.11.17}
\end{equation}
With this and \re{2.11.4} rewritten as
\begin{equation}
\frac{gr^3}{L^2} =v (\theta) \label{2.11.18}
\end{equation}
by the replacement of $\dot{\theta} $ by $L/r^2 $,
we obtain an alternative expression for $g$, \viz
\begin{equation}
g =\frac{1}{r^3}N^2 (h,r,\theta)v (\theta).\label{2.11.19}
\end{equation}
Equations \re{2.11.14} and \re{2.11.19} establish one
possible form of \re{2.10.3} for which the conserved vectors $\bfK $, \re{2.11.9},
and $\bfJ $, \re{2.11.10}, and the orbit equation, \re{2.11.13}, exist.

For the model of the motion of a low altitude satellite Jezewski
and Mittleman \cite{Jezewski 83} obtained an energylike first
integral by a complicated calculation.  For the more general problem
discussed here it is also possible to obtain a scalar energylike first integral as
\begin{gather}
I  = \half\bfK\cdot \bfK =\half\bfJ\cdot \bfJ\nonumber\\
\phantom{I}{} = \half\frac{\dot{\bfr}\cdot
\dot{\bfr}}{L^2} +\frac{1}{L}\lb\dot{r}z'\(\theta\) -r\dot{\theta}z (\theta)\rb
+\half\lb z^2 (\theta) +z'{}^2 (\theta)\rb.\label{2.11.20}
\end{gather}

The radial motion in time may be found by differentiation of \re{2.11.13}
and use of the orbit equation to remove the $\sin (\theta -\theta_0) $ term.  We obtain
\begin{equation}
\dot{r} = -L\(z'(\theta) -\frac{1}{r}\((J^2-z^2 (\theta))r^2+ 2z (\theta)r- 1\)^{\half}\),\label{2.12.1}
\end{equation}
where $L =r^2\dot{\theta} $ has been used to eliminate $\dot{\theta} $.
The only possibility that \re{2.12.1} is solvable in closed form is that $z (\theta) $,
$z'(\theta) $ and $L $ can be expressed as functions of~$r $,
apart from any constants of the motion.  If this be possible
and the subsequent quadrature performed, we obtain
\begin{equation}
t =f_1 (r)\label{2.12.2}
\end{equation}
and are still left with the problem of the inversion to obtain $r (t) $.
We treat one class of such problems below.

The angular motion in time is obtained from the two-dimensional
angular momentum equation $L =r^2\dot{\theta} $ and
the elimination of $r $ with \re{2.11.13}.  We obtain
\begin{equation}
\dot{\theta} = L\(z (\theta) +J\cos (\theta -\theta_0)\)^2\label{2.12.3}
\end{equation}
and, provided $L $ can be expressed in terms of $\theta $,
it may be possible to perform the quadrature to obtain
\begin{equation}
t =f_2 (\theta).\label{2.12.4}
\end{equation}
The areal velocity is obtained from
\begin{equation}
\dot{A} =\half L (t),\label{2.12.5}
\end{equation}
where $L (t) $ is obtained as an explicit function of
time through the inversion of \re{2.12.2} and \re{2.12.4} to replace $r $
and $\theta $ by functions of time.

We see that there is much potential for not to be able to obtain explicit
expressions.  However, the establishment of the existence of the conserved
quantities enables us to carry out numerical computations with confidence.


These examples are described in Leach \cite{Leach 87a}
and Gorringe and Leach \cite{Gorringe 88a}.  Mavraganis
\cite{Mavraganis 91} considered equations of the form \re{2.10.3} in which the
resistive term varied slowly with time.  With the use of a Taylor series
expanded about the initial velocity he obtained expressions for the orbit
equation and the Hamilton and Laplace--Runge--Lenz vectors which are similar in
structure to those described by Jezewski and Mittleman \cite{Jezewski 83} and
Leach \cite{Leach 87a} for the Danby \cite{Danby 62} problem.

\begin{example}
The equation of motion for the Danby problem is
\begin{equation}
\ddot{\bfr} +\frac{\alpha\dot{\bfr}}{r^2} +\frac{\mu\bfr}{r^3} = 0.\label{2.13.1}
\end{equation}
From a comparison of \re{2.13.1} with \re{2.10.3} we have
\begin{equation}
f =\frac{\alpha}{r^2},\qquad g =\frac{\mu}{r^3}.\label{2.13.2}
\end{equation}
From \re{2.10.4} the magnitude of the angular momentum satisfies the equation
\begin{equation}
-\frac{\dot{L}}{L} =\frac{\alpha}{r^2} =\frac{\alpha\dot{\theta}}{L}\label{2.13.3}
\end{equation}
which has the solution
\begin{equation}
L =k-\alpha\theta,\label{2.13.4}
\end{equation}
where $k $ is some arbitrary constant, in agreement with the result of
Jezewski and Mittleman~\cite{Jezewski 83}.  We also find that
\begin{gather}
v (\theta)  = \frac{\mu}{(k-\alpha\theta)^2},\label{2.13.5}\\
z (\theta)=\mu\int_{\theta_0}^{\theta}\frac{\sin (\theta -\eta)\d\eta}{(k-\alpha\eta)^2}.\label{2.13.6}
\end{gather}
In terms of
\begin{equation}
u =\(\frac{k}{\alpha} -\theta\),\qquad u_0 =\(\frac{k}{\alpha} -\theta_0\)\label{2.13.9}
\end{equation}
and some standard integrals (\cite[2.641.1--4]{GAR}) we obtain
\begin{gather}
z (\theta)  = \frac{\mu}{\alpha^2}\lb\frac{\sin (u-u_0)}{u_0} -
\sin u\(\mbox{\rm si}(u)-\mbox{\rm si}(u_0)\) -\cos u\(\mbox{\rm Ci}(u)
-\mbox{\rm Ci}(u_0)\)\rb,\label{2.13.8}\\
z'(\theta)  =  -\frac{\mu}{\alpha^2}\lb\frac{\cos (u-u_0)}{u_0}
-\frac{1}{u} +\sin u\(\mbox{\rm Ci}(u) -\mbox{\rm Ci}(u_0)\)
-\cos u\(\mbox{\rm si}(u) - \mbox{\rm si}(u_0)\)\rb,\nonumber
\end{gather}
where the sine and cosine integrals are given by
\begin{gather}
\mbox{\rm si}(x)  =  -\half\pi +\int_0^x\frac{\sin t}{t}\d t,\label{2.13.10}\\
\mbox{\rm Ci}(x)  = \gamma +\log x+\int_0^x\frac{\cos t- 1}{t}\d t\label{2.13.11}
\end{gather}
and $\gamma = 0.57721566490\ldots $ is Euler's constant.

The explicit expressions for the conserved vectors are
\begin{gather}
\bfK  = \frac{\dot{\bfr}}{L} +\frac{\mu}{\alpha^2}\(\frac{1}{u} -
\sin u \mbox{\rm Ci}(u) +\cos u \mbox{\rm\, si}(u)\)\hat{\bfr}
\nonumber\\
\phantom{\bfK  =}{} +\frac{\mu}{\alpha^2}\(\sin u \mbox{\rm\, si}(u) +
\cos u \mbox{\rm Ci}(u)\)\bht\nonumber\\
\phantom{\bfK  =}{}
-\frac{\mu}{\alpha^2}\(\frac{\hat{\bfr}_{\theta_0}}{u_0}
+ \mbox{\rm\, si}(u_0)\hat{\bfr}_{k/\alpha} +\mbox{\rm Ci}(u_0)
\bht_{k/\alpha}\),\label{2.13.13}\\
\bfJ  = \frac{\dot{\bfr}\times\hat{\bfL}}{L} +\frac{\mu}{\alpha^2}\(\sin u \mbox{\rm\, si}(u)
+\cos u \mbox{\rm Ci}(u)\)\hat{\bfr}
\nonumber\\
\phantom{\bfJ  =}{} -\frac{\mu}{\alpha^2}\(\frac{1}{u} -\sin u \mbox{\rm Ci}(u)
+\cos u \mbox{\rm\, si}(u)\)\bht\nonumber\\
 \phantom{\bfJ  =}{} +
\frac{\mu}{\alpha^2}\(\frac{\bht_{\theta_0}}{u_0} + \mbox{\rm\, si}(u_0)\bht_{k/\alpha}
- \mbox{\rm Ci}(u_0)\hat{\bfr}_{
k/\alpha}\).\label{2.13.14}
\end{gather}
Since the final term in both \re{2.13.13} and \re{2.13.14}
is a constant vector, it can be ignored in order to simplify matters.  The simplified
version coincides with that obtained by Jezewski and Mittleman
\cite{Jezewski 83} who obtained their result by means of an integrating factor.
For either version the equation for the orbit comes from the scalar product of $\bfr $ and $\bfJ $.

A typical orbit, for which see Gorringe and Leach \cite{Gorringe 88a},
spirals inwards which is not surprising considering what the equation
of motion was intended to model.
\end{example}

\begin{example}
The density of the atmosphere is not constant for all $\theta $.
A more realistic resistive force for the motion of a low altitude
satellite which incorporates a variation in the density of the atmosphere is given by
\begin{equation}
f =\frac{a\cos\theta +b}{r^2},\qquad g =\frac{\mu}{r^3}.\label{2.13.22}
\end{equation}
In this case
\begin{equation}
z (\theta) =\mu\int_{\theta_0}^{\theta}\frac{\sin (\theta -\eta)\d \eta}
{(k-a\sin\eta -b\eta)^2}.\label{2.13.23}
\end{equation}
The nature of the orbit depends upon the relative values
of the constants $a $ and $b $.  For values which give a
resistive force for all values of $\theta $, the orbit spirals inwards,
as expected, but the rate at which it spirals inwards varies with $\theta $
to give almost the impression of a~precession of the orbit.
\end{example}

\begin{example}
There is no necessity to maintain an air of reality in the model if one
wants to explore the variety of orbits which can be obtained.  If we make the choice
\begin{equation}
f =\frac{a\cos\theta +b}{Lr^2},\qquad g =\frac{\mu}{r^3},\label{2.13.24}
\end{equation}
we find that
\begin{equation}
z (\theta) =\half\mu\int_{\theta_0}^{\theta}\frac{\sin (\theta -\eta)\d\eta}{k-a\sin\eta -b\eta}.\label{2.13.25}
\end{equation}
The ``precession'' of the orbit is even more pronounced.
In the case that the force is not always resistive the angular momentum
increases and eventually the orbit appears to depart asymptotically.
\end{example}

\begin{example}
In the case that
\begin{equation}
f =\frac{-a\e^{-(\theta -\theta_0)}L^3}{2r^2},\qquad g =\frac{\mu}{r^3}\label{2.13.26}
\end{equation}
for which
\begin{equation}
z (\theta) =\half\mu a\(\e^{- (\theta -\theta_0)} +\sin (\theta -\theta_0)
-\cos (\theta -\theta_0)\) +
\mu b\(1-\cos (\theta -\theta_0)\)\label{2.13.27}
\end{equation}
the ``resistive'' force decreases to zero as $\theta $ increases.
 Even for small values of $\theta $ the value of $f $ is small.
The orbit approaches a normal Keplerian orbit asymptotically.  For illustrations
of this and other orbits see Gorringe and Leach \cite{Gorringe 88a}.
\end{example}

\subsection{Force laws admitting Keplerian orbits}

Commencing with \re{2.10.3} we construct the most general Keplerian orbits,
\ie conic sections centred on a focus.  This is motivated by the results of
Bertrand's theorem~\cite{Bertrand 73} which applies specifically to central
force orbits.  We are not particularly concerned with the inverse problem of
finding the most general equation of motion for a given orbit.  This has a
long history for which see Whittaker~\cite{Whittaker 44} and more recently
Broucke~\cite{Broucke 79} and the references cited therein.

The Keplerian orbit has its origin at one of the foci and from \re{2.11.13} we see that,
since $z (\theta_0) = 0 $,
\begin{equation}
z (\theta) = A\(1-\cos (\theta -\theta_0)\),\label{2.14.1}
\end{equation}
where $A $ is some constant.  With this and \re{2.11.8} and \re{2.11.19} we obtain
\begin{equation}
v (\theta) = A\qquad\mbox{\rm and}\qquad g =\frac{AL^2}{r^3}.\label{2.14.4}
\end{equation}
We solve this for $L^2 $ and differentiate with respect to time to obtain
\begin{equation}
2L\dot{L} =\frac{\dot{g}r^3}{A} +\frac{3r^2\dot{r}g}{A}.\label{2.14.5}
\end{equation}
Hence we find that
\begin{equation}
f = -\half\(\frac{\dot{g}}{g} +3\frac{\dot{r}}{r}\).\label{2.14.6}
\end{equation}
Thus the most general equation of motion possessing the conserved vectors
$\bfK $ and $\bfJ $ and, in addition, an orbit equation which is a conic section is given by
\begin{equation}
\ddot{\bfr} -\half\(\frac{\dot{g}}{g} +3\frac{\dot{r}}{r}\)\dot{\bfr} +g\bfr = 0.\label{2.14.7}
\end{equation}

Under the change of timescale $t\rightarrow\rho (t) $ \re{2.14.7} becomes
\begin{equation}
\dot{\rho}^2\dd{\bfr}{\rho} +\(\ddot{\rho} -\half\(\frac{\dot{g}}{g}
+3\frac{\dot{r}}{r}\)\dot{\rho}\)
\dm{\bfr}{\rho} +g\bfr = 0.\label{2.14.8}
\end{equation}
 We define the transformation by setting
the middle term to zero to obtain the equation of motion
\begin{equation}
\dd{\bfr}{\rho} +\frac{g}{\dot{\rho}^2}\bfr = 0,\label{2.14.9}
\end{equation}
with the transformation defined by
\begin{gather}
\ddot{\rho} -\half\(\frac{\dot{g}}{g} +3\frac{\dot{r}}{r}\)\dot{\rho} =
0\label{2.14.100}\\
\qquad \Longrightarrow \quad \dot{\rho}^2 =\frac{gr^3}{\mu} =\frac{AL^2}{\mu} =\frac{4 A}{\mu}\dot{S}^2
=\frac{4}{\bar{L}^2}\dot{S}^2,\label{2.14.10}
\end{gather}
where $\mu $ is the arbitrary constant in the Kepler problem
\begin{equation}
\dd{\bfr}{\rho} +\frac{\mu}{r^3}\bfr = 0, \label{2.14.11}
\end{equation}
$\bar{L} $ is the angular momentum for \re{2.14.11}
and $\dot{S} =\half L$ is the areal velocity of
\re{2.14.7}.  Equation \re{2.14.7} has been transformed into an equivalent
Kepler problem \re{2.14.11} using the area swept out in the orbit described by \re{2.14.7}.
Equation \re{2.14.10} is consistent with that obtained by
taking the area $S (t) $ swept out in the orbit described by \re{2.14.7} in time~$t $
with the same area swept out in the Kepler problem at time $\rho (t) $
as given by Kepler's Second Law, \ie $\rho (t) =
2S (t)/\bar{L} $.  The conserved vectors, $\bfK $ and $\bfJ $, of \re{2.14.7}
are form invariant under the transformation $t\rightarrow\rho (t) $
since the term $\dot{\bfr}\times\bfL/L^2 =\bfr'\times\bfL_t/L_t^2 $,
where the prime denotes differentiation with respect to $\rho $ and $L_t =r^2\theta'$, and so
\begin{gather}
\bfK  = \frac{\bfr'}{L_t} - A\(\bht -\bht_{\theta_0}\),\nonumber\\
\bfJ = \frac{\bfr'\times\bfL_t}{L_t} - A\({\bfr} -\hat{\bfr}_{\theta_0}\),\label{2.14.30}
\end{gather}
where $\hat{\bfr}_{\theta_0} =\bhi\cos\theta_0+\bhj\sin\theta_0 $ and $\bht_{\theta_0} =
-\bhi\sin\theta_0+\bhj\cos\theta_0 $.

The angular momentum for the equation of motion \re{2.14.7} is
obtained by taking the vector product with $\bfr $ to obtain
\begin{equation}
\dot{\bfL} -\half\(\frac{\dot{g}}{g} + 3\frac{\dot{r}}{r}\)\bfL = 0\label{2.14.14}
\end{equation}
and, since $\hat{\bfL} $ is constant, the scalar part of \re{2.14.14} gives on integration
\begin{equation}
L =kg^{\half}r^{\tha},\label{2.14.50}
\end{equation}
where $k $ is the constant of integration.  With the explicit form $L =r^2\dot{\theta} $
\begin{equation}
k =\(\frac{r}{g}\)^{\half}\dot{\theta}\label{2.14.60}
\end{equation}
is also a constant of the motion and $\bfk =k\hat{\bfL} $
can be regarded as a conserved generalised angular momentum.

We obtain the scalar energylike first integral for \re{2.14.7}
by simplifying \re{2.11.20} using \re{2.14.1} to substitute for $z (\theta) $
and its derivatives.  The integral is
\begin{equation}
I =\half\bfJ\cdot  \bfJ =\half A\frac{\dot{\bfr}\cdot
\dot{\bfr}}{gr^3} -\frac{A}{r} + AJ.\label{2.14.70}
\end{equation}
An alternate route to obtain an energylike first integral for \re{2.14.7}
is to take the scalar product with
$(gr^3)^{- 1}\dot{\bfr} $, which is an integrating factor, and to obtain the conserved quantity
\begin{equation}
E =\half\frac{\dot{\bfr}\cdot \dot{\bfr}}{gr^3} -\frac{1}{r} =\half{J^2}{A} - J.\label{2.14.18}
\end{equation}

The differential equation for the radial motion in time
comes from \re{2.14.18} and \re{2.14.50} and is
\begin{equation}
\dot{r} =\lb gr^2\(2Er -k^2r- 2\)\rb^{\half}.\label{2.14.19}
\end{equation}
The integration of \re{2.14.19} can only be possible if $g $ be expressible
solely as a function of~$r $ and, even then, in general the quadrature would be formal.
The angular motion in time is found from the integration of
\begin{equation}
\dot{\theta} =\lb\frac{g}{A}\(A + (J- A)\cos (\theta -\theta_0)\)\rb^{\half}.\label{2.14.20}
\end{equation}
Now the possibility of quadrature requires that $g $ be expressible
in terms of $\theta $ only.  The aerial velocity is
\begin{equation}
\dot{A} =\half L =\half k \(gr^3\)^{\half}.\label{2.14.21}
\end{equation}
The quadrature to obtain $A (t) $ requires that the right hand
side of \re{2.14.21} be expressible as a function of time.
In general this would require not only the ability to perform the quadrature
in either \re{2.14.19} or
\re{2.14.20} but also the ability to invert the result.

\subsection{The geometry of the generalised Kepler problem}

In general an expression for the area swept out by the radius vector in time
cannot be found explicitly due to problems with the inversion of functions at
more than a local level.  However, in the case that $g =\mu
r^{\alpha} $, expressions for the periodic time can be found rather elegantly
and also informatively\footnote{Interestingly there
is a change in the Lie symmetry properties of the equation
of motion~\cite{Pillay 99}, for which see \S~5.8.2.}~\cite{Gorringe 93a}.
The equation of motion, \re{2.14.7}, is now
\begin{equation}
\ddot{\bfr} -\half (\alpha + 3)\frac{\dot{r}}{r}\dot{\bfr} +\mu r^{\alpha}\bfr = 0.\label{2.15.1}
\end{equation}
To be consistent with the conserved vectors given for the standard Kepler
problem in Chapter 2 we delete the additional constant terms in the conserved
vectors given by \re{2.13.14}.  By means of the usual operations and techniques
of this Chapter we obtain
\begin{equation}
\dot{\bfL} -\half (\alpha + 3)\frac{\dot{r}}{r}\bfL = 0\qquad\Longrightarrow\qquad
L =kr^{\half (\alpha + 3)}.
\label{2.15.3}\end{equation}
Since the motion is planar, $L =r^2\dot{\theta} $ and so we have
\begin{equation}
k =r^{-\half (\alpha - 1)}\dot{\theta}\label{2.15.4}
\end{equation}
is a constant of the motion.  The Hamilton's vector,
\begin{equation}
\bfK =\frac{\dot{\bfr}}{L} -\frac{\mu}{k^2}\bht,\label{2.15.5}
\end{equation}
follows from the integration of the combination of \re{2.15.1}
divided by $L $ and \re{2.15.4}.  The Laplace--Runge--Lenz vector is
\begin{equation}
\bfJ =\bfK\times\hat{\bfL} =
\frac{\dot{\bfr}\times\bfL}{L^2} -\frac{\mu}{k^2}\hat{\bfr}\label{2.15.6}
\end{equation}
and from this we obtain the equation of the orbit
\begin{equation}
r =\frac{1}{\mu/k^2+J\cos (\theta -\theta_0)},\label{2.15.8}
\end{equation}
where $\theta_0 $ is the angle between $\bfJ $ and $\bhi$.

The scalar product of $r^{- (\alpha + 3)}\dot{\bfr} $ with \re{2.15.1} gives the energylike integral
\begin{equation}
E =\half\frac{\dot{\bfr}.\dot{\bfr}}{r^{\alpha + 3}} -\frac{\mu}{r}\label{2.15.9}
\end{equation}
after integration with respect to time.  The conserved quantities $J $, $k $
and $E $ are related according to
\begin{equation}
J^2 =K^2 =\frac{2E}{k^2} +\frac{\mu^2}{k^4}.\label{2.15.10}
\end{equation}
The period of the motion is found from \re{2.15.4} and \re{2.15.8}
to be expressible as the quadrature
\begin{equation}
T =\frac{2}{k}\int_{\theta_0}^{\pi +\theta_0}\(\frac{\mu}{k^2} +J\cos (\theta -\theta_0)\)^{\half (\alpha - 1)}\d\theta.\label{2.15.11}
\end{equation}
Using \re{2.15.10} we may rewrite the term within the large parentheses in the
integrand of \re{2.15.11} as
\begin{equation}
\(\frac{-2E}{k^2}\)^{\half}\left\{\frac{\mu}{k (- 2E)^{\half}} +\lb\(\frac{\mu}{k (- 2E)^{\half}}\)^2-1\rb^{\half}
\cos (\theta -\theta_0)\right\}.\label{2.15.12}
\end{equation}
The definition of the Legendre function of the first kind is (\cite{GAR} [8.822.1])
\begin{equation}
P_{\nu} (z) =\frac{1}{\pi}\int_0^{\pi}\frac{\d\xi}{\lb z + \(z^2-1\)^{\half}\cos\xi\rb^{\nu + 1}}
=\frac{1}{\pi}\int_0^{\pi}\lb z+\(z^2-1\)^{\half}\cos\xi\rb^{\nu}\d\xi.\label{2.15.13}
\end{equation}
When $\nu $ is an integer, this becomes the Legendre polynomial.

The energylike integral, \re{2.15.9}, is negative for an elliptical orbit.
With \re{2.15.12} and \re{2.15.13} the quadrature of the right-hand side
of \re{2.15.11} can be performed to give the period as
\begin{equation}
T =\frac{2\pi}{k}\(\frac{- 2E}{k^2}\)^{\oqr (\alpha- 1)}P_{\half (\alpha - 1)} (z),\label{2.15.14}
\end{equation}
where the argument of the Legendre function of the first kind is given by
\begin{equation}
z =\frac{\mu}{k (- 2E)^{\half}}.\label{2.15.15}
\end{equation}
The relationship between the semilatus rectum, the semimajor
axis and the eccentricity of an ellipse is
\begin{equation}
l =R \(1-e^2\),\label{2.15.17}
\end{equation}
where the symbols are in the same order.  From the orbit equation \re{2.15.8}
the semilatus rectum and the semimajor axis are given by
\begin{align}
& l   = r\(\half\pi\)\qquad && R  =\half\lb r (0) +r (\pi)\rb&&&\nonumber\\
&\phantom{I}{} =  \dsp{\frac{k^2}{\mu}} && \phantom{K}{} =\dsp{\frac{\mu/k^2}{\mu^2/k^4-J^2}}&&&\nonumber\\
&   & & \phantom{K}{}=\dsp{\frac{\mu}{- 2E}},&&&
\label{2.15.18}
\end{align}
where we have used \re{2.15.10} in the third line of the second column, so
that $z $, the argument of the generalised Kepler's Third Law, becomes
\begin{equation}
z = \(1-e^2\)^{-\half}\label{2.15.19}
\end{equation}
and the law itself can be written as
\begin{equation}
T^2R^{\alpha} = \frac{4\pi^2}{\mu}\(1-e^2\)^{-\half(\alpha+1)}
P_{\half(\alpha-1)}^2
\lb\(1-e^2\)^{-\half}\rb\label{2.15.20}
\end{equation}
when \re{2.15.14} is squared.

The differential equation \re{2.15.1} is invariant under the similarity transformation
\begin{equation}
(t,\bfr) \longrightarrow \(\bar{t},\bar{\bfr}:t =\gamma\bar{t},\bfr =\gamma^{\frac{2}{\alpha}}\bar{\bfr}\).
\label{2.15.21}
\end{equation}
Since
\begin{equation}
E =\gamma^{\frac{2}{\alpha}}\bar{E}\quad\Longrightarrow
\quad R =\gamma^{-\frac{2}{\alpha}}\bar{R},
\label{2.15.23}\end{equation}
it follows that
\begin{equation}
\frac{R^{-\frac{\alpha}{2}}}{t} =\frac{\(\gamma^{-\frac{2}{\alpha}}\bar{R}\)^{-\frac{\alpha}{2}}}{\gamma
\bar{t}} =\frac{\gamma\bar{R}^{-\frac{\alpha}{2}}}{\gamma\bar{t}} =\frac{\bar{R}^{-\frac{\alpha}{2}}}{
\bar{t}}\label{2.15.24}
\end{equation}
is invariant under the transformation.  This implies that $TR^{\frac{\alpha}{2}} $ is invariant.  In the case of the angular momentum
\begin{equation}
L =\gamma^{-\frac{4}{\alpha} - 1}\bar{L}\quad\Longrightarrow
\quad l =\gamma^{-\frac{2}{\alpha}}\bar{l}
\label{2.15.26}\end{equation}
so that the ratio
\begin{equation}
\frac{\bar{l}}{\bar{R}} =\frac{\gamma^{\frac{2}{\alpha}}l}{\gamma^{
\frac{2}{\alpha}}R} =\frac{l}{R} =
\(1-e^2\)\label{2.15.27}
\end{equation}
is invariant and hence the eccentricity, $z $, is also invariant.
Alternatively we can use the first extension of the infinitesimal generator
of the transformation, \viz
\begin{equation}
G =t\upt -\frac{2}{\alpha}r\upr\label{2.15.28}
\end{equation}
to show the invariance of \re{2.15.20}.  The task is simplified
somewhat by rewriting $G^{\lb 1\rb} (t,r,$ $\dot{r},
\dot{\theta}) $ in terms of the relevant conserved quantities,
\ie as $G^{\lb 1\rb} (T,R,l) $.  For then we have
\begin{gather}
 G^{\lb 1\rb}  = t\upt -\frac{2}{\alpha}r\upr -\frac{\alpha + 2}{\alpha}\dot{r}\up{\dot{r}}
-\dot{\theta}\up{\dot{\theta}}\nonumber\\
\qquad \Longrightarrow \quad G^{\lb 1\rb}  = T\up{T} -\frac{2}{\alpha}R\up{R} -\frac{2}{\alpha}l\up{l}\label{2.10.30}
\end{gather}
when acting upon functions of $T $, $R $ and $l $.
In the case of \re{2.15.20} it is easy to show that
\begin{equation}
G^{\lb 1\rb}\(T^2R^{\alpha}\) =G^{\lb 1\rb}\(p\(\(1-e^2\)^{-\half}\)\)
=G^{\lb 1\rb}\(p\(\(\frac{R}{l}\)^{\half}\)\)
= 0,\label{2.15.31}
\end{equation}
where $p (\cdot) $ denotes the function
on the right side of \re{2.15.20}, which illustrates the invariance.

In summary the equation of motion
\begin{equation}
\ddot{\bfr} -\half (\alpha + 3)\frac{\dot{r}}{r}\dot{\bfr} +\mu r^{\alpha}\bfr = 0\label{2.15.32}
\end{equation}
describes motion in the plane which has the following properties:
\begin{description}
\item[$(i)$] the orbit is a conic section with the origin
at a focus, \ie of the type of the Kepler problem,
\item[$(ii)$] the areal velocity is $\half kr^{\half (\alpha + 3)} $ and
\item[$(iii)$] in the case of an elliptical orbit the period and semimajor axis are related by
\begin{equation}
T^2R^{\alpha} =
\frac{4\pi^2}{\mu}\(1-e^2\)^{-\half (\alpha + 1)}P_{\half (\alpha - 1)}^2\lb\(1-e^2\)^{-\half}\rb.
\end{equation}
\end{description}

For $\alpha = - 3 $ in \re{2.15.32} we obtain the usual Kepler's laws of
planetary motion and, in particular, we see that $T^2\propto R^3 $
irrespective of the value of the eccentricity of the ellipse.  In general the areal velocity
is not constant.  The equation \re{2.15.20} is the generalised Kepler's
Third Law for a power law central force plus a resistive term.  We note
that the constant of proportionality depends explicitly upon the eccentricity of the orbit.
When $\alpha = - 3 $, the selfsimilar transformation \re{2.15.28}
maps solutions into solutions of the same eccentricity (see Prince
and Eliezer \cite{Prince 81}).  This is not reflected in the relationship
between the period and the semimajor axis.  However, for general $\alpha $
not only does the symmetry
maps solutions into solutions of the same eccentricity but in general
the relationship between the period and the semimajor axis
now depends explicitly upon the eccentricity.
We note that several
cases arise in which the constant of proportionality in \re{2.15.20}
does not depend upon the eccentricity.  This occurs for the Kepler
problem for which $\alpha = - 3 $ and when $\alpha = - 1 $.
In the cases $\alpha = 1 $ and $\alpha = 3 $ the constant
of proportionality can be made independent of the eccentricity if the
semilatus rectum, $l =R\(1-e^2\) $, is used instead of the semimajor axis length.  We obtain
\begin{equation}
T^2l =\frac{4\pi^2}{\mu}\qquad\mbox{\rm and}\qquad T^2l^3 =\frac{4\pi^2}{\mu}\label{2.15.36}
\end{equation}
respectively.  (Note that the units of $\mu$ depend upon the value of $\alpha$.)
In these two instances the implication is that all orbits with the same length of
semilatus rectum have the same period.  In the case that $\alpha = 0 $
in \re{2.15.32} we have an oscillator with an additional velocity-dependent
force which has the effect of changing the usual geometric-centred oscillator
ellipse into a focus-centred Kepler ellipse.  The isochronism
of the oscillator is now only preserved for all orbits of the same eccentricity.

From the energy equation, \re{2.15.9}, we obtain
\begin{equation}
\dot{r} =r^{\half (\alpha + 1)}\(2Er^2+ 2\mu r-k^2\)^{\half}.\label{2.15.37}
\end{equation}
This equation can be integrated in closed form for many values of $\alpha $.
 In particular for $\alpha =\pm 1 $ we can not only perform the quadrature
in closed form but also invert it (\cite[formul\ae\ 2.26ff]{GAR})
  to obtain $r (t) $.  Consequently it is possible to obtain expressions for
both $\theta (t) $ and $S (t) $.  We have

1. $\alpha = - 1 $
\begin{gather}
r (t)  = \frac{\mu}{(- 2E)} -\frac{k^2J}{(-2E)}\cos\((- 2E)^{\half} (t-t_0)\),\nonumber\\
\theta (t) = \theta_0+ 2\arctan\(\frac{k (- 2E)^{\half}}{(\mu -k^2J)}\tan\(\half (- 2E)^{\half} (t-t_0)\)\),
\nonumber\\
S (t)  = \half\lb\frac{\mu k (t-t_0)}{ (- 2E)} -\frac{k^3J}{(- 2E)^{\tha}}\sin\((- 2E)^{\half} (t-t_0)\)\rb.
\label{2.15.40}
\end{gather}

2. $\alpha = 1 $
\begin{gather}
r (t)  = \frac{1}{\mu/k^2+J\cos\(k (t-t_0)\)},\nonumber\\
\theta (t) = \theta_0+ k(t-t_0),\nonumber\\
S (t) = \half\lb\frac {2\mu k}{(- 2E)^{\tha}}
\arctan\(\frac{k (- 2E)^{\half}\tan (\half k (t-t_0))}{\mu +k^2J}\)\right.
\nonumber\\
\phantom{S(t)=}{}\left.
-\frac{k^2J\sin \(k (t-t_0)\)}{(- 2E) \(\mu/k^2+J\cos\(k (t-t_0)\)\)}\rb.\label{2.15.43}
\end{gather}
In this case for $S (t) $ the formul\ae\ of \cite[2.554.3 and 2.2553.3]{GAR} were used.

In the second case, that of $\alpha = 1 $, it is rather interesting that
$\theta $ increases linearly with time.  We note that the function $\rho (t) $
introduced in \re{2.14.8} can be obtained explicitly in these two cases from
the relationship $\rho (t) = 2S (t)/\bar{L}$.  Since the orbit is elliptical
and the area swept out in the corresponding Kepler problem is simply
a constant multiple of time, one would expect intuitively that the time
transformation required to transform \re{2.15.32} to the equivalent
Kepler equation of motion should involve the area swept out in the original elliptical orbit.

\section{Motion with conserved $\boldsymbol{L}$}

\subsection[A conserved Laplace--Runge--Lenz vector for
$\ddot{\bfr} +f\bfL +g\hat{\bfr} = 0$]{A conserved Laplace--Runge--Lenz vector for
$\boldsymbol{\ddot{\bfr} +f\bfL +g\hat{\bfr} = 0}$}

In the previous chapter we considered equations
of motion for which the direction of the angular momentum vector was constant.
In this chapter we require that the magnitude of the angular momentum vector, $\bfL $,
be constant.  We write
\begin{equation}
L^2 =\bfL\cdot \bfL\label{3.1.1}
\end{equation}
and upon differentiation of this obtain
\begin{equation}
\dot{\bfL}\cdot \bfL = 0.\label{3.1.2}
\end{equation}
Equation \re{3.1.2} implies that $\dot{\bfL} $ and $\bfL $ are orthogonal.
Equations of motion arising from~$\dot{\bfL} $ and combinations of other
vectors orthogonal to $\bfL $ describe motion for which $L $ is conserved.
One such orthogonal vector is $\bfr\times\bfL $ and so the equation
\begin{equation}
\dot{\bfL} +f\bfr\times\bfL = 0\label{3.1.3}
\end{equation}
describes one possible, not to be regarded as the most general,
class of equations meeting the requirement that the magnitude
of the angular momentum be conserved.  Since $\dot{\bfL} =\bfr\times\dot{\bfr} $
and $\bfr\times\hat{\bfr} = 0 $, we may write \re{3.1.3} as
\begin{equation}
\bfr\times\ddot{\bfr} +f\bfr\times\bfL +g\bfr\times\hat{\bfr} = 0\label{3.1.4}
\end{equation}
from which it follows that the equation of motion
\begin{equation}
\ddot{\bfr} +f\bfL +g\hat{\bfr} = 0,\label{3.1.5}
\end{equation}
where for the present $f $ and $g $ are arbitrary functions with
their variables to be specified below when it is necessary, describes motion subject to the
constraint that $L $ be constant.  We note that it is possible to introduce
additional terms of the form $e\dot{\bfr}\times\bfL $ into \re{3.1.3},
but we do not consider them here.  With the techniques described
earlier in this work we now construct Laplace--Runge--Lenz vectors for
\re{3.1.5} by the means of the imposition of restrictions
upon the so far arbitrary functions $f $ and $g $.

The vector product of \re{3.1.5} with $\bfL $ gives
\begin{equation}
\ddot{\bfr}\times\bfL +g\hat{\bfr}\times\bfL = 0.\label{3.1.7}
\end{equation}
With the aid of \re{3.1.3} and making the vector triple expansion of $\dot{\bfr}
\times (\bfL\times\bfr) $ we rewrite \re{3.1.7} in the equivalent form
\begin{equation}
(\dot{\bfr}\times\bfL)^{\dsp{{\bf .}}}  -fr\dot{r}\bfL +\frac{g}{r}\bfr\times\bfL = 0.\label{3.1.8}
\end{equation}
The ultimate term in the left hand side of \re{3.1.8} may be written in the two alternate forms
\begin{equation}
\frac{g}{r}\bfr\times\bfL = -\frac{g}{rf}\dot{\bfL}\qquad\mbox{\rm and}
\qquad \frac{g}{r}\bfr\times\bfL =
-gr^2\dot{\hat{\bfr}}\label{3.1.11}
\end{equation}
thereby prompting the suggestion that we write
\begin{equation}
g =g_1+g_2\label{3.1.12}
\end{equation}
so that \re{3.1.8} becomes
\begin{equation}
(\dot{\bfr}\times\bfL)^.  -fr\dot{r}\bfL -\frac{g_1}{rf}\dot{\bfL} -g_2r^2\dot{\hat{\bfr}} = 0.\label{3.1.13}
\end{equation}
 We now impose constraints on the as yet arbitrary functions $f $, $g_1 $
and $g_2 $ to make \re{3.1.13} trivially integrable.  We set
\begin{equation}
f =\frac{h'(r)}{r},\qquad g_1 =h (r)h'(r)\qquad\mbox{\rm and}\qquad g_2 =\frac{k}{r^2}.\label{3.1.10}
\end{equation}
Hence the system described by the equation of motion~\cite{Leach 88}
\begin{equation}
\ddot{\bfr} +\frac{h'(r)}{r}\bfL +\(h (r)h'(r) +\frac{k}{r^2}\)\hat{\bfr} = 0\label{3.1.15}
\end{equation}
possesses the conserved Laplace--Runge--Lenz vector
\begin{equation}
\bfJ =\dot{\bfr}\times\bfL -h (r)\bfL -k\hat{\bfr}.\label{3.1.14}
\end{equation}
It does not seem possible to construct a Hamilton's
vector for \re{3.1.14} for general $h (r) $.  The scalar product of $\dot{\bfr} $
with \re{3.1.15} yields the energylike integral
\begin{equation}
I =\half\dot{\bfr}\cdot \dot{\bfr} +\half h^2 (r) -\frac{k}{r}.\label{3.1.16}
\end{equation}
The conserved scalars are not independent since
\begin{equation}
J^2 = 2L^2I +k^2.\label{3.1.17}
\end{equation}

If we measure the polar angle $\theta $ from $\bfJ $,
the scalar product of \re{3.1.14} with $\bfr $ gives
\begin{equation}
r =\frac{L^2}{k+J\cos\theta}\label{3.2.2}
\end{equation}
after some rearrangement.  However, as the direction of the angular
momentum vector is not constant, the motion is truly three-dimensional
and we need not one but two equations to describe the orbit.  We
observe that \re{3.2.2} is independent of the azimuthal
angle, $\phi $, \ie we have the radial distance, $r $, as a function of the polar
angle only.  Consequently any surface upon which the position
vector moves will be a surface of revolution about the polar axis.
For the second equation we need to involve the azimuthal
angle, $\phi $.  We achieve this by taking the scalar product of $\bfJ $
with $r^{- 2}\bhf $.  This gives
\begin{equation}
\dot{r}\dot{\phi}\sin\theta +h\dot{\theta} = 0
\quad \Longrightarrow \quad \dot{\phi} = -\frac{h\dot{\theta}}{\dot{r}
\sin\theta}.\label{3.2.4}
\end{equation}
By differentiating \re{3.2.2} with respect to time we obtain
\begin{equation}
\dot{r} =\frac{Jr^2\sin\theta\dot{\theta}}{L^2}
\quad \Longrightarrow\quad \dot{\phi} = -\frac{hL^2}{Jr^2\sin 2\theta}.
\label{3.2.6}
\end{equation}
In terms of spherical polar coordinates
\begin{equation}
L^2 =r^4\(\dot{\theta}^2+\dot{\phi}^2\sin^2\theta\)\label{3.2.7}
\end{equation}
and this in conjunction with \re{3.2.6} enables us to write
\begin{equation}
\dot{\theta}^2 =\frac{L^2\(J^2\sin^2\theta -h^2L^2\)}{J^2r^4\sin^2\theta}.\label{3.2.8}
\end{equation}
We can then express the azimuthal angle as
\begin{gather}
\phi  = \int_{\theta_0}^{\theta}\frac{\dot{\phi}}{\dot{\theta}}\d\theta\nonumber\\
\phantom{\phi}{} =  -\int_{\theta_0}^{\theta}\frac{hL}{\sin\theta}\(J^2\sin^2\theta -h^2L^2\)^{-\half}\d\theta.
\label{3.2.9}
\end{gather}
From the orbit equation \re{3.2.2} $h (r) $ can be expressed in terms of
$\theta $.  Equations \re{3.2.2} and \re{3.2.9} fully describe the orbit.
Generally we would not expect to be able to express the quadrature in \re{3.2.9}
in closed form.  The relationship between $\phi $ and $\theta $
expressed through \re{3.2.9} is critical for the existence of closed
rather than surface filling orbits.  The existence of the term, $\sin\theta $,
in the denominator means that in general the pole is avoided.  From the second
term in the denominator it is also possible that a singularity will occur for an
angle in $0 < \theta_0<\theta<\pi $ and so the motion be confined to a zone of
the surface of revolution.

From \re{3.2.8} we can write
\begin{equation}
t =\int_{\theta_0}^{\theta}\frac{JL^3\sin\theta\d\theta}{(k+J\cos\theta)^2\(J^2\sin^2\theta -h^2 (r (\theta))
L^2\)^{\half}}.\label{3.3.2}
\end{equation}
If this can be inverted to obtain $\theta $ as a function of $t $,
the orbit equation \re{3.2.2} can be written as
\begin{equation}
r (t) =\frac{L^2}{k+J\cos (\theta (t))}.\label{3.3.3}
\end{equation}
Similiarly $\phi (t) $ is obtained from \re{3.2.9} by
the replacement of $\theta$ with the inverse of \re{3.3.2}.
In practice, however, it is unlikely that the quadratures could be achieved
in closed form and, even if the unlikely were to happen,
one then has the problem of the inversion.  Nevertheless everything
is sufficiently well-defined for numerical integrations to be used.
Finally we note that, although the direction of the areal velocity is
not constant, its magnitude is and so we have equal areas swept out in equal times, \ie
\begin{equation}
A =\half Lt.\label{3.3.6}
\end{equation}
We have already noted that the motion occurs on a surface of revolution.
If this be a recognisable surface, by expressing the area in terms of well-known
geometric quantities, one could possibly obtain an expression relating
the period and the geometry of the orbit.  One could of course always
use \re{3.3.2} and \re{3.2.9} (in the form which gives $\phi (t) $) to determine
the period and in the case of complex geometry there would be no choice.

\begin{example}
The possible orbits vary considerably depending upon the choice of the function
$h (r) $ and, indeed, even on the choice of the parameters in the chosen
function.  In the case of the simple choice $h (r) =\lambda r $ we have the equation

\begin{equation}
\ddot{\bfr} +\frac{\lambda}{r}\bfL +\(\lambda^2r+\frac{k}{r^2}\)\hat{\bfr} = 0\label{3.3.7}
\end{equation}
which is the three-dimensional isotropic harmonic oscillator with the additional forces,
$\lambda\bfL/r $ and $
k\hat{\bfr}/r^2 $.  (Alternatively we could consider this as the Kepler problem
with two additional forces dependent on the parameter $\lambda $.)
The formula \re{3.2.9} for the azimuthal angle takes the explicit form
\begin{equation}
\phi = -\int_{\theta_0}^{\theta}\lb\sin^2\theta\(a^2\sin^2\theta (k+J\cos\theta)^2-1\)\rb^{-\half}\d\theta
\label{3.4.1}
\end{equation}
where $a^2 =J^2\lambda^{- 2}L^{- 6} $.  The performance of the
quadrature in closed form does not appear to be possible and so we can
only evaluate $\phi (\theta) $ numerically.  Given the
comment just after \re{3.2.9} it is not surprising that the orbits close only
for certain choices of the parameter~$k $.  As one increases the value of $k $
from zero with values selected to close the orbit, there is an increase
in the amount of rotational symmetry.  The motions occur on a zone of a~sphere.
\end{example}

\begin{example}
The equation of motion
\begin{equation}
\ddot{\bfr} =\frac{\lambda\dot{\bfr}\times\bfr +F (r)\bfr}{r^3}\label{3.4.2}
\end{equation}
was studied in some detail by Thompson \cite{Thompson 87}.
He established conditions on $F (r) $ in order that \re{3.4.2} possess
an autonomous second first integral quadratic in the momentum
independent of the energy.  In summary his results were that
\begin{description}
\item[$(i)$] $\lambda = 0 $ and $F (r) =k_1 $, a constant, which is the Kepler--Coulomb
problem for $k_1\neq 0 $ and the free particle otherwise,
\item[$(ii)$] $\lambda = 0 $ and $F (r) =k_1r^3 $
which is the three-dimensional isotropic harmonic oscillator or
\item[$(iii)$] $\lambda = 1 $ and $F (r) =k_1- 1/r $.
\end{description}

For case (iii) Thompson found the conserved vector
\begin{equation}
\bfJ =\bfL\times\dot{\bfr} +\frac{\bfL}{r} -k_1{\bfr}\label{3.4.3}
\end{equation}
which is the Laplace--Runge--Lenz vector for this problem.  Case (iii) of
equation \re{3.4.2} is, of course, a special case of \re{3.1.15} with
$h = - 1/r $ and $k = -k_1 $.  It appears that the expression which
Thompson uses for $F (r) $ is incorrect and should be $F (r) =k_1+ 1/r $.
The error seems to stem from the incorrect solution of an equation in his
paper \cite[(4.12)]{Thompson 87}.  As a consequence
the sign of the angular momentum term in \re{3.4.3} is incorrect.  Thompson's Case (iii)
is an apparently specific example of the more general equation
\begin{equation}
\ddot{\bfr} +\frac{\lambda\bfL}{r^3} +\(\frac{k}{r^2} -\frac{\lambda^2}{r^3}\)\hat{\bfr} = 0\label{3.4.4}
\end{equation}
which has attracted a great deal of interest in connection with monopoles.
 In fact it is not a special case since \re{3.4.4} contains only one essential
parameter which can be taken to be $k $ by a simple rescaling of distance and time.
This equation is treated in greater detail below.  For $\lambda = 1 $ and $F (r) = 0 $
 \re{3.4.2} becomes the equation of motion describing an electric charge interacting
 with a magnetic monopole fixed at the origin for which it has been shown
that analogues of the angular momentum first integrals
exist as well as some additional first integrals which are quadratic in the
momentum, but which explicitly depend upon time (see Moreira \cite{Moreira
85b}).  The monopole system has no other time-independent scalar integrals
apart from the conserved energy.  It does seem that the vector approach,
which we have been using, seems to offer results which are more general than
the approaches adopted by other writers and with the benefit of far less
mathematical complexity.  It is the existence of the first integrals/invariants
which determines the integrability of a given equation of motion,
{\bf not} the precise assumed form of those first integrals/invariants.

The equation of motion \re{3.4.4} describes the motion of
an electric charge interacting with a magnetic monopole fixed at the origin
with the additional centripetal forces $k\hat{\bfr}/r^2 $ and $-\lambda^2\hat{\bfr}/r^3 $.

When $h (r) = -\lambda/r $ and $k = 0 $, from \re{3.2.2} we see that
\begin{equation}
r\cos\theta =\frac{L^2}{J}.\label{3.4.5}
\end{equation}
This means that $\bfr $ has a constant projection onto $\bfJ $ of length
$L^2/J $.  From \re{3.2.9} we easily determine that
\begin{equation}
\phi =\arcsec\left.\(\(1+\frac{L^2}{\lambda^2}\)^{\half}\sin\theta\)
\right|_{\theta_0}^{\theta}.\label{3.4.6}
\end{equation}
From \re{3.4.5} it is evident that $\theta $ is in the first quadrant and,
since the argument of the arcsec is greater than or equal to one, we have
that
\begin{equation}
\arcsin\frac{1}{(1+L^2/\lambda^2)^{\half}}\leq\theta\leq\half\pi.\label{3.4.7}
\end{equation}
The geometry of this orbit and those to be described below are intimately
connected with the orbits of the Kepler problem.

In the general case that $h (r) = -\lambda/r $ and $k\neq 0 $ we have a
truly three-dimensional motion for which there exists a Laplace--Runge--Lenz
vector.  The orbits may or may not be closed depending upon the relative
values of $k $ and $J $.  The integration of \re{3.2.9} is not as elementary
as in the case that $k = 0 $ and the expressions are somewhat more complicated.
We find the three distinct cases

$J\neq k $ or $-k $:
\begin{gather*}
\phi  = \half\arctan\left.\lb\frac{\frac{J^2L^2}{\lambda^2} -k^2-Jk -
J\(\frac{JL^2}{\lambda^2} +J+k\)\cos\theta}{(J+k)\(\frac{J^2L^2}{\lambda^2}
\sin^2\theta - (k+J\cos\theta)^2\)^{\half}}\rb
\right|_{\theta_0}^{\theta} \nonumber
\end{gather*}
\begin{gather}
\phantom{\phi=}{} + \half\arctan\left.\lb\frac{\frac{J^2L^2}{\lambda^2} +k^2-Jk +
J\(\frac{JL^2}{\lambda^2} +J+k\)\cos\theta}{(J- k)\(\frac{J^2L^2}{\lambda^2}
\sin^2\theta - (k+J\cos\theta)^2\)^{\half}}\rb
\right|_{\theta_0}^{\theta},
\label{3.4.8}
\end{gather}

$J =k $:
\begin{equation}
\phi =\half\arctan\left.\lb \frac{\frac{L^2}{\lambda^2} - 2-
\(\frac{L^2}{\lambda^2} + 2\)\cos\theta}{2\(\frac{L^2}{\lambda^2}\sin^2\theta
- (1+\cos\theta)^2\)^{\half}}\rb\right|_{\theta_0}^{\theta},\label{3.4.9}
\end{equation}

$J = -k $:
\begin{equation}
\phi = -\half\arctan\left.\lb \frac{\frac{L^2}{\lambda^2} - 2-
\(\frac{L^2}{\lambda^2} + 2\)\cos\theta}{2\(\frac{L^2}{\lambda^2}\sin^2\theta
- (1+\cos\theta)^2\)^{\half}}\rb\right|_{\theta_0}^{\theta}.\label{3.4.10}
\end{equation}
\end{example}

In the illustrative examples we have concentrated on the monopole because of
its phy\-si\-cal interest.  We could have used any differentiable function $h (r) $.
 The point is that Laplace--Runge--Lenz vectors exist for truly three-dimensional
motions for which only the magnitude of the angular momentum is conserved.
To obtain the orbit  we used the Laplace--Runge--Lenz vector
in two separate vectorial combinations to obtain the two equations
required to specify the orbit in three dimensions.

\subsection{The classical MICZ problem}

The motion of a spinless test particle in the field of a Dirac monopole plus
Coulomb potential with an additional centrifugal potential has been called
the MIC problem by Mladenov and Tsanov \cite{Mladenov 88} and the MICZ problem
by Cordani \cite{Cordani 88} after the studies of this problem by McIntosh and Cisneros
\cite{McIntosh 70} and also by Zwanziger \cite{Zwanziger 68}.
The MICZ system is related to the problem of the asymptotic scattering
of two self-dual monopoles by a canonical transformation \cite{Cordani 88}
which gives the reduced Hamiltonian of a particle in an Euclidean Taub-NUT space.
This in turn is related to the scattering of slowly moving Bogomol'nyi--Prasad--Sommerfield
(BPS) monopoles \cite{Atiyah 85a, Atiyah 85b, Manton 85}.

The classical MICZ problem is described by the equation of motion
\begin{equation}
\ddot{\bfr} +\frac{\lambda}{r^3}\bfL +\(\frac{\mu}{r^2} -\frac{\lambda^2}{r^3}\)\hat{\bfr} = 0,\label{3.5.1}
\end{equation}
in which the mass is taken as unity, $\lambda $ is the strength of the monopole and
$\mu $ is the strength of the Coulombic field.  The Hamiltonian for \re{3.5.1} is
\begin{equation}
H =\half\(\bfp\cdot \bfp +\frac{\lambda^2}{r^2}\) -\frac{\mu}{r},\label{3.5.2}
\end{equation}
where $\bfp =\dot{\bfr} $ is not the canonical momentum since, although
$\lb x_i,x_j\rb_{PB} = 0 $ and $\lb x_i,p_j\rb_{PB} =\delta_{ij} $
$(x_ix_i =r^2) $, the Poisson Bracket of the components of the mechanical
momentum, $\bfp $, is
\begin{equation}
\lb p_i,p_j\rb_{PB} =\lambda\varepsilon_{ijk}\frac{x_k}{r^3},\label{3.5.3}
\end{equation}
where $\delta_{ij} $ and $\varepsilon_{ijk} $ are the Kronecker delta
and Kronecker epsilon respectively.  The MICZ system possesses
two conserved vectors which can be easily derived.  We take the vector product
of $\bfr $ with \re{3.5.1} and integrate the result to obtain
\begin{equation}
\bfP =\bfL -\lambda\hat{\bfr},\label{3.5.4}
\end{equation}
which is known as Poincare's vector \cite{Poincare 96}.  The scalar product of
\re{3.5.4} with itself gives
\begin{equation}
P^2 =L^2+\lambda^2,\label{3.5.5}
\end{equation}
which indicates that not only $P $ is a constant but also $L $.  The vector
product of \re{3.5.1} with $\bfL $ is easily integrated to give the second conserved vector
\begin{equation}
\bfJ =\dot{\bfr}\times\bfL +\frac{\lambda}{r}\bfL -\mu\hat{\bfr}.\label{3.5.6}
\end{equation}
Hamilton's vector for the system \re{3.5.1} is given by
\begin{equation}
\bfK =\bfJ\times\bfP = (\dot{\bfr}\times\bfP)\times\bfP -
\frac{\mu}{r}\bfr\times\bfP.\label{3.5.8}
\end{equation}
The integrals are related according to
\begin{equation}
J^2 = 2L^2H+\mu^2\qquad\mbox{\rm and}\qquad\bfJ\cdot \bfP =\lambda\mu.\label{3.5.9}
\end{equation}
The presence of two conserved vectors for this three-dimensional motion gives
rise to a~rather special geometry.

If the velocity $\dot{\bfr} $ is replaced by the momentum
$\bfp $ in $\bfP $ and $\bfJ $, the Poisson Bracket relations are given by
\cite{Mladenov 87}
\begin{equation}
\lb P_i,P_j\rb_{PB} =\varepsilon_{ijk}P_k,\qquad\lb P_i,J_j\rb_{PB}=\varepsilon_{ijk}J_k,\qquad
\lb J_i,J_j\rb_{PB} = - 2H\varepsilon_{ijk}P_k.  \label{3.5.10}
\end{equation}
The Lie algebra of the first integrals under the operation of taking the Poisson Bracket
is $so (4) $, $e (3) $ or $so (3,1) $ depending upon whether $H $ is negative,
zero or positive.  This is the same result as is found for the standard
Kepler-Coulomb problem.  We note, however, that the Lie algebra of the Lie
point symmetries of the equation of motion, \re{3.5.1}, is $A_1\oplus so (3) $
whereas for the Kepler--Coulomb problem it is
$A_2\oplus so (3) $, where $A_1 $ and $A_2 $ are the abelian subalgebra of
dimension one and the nonabelian subalgebra of dimension two respectively~\cite{Leach 90}.

The motion of a particle described by the equation of motion \re{3.5.1} is
known to be a conic section.  The only detailed study of the geometry of the
orbit is to be found in the work of Gorringe \cite{Gorringet} and this is a
very detailed study indeed.  McIntosh and Cisneros \cite{McIntosh 70}
state that they did not find any simple algebraic expressions for the parameters
of the orbit in terms of constants of the motion.  Gorringe comments
``Using the first integrals of the MICZ system it has been possible to
describe, using very elementary methods, the complete geometry of the orbits.
Despite the comment of McIntosh and Cisneros \cite{McIntosh 70} referred to
earlier, the orbital quantities can be expressed fairly simply in terms of the
conserved quantities and orbit parameters.'' \cite{Gorringet}.  Unfortunately
the elementariness and simplicity of the derivations is not matched by brevity
and it would be out of place in terms of the balance of this work to include
it.  Consequently we present only a brief summary of the general features.

The orbit of the MICZ problem is described by the intersection of a plane and
a right circular cone.  Consequently the orbit is a conic section.  The projection
of the orbit onto the azimuthal plane is a focus-centred conic section which
corresponds naturally with the orbit of the standard Kepler problem.
In the case of negative energy, Kepler's three laws of motion for the
standard Kepler problem have a natural extension.  They
become\footnote{Bates \cite{Bates 90} has approached
this problem from a different direction using differential geometry.
In the case of the Third Law he was able to obtain only the relationship
between the period and the energy, which is the same as for the Kepler
problem in the plane, since, without
a knowledge of the geometry of the orbit, he was unable to
relate this to the semimajor axis of the orbit.}
\begin{description}
\item[$(i)$] The orbit is an ellipse.  However, the origin is not coplanar with the ellipse.
\item[$(ii)$] Equal areas are swept out on the surface of the cone in equal times.
\item[$(iii)$] The relationship between the period of the motion and the
semimajor axis length, \viz
\begin{equation}
\frac{T^2}{R^3} =\frac{4\pi^2}{\mu},\label{3.9.53}
\end{equation}
persists.
\end{description}

The orientation of the right circular cone is determined by Poincare's vector, $\bfP $.
 The line of the
Laplace--Runge--Lenz vector passes through the geometric centre of the orbit.
Consequently the plane determined
by $\bfP $ and $\bfJ $ bisects the right circular cone and the intersection
of this plane with the orbit gives the ends of the major axis.
A linear combination of $\bfP $ and $\bfJ $ naturally gives rise
to two vectors one of which is parallel
to the normal to the orbital plane and the other of which is parallel to the orbital plane.
 The second
vector plays the role of a Laplace--Runge--Lenz vector
in that it provides the orbital equation in a natural way using a suitable scalar product.

Let $\pi -\alpha $ and $\beta $ be the angles between $\bfP $
and $\bfr $ and $\bfP $ and $\bfJ $ respectively.  The scalar product
of $\bfP $, \re{3.5.4}, and $\bfr $ and of $\bfP $ and $\bfJ $, \re{3.5.6}, give
\begin{equation}
\cos\alpha =\frac{\lambda}{P}\qquad\mbox{\rm and}
\qquad \cos\beta =\frac{\lambda\mu}{PJ}\label{3.6.2}
\end{equation}
respectively.  If the angle between $\bfJ $ and $\bfr $ is $\pi -\psi $,
the scalar product of $\bfJ $ and $\bfr $ can be rearranged to give
\begin{equation}
r =\frac{L^2}{\mu -J\cos\psi}\label{3.6.3}
\end{equation}
which is one of the equations necessary to specify the orbit.
The equation \re{3.6.3} describes a surface of revolution with the line
of $\bfJ $ being the axis of symmetry and one focus of the surface is
 centred at the origin.  The constancy of the angle between $\bfP $
and $\bfr $ implies that the particle moves on the surface of
a cone with vertex at the origin.  The
curve of intersection of the cone and the surface of revolution, known as a
conicoid, is the orbit of the particle and also a plane conic section.
 By virtue of \re{3.5.4} $\bfP $, $\bfL $ and $\bfr $ are always coplanar.
 However, the scalar triple product,
\begin{equation}
\lb\bfP,\bfJ,\bfr\rb =L^2r\dot{r},\label{3.6.4}
\end{equation}
implies that $\bfP $, $\bfJ $ and $\bfr $ are coplanar only when $\dot{r} $ is zero, \ie at the
extremities of the motion.  We define the conserved vector
\begin{equation}
\bfN =\bfP -\frac{\lambda}{\mu}\bfJ.\label{3.6.5}
\end{equation}
The scalar product of $\bfN $ with $\bfr $ gives
\begin{equation}
\bfN\cdot \bfr = -\frac{\lambda}{\mu}L^2,\label{3.6.6}
\end{equation}
which confirms that the orbit lies on a plane.
However, the origin does not lie on this plane since the scalar
product produces a nonzero value.  The vector $\bfN $ is parallel
to the normal to the orbital plane.  A vector $\bfS $, which
is parallel to the orbital plane, is given by
\begin{equation}
\bfS =\bfP +\frac{\mu}{2\lambda H}\bfJ\label{3.6.8}
\end{equation}
and is obtained by using the requirement that $\bfS \cdot\bfN = 0 $.
 The angle, $\xi $, between $\bfP $ and~$\bfN $ is found by taking
the scalar product of these two vectors.  This gives
\begin{equation}
\cos\xi =\sin\gamma =\frac{\mu L}{P (2H\lambda^2+\mu^2)^{\half}},\label{3.6.9}
\end{equation}
where $\gamma =\half\pi -\xi $ is the angle between the orbital plane and the
vector $\bfP $ and can be used to find the cartesian equation of the plane.
Similiarly the angle between $\bfJ $ and $\bfN $ is found from the
scalar product between these two vectors and is given by
\begin{equation}
\cos\eta =\frac{- 2HL\lambda}{(2HL^2+\mu^2)^{\half} (2H\lambda^2+\mu^2)^{\half}}.\label{3.6.10}
\end{equation}
The projection of $\bfr $ onto $\bfN $, which defines
the orbital plane, is given, with the use of \re{3.6.6}, by
\begin{equation}
\bfr\cdot \hat{\bfN} =r\cos\zeta =\frac{\lambda L}{(2H\lambda^2+\mu^2)^{\half}},\label{3.6.11}
\end{equation}
where $\zeta $ is the angle between $\bfr $ and $\bfN $.  If we fix $\bfP $ to
be along the $-\bfk $ direction and rotate the orbit around $\bfP $ so that
the major axis of the orbit lies in the $xz $-plane, the unit normal is
given by $\hat{\bfN} = (\cos\gamma, 0,\sin\gamma)$
and correspondingly the equation of the orbital plane described by \re{3.6.11}
can be expressed in terms of cartesian components as
\begin{equation}
x\cos\gamma +z\sin\gamma =\frac{\lambda P}{\mu}\sin\gamma,\label{3.6.12}
\end{equation}
where $\gamma $ is the angle between $\bfP $ and the plane of the orbit.
The cartesian representation for the trajectory in spherical polar coordinates is given by
\begin{equation}
x =r\sin\alpha\cos\phi,\qquad y =r\sin\alpha\sin\phi,\qquad z =r\cos\alpha.\label{3.6.13}
\end{equation}
The substitution of these expressions into \re{3.6.12} gives
\begin{equation}
r =\frac{\lambda P\sin\gamma/\mu}
{\cos\alpha\sin\gamma +\sin\alpha\cos\gamma\cos\phi}\label{3.6.14}
\end{equation}
which can be equated with the orbit equation, \re{3.6.3}, to give
\begin{equation}
\psi =\arccos\(\frac{\lambda^2\mu -L^2 (2HP^2+\mu^2)^{\half}\cos\phi}{P^2J}\).\label{3.6.15}
\end{equation}
The general features of the motion are that the particle moves on a plane section of a
cone with axis of symmetry along the line of $\bfP $ and semivertex angle determined
by $\lambda/P $.  The ratio $J/\mu $ determines whether the orbit is an ellipse,
a parabola or an hyperbola.  As $\bfr $ moves over the
cone, $\bfL $, which is coplanar with $\bfr $ and $\bfP $, describes
a circle on its own cone which has semivertex angle $\half\pi -\alpha $
with $\bfP $ its axis of symmetry.  The vector $\bfN $ is parallel to the
 normal to the orbital plane and the vector $\bfS $ is parallel to the orbital plane.
From these basic considerations the geometric attributes of the orbit may
be described in terms of the parameters in and the conserved
quantities of the equation of motion \re{3.5.1}
\cite[pp.~150--206]{Gorringet}.

\section{Symmetries and conserved vectors}

\subsection{Introduction}

A second order ordinary differential equation,
\begin{equation}
\bfE (t,x,\dot{x},\ddot{x}) = 0,\label{4.1.1}
\end{equation}
 possesses a Lie symmetry,
\begin{equation}
G =\tau\upt +\eta_i\up{x_i},\label{4.1.2}
\end{equation}
if
\begin{equation}
G^{\lb 2\rb}\bfE (t,x,\dot{x},\ddot{x})_{\displaystyle{\left|_{\bfE = 0}\right.}} = 0,\label{4.1.3}
\end{equation}
where $G $ has been twice extended to be able to operate
on the first and second derivatives in the differential equation.
The coefficient functions, $\tau $ and $\eta_i $, can
be anything provided they are twice differentiable.

If the coefficient functions depend upon $t $ and $x $
only, the symmetry is a Lie point symmetry.  If in addition
they depend upon $\dot{x} $, the symmetry is a generalised one.
If the coefficient functions contained terms which are expressed
as integrals that cannot be evaluated without a knowledge
of the solution of the differential equation, the symmetry is a nonlocal symmetry.

Each of the three types of symmetry has a role to play in the discussion of
the equations of motion which we have been considering.  Traditionally the
emphasis has been placed upon Lie point symmetries, but generalised symmetries
-- particularly in the context of Noether's theorem -- have been found in many
instances to be necessary to be able to relate the existence of a first
integral/invariant and the symmetry underlying its existence.  More recently
nonlocal symmetries have been found necessary to explain the existence of
integrating factors for certain second order equations~\cite{Bouquet 2000},
to account for one of the integrable cases of the H\'enon--Heiles problem
\cite{Pillay 97} and to provide the symmetries necessary to construct the
complete symmetry groups of some differential equations
\cite{Krause 94, Nucci 2000, Cotsakis 2000}.

The application of \re{4.1.3} to the standard Kepler problem gives
the following Lie point symmetries~\cite{Prince 81}
\begin{align}
& G_1 =\upt, &&G_3 =z\upy -y\upz,&&&\nonumber\\
& G_2 =t\upt +\tth \(x\upx +y\upy +z\upz\), &&G_4 =z\upx -x\upz,&&&\nonumber\\
&&&G_5 =x\upy -y\upx&&&
\label{4.1.6}
\end{align}
which have the nonzero Lie Brackets
\begin{gather}
\lb G_1,G_2\rb =G_1,\nonumber\\
\lb G_3,G_4\rb =G_5,\qquad \lb G_4,G_5\rb =G_3,\qquad \lb G_5,G_3\rb =G_4
\label{4.1.7}
\end{gather}
from which it is evident that the Lie algebra is the direct sum $A_2\oplus so
(3) $.  The three symmetries, $G_3 $, $G_4 $ and $G_5 $, correspond to the
invariance of \re{1.1.0}
under rotation, $G_1 $ represents invariance under time translation and $G_2 $,
the symmetry often associated with the Laplace--Runge--Lenz vector,
reflects invariance under the similarity transformation
$(t,\bfr)\longrightarrow (\bar{t},\bar{\bfr}:t =\gamma\bar{t},\bfr
=\gamma^{\tth}\bar{\bfr}$).  One must realise that any of the first integrals
can in fact be constructed from more than one of the symmetries in \re{4.1.6}
\cite{Leach 81} and so the association is not strictly accurate in the
mathematical sense although it is in a geometric sense.  The two-dimensional
algebra, $A_2 $, of $G_1 $ and $G_2 $, representing invariance under time
translation and rescaling, is a common combination in physical problems and
has attracted attention in applications to a number of systems
\cite{Bouquet 91, Leach 88a, Leach 2000}.

Since the Kepler problem and the time-dependent Kepler
problem are related by a~point transformation, they have
the same algebraic structure \cite{Leach 85}.

By way of comparison the Lie point symmetries of the
charge-monopole described by the equation of motion
\begin{equation}
\ddot{\bfr} = -\frac{\mu}{r^3}\bfL,\label{4.1.8}
\end{equation}
which were studied in some detail by Moreira \etal \cite{Moreira 85b}, are
\begin{align}
& G_1 =\upt, &&G_4 =z\upy -y\upz,&&&\nonumber\\
& G_2 =t\upt +\half \(x\upx +y\upy +z\upz\), &&G_5 =z\upx -x\upz,&&&\nonumber\\
& G_3 =t (t\upt +x\upx +y\upy +z\upz), &&G_6 =x\upy -y\upx,&&&
\label{4.1.9}
\end{align}
and have the nonzero Lie Brackets
\begin{align}
& \lb G_1,G_2\rb =G_1, &&\lb G_2,G_3\rb =G_3, &&\lb G_3,G_1\rb = - 2G_2,&&&\nonumber\\
& \lb G_4,G_5\rb =G_6, &&\lb G_5,G_6\rb =G_4, &&\lb G_6,G_4\rb =G_5.&&&
\label{4.1.10}
\end{align}
The algebra is the direct sum of the subalgebras $\{G_1,G_2,G_3\} $
and $\{G_4,G_5,G_6\} $ and is $sl (2,R)
\oplus so (3) $\footnote{Moreira \etal \cite{Moreira 85b} use $so (2,1) $
rather than $sl (2,R) $.  However, we believe that the $sl (2,R) $ is more
appropriate since the three symmetries with this algebra are common to all
scalar ordinary differential equations of maximal (point) symmetry of order
greater than one and the algebra is characteristic of Ermakov systems
\cite{Ermakov 80} of which \re{4.1.8} is an instance \cite{Govinder 93}.}.
Since the algebraic structure of \re{4.1.8} is more complex than that of the
Kepler problem \re{1.1.0}, as Thompson \cite{Thompson 87} has already noted,
the charge-monopole problem is not related to the Kepler problem by a point
transformation.

In this chapter we look at the symmetries of the various systems which have
been considered above.  In the first instance we look at the point symmetries.
To explain the persistence of features such as the Laplace--Runge--Lenz
vector for problems which do not have the appropriate number of point
symmetries, in particular some rescaling symmetry such as $G_2 $
in \re{4.1.6} which is traditionally associated with the vector,
we find it necessary to introduce a nonlocal symmetry.
Following Krause \cite{Krause 94} we look at the different
systems from the point of view of their complete symmetry
groups for which, in the case of these systems, it is also necessary to introduce nonlocal
symmetries\footnote{Nonlocal symmetries are not always necessary to specify
completely an ordinary differential equation, \eg\ in the paper of Andriopoulos \etal
\cite{Andriopoulos 00} it is shown that a linear second order equation has a
complete symmetry group based on three point symmetries.}.  In the context of
the complete symmetry group we find a remarkable unity in these
diverse problems with a very interesting connection between the
scalar simple harmonic oscillator and the
Ermanno--Bernoulli constants~\cite{Nucci 2000}.

\subsection[The Lie point symmetries
 of $\ddot{\bfr} +f (r)\bfL +g (r)\bfr = 0$]{The Lie point
symmetries of $\boldsymbol{\ddot{\bfr} +f (r)\bfL +g (r)\bfr = 0}$}

We consider the more general equation
\begin{equation}
\ddot{\bfr} +f (r)\bfL +g (r)\bfr = 0 \label{4.2.1}
\end{equation}
since, when $f = 0 $ and $g =\mu/r^3 $, we regain the Kepler problem and, when
$f =\mu/r^3 $ and $g = 0 $, we obtain the charge-monopole problem.
More generally \re{4.2.1} contains the class of equations
\begin{equation}
\ddot{\bfr} +\frac{h'(r)}{r}\bfL +\(h (r)h'(r) +\frac{k}{r^2}\)\hat{\bfr} = 0 \label{4.2.2}
\end{equation}
which was shown in the previous chapter to possess a Laplace--Runge--Lenz vector.

The system \re{4.2.1} must be expressed in terms of components
for the purposes of calculating its symmetries.  Since the angular momentum
has a regular structure in a cartesian basis, it is sensible to perform the calculation in
this basis to provide a computational check through certain
regularities in the determining equations.  We write \re{4.2.1} as
\begin{gather}
\ddot{x} +f (y\dot{z} -z\dot{y}) +gx = 0,\nonumber\\
\ddot{y} +f (z\dot{x} -x\dot{z}) +gy = 0,\nonumber\\
\ddot{z} +f (x\dot{y} -y\dot{x}) +gz = 0.\label{4.2.3}
\end{gather}
Even with the aid of a symmetry computing package such as Alan Head's
Program
LIE \cite{Head 93, Head 98} one cannot do much more than obtain
the determining equations using symbolic manipulation due to
the unknown functions $f $ and $g $ in~\re{4.2.3}\footnote{It would
be interesting to see how far the very flexible
and well-known interactive package of Nucci
\cite{Nucci 90, Nucci 96a} would be able to take the calculation.
Unfortunately we did not have the package at the time.}.
There are forty-eight determining equations, twelve of which are duplicates.
The solution of the thirty-six independent equations is a tedious
business and we simply summarise the results.  We find that two cases occur, $f = 0 $
and $f\neq 0 $.

\subsubsection[The case $f = 0$]{The case $\boldsymbol{f = 0}$}

In all cases we obtain the three symmetries associated with the algebra $so (3) $ given by
\begin{gather}
G_1 =z\upy -y\upz,\nonumber\\
G_2 =x\upz -z\upx,\nonumber\\
G_3 =y\upx -x\upy.\label{4.2.4}
\end{gather}

Additional point symmetries depend upon the specific form of $g (r) $.
\begin{enumerate}
\item
$g (r) $ is unspecified: there is the single additional symmetry
\begin{equation}
G_4 =\upt.\label{4.2.5}
\end{equation}
The algebra is $A_1\oplus so (3) $.
\item
$g =\mu r^{\alpha} $: there are the two additional symmetries
\begin{gather}
G_4 =\upt,\nonumber\\
G_5 =\alpha t\upt -2r\upr.\label{4.2.6}
\end{gather}
The algebra is $A_2\oplus so (3) $ as was reported for the Kepler problem,
\re{1.1.0}, by
Prince and Eliezer
\cite{Prince 81} and Leach \cite{Leach 81}.  The Kepler problem is recovered when $\alpha = - 3 $.
\item
$g =\mu r^{-4} $: the additional symmetries are
\begin{gather}
 G_4 =\upt,\nonumber\\
G_5 = 2t\upt +r\upr,\nonumber\\
G_6 =t\(t\upt +r\upr\) .\label{4.2.7}
\end{gather}
The algebra is $sl (2,R)\oplus so (3) $
 which is the same as that found for the charge-monopole problem, \re{4.1.8},
by Moreira \etal \cite{Moreira 85b}.
\item
$g =\mu r^{- 4}-\epsilon $: in this case the additional symmetries are
\begin{gather}
 G_4 =\upt,\nonumber\\
G_5 = \e^{ 2t\epsilon^{1/2}}\(\epsilon^{- 1/2}\upt +r\upr\),\nonumber\\
G_6 =\e^{- 2 t\epsilon^{1/2}}\(\epsilon^{- 1/2} \upt -r\upr\) .\label{4.2.8}
\end{gather}
The algebra is again $sl (2,R)\oplus so (3) $.  In the case that $\epsilon $
is a negative number, say $- \omega^2 $, one may prefer to replace $G_5 $ and
$G_6 $ with
\begin{gather}
X_5 = -\half i\(G_5+G_6\) =\frac{\cos 2\omega t}{\omega}\upt +r\sin 2
\omega t\upr,\nonumber\\
X_6 =\half\( G_5-G_6\) = -\frac{\sin 2\omega t}{\omega}\upt +r\cos 2\omega t
\upr,\label{4.2.9a}
\end{gather}
although the former form is probably preferable for applications in quantum
mechanics.
\item
$g =\epsilon $: the equation of motion \re{4.2.1} is now that of a simple
three-dimensional linear system, the free particle for $\epsilon = 0 $,
the isotropic simple harmonic oscillator for $\epsilon > 0 $
and the isotropic repulsor for $\epsilon < 0 $.  In all cases there
are twenty-four symmetries and the algebra is $sl (5,R) $~\cite{Leach 81}.
\end{enumerate}

Despite the existence of a Laplace--Runge--Lenz vector
in the classical Kepler problem we see that this does little to influence
the construction of the symmetry algebra.  Apart from the three-dimensional
isotropic harmonic oscillator and the Newton--Cotes
potential all power law central force problems possess the same
number of symmetries and the same symmetry algebra,
yet global (or even local) representations for a Laplace--Runge--Lenz vector are
rarely possible.  Although the Kepler problem is regarded as
unusual because of the existence of the
Laplace--Runge--Lenz vector, in terms of the point symmetry
algebra of its equation of motion the Kepler problem is just
one of an infinite number of problems.

\subsubsection[The case $f\neq 0$]{The case $\boldsymbol{f\neq 0}$}

The generic result is that there exist the three
generators of $so (3) $ given in \re{4.2.4} and invariance under
time translation given in \re{4.2.4}.  The exceptional cases are
\begin{enumerate}
\item
$f =\lambda r^{- 3}$, $g =\mu r^{- 4} $: in addition to the usual
generators of $so (3) $ there are the three additional symmetries
given in \re{4.2.7}.  The algebra is again $sl (2,R)\oplus so (3) $.  The equation of motion is
\begin{equation}
\ddot{\bfr} +\frac{\lambda}{r^3}\bfL +\frac{\mu}{r^4}\bfr = 0\label{4.2.9}
\end{equation}
and the vector product of $\bfr $ with this gives Poincare's vector
\begin{equation}
\bfP =\bfL -\lambda\hat{\bfr}\label{4.2.10}
\end{equation}
when integrated with respect to time.  The vector product of the
equation of motion, \re{4.2.9}, with Poincare's vector, $\bfP $, gives
\begin{equation}
\dm{(\bfr\times\bfP)}{t} +\(\frac{L^4 (\lambda^2+\mu)}{| (\bfr\times\bfP) |^4}\) (\bfr\times\bfP) = 0
\label{4.2.11}
\end{equation}
which is just the equation of motion in Case 3 above in
terms of the new variable $\bfr\times\bfP $.
\item
$f =\lambda r^{- 3}$, $g =\mu r^{- 4} -\epsilon $:
the additional symmetries are given by \re{4.2.8} in Case 4 above.  The equation of motion
\begin{equation}
\ddot{\bfr} +\frac{\lambda}{r^3}\bfL +\(\frac{\mu}{r^4} -\epsilon\)\bfr = 0\label{4.2.12}
\end{equation}
also has the Poincare's vector \re{4.2.10} and, as in the previous case,
the equation of motion in the variable
$\bfr\times\bfP $ is just that of Case 4 above.
\end{enumerate}

\subsubsection{Comment}

The equation of motion
\begin{equation}
\ddot{\bfr} +f (r)\bfL +g (r)\bfr = 0\label{4.5.1}
\end{equation}
has four Lie point symmetries for general $f $ and $g $ with the
algebra $A_1\oplus so (3) $ with $A_1 $ representing
invariance under time translation and $so (3) $ the usual rotational invariance.
In the special cases
\begin{align}
& \ddot{\bfr} +\mu r^{\alpha}\bfr = 0, &&&&&\nonumber\\
& \ddot{\bfr} +\dsp{\frac{\mu}{r^4}}\bfr = 0, &&\ddot{\bfr} +
\(\dsp{\frac{\mu}{r^4}} -\epsilon\)\bfr = 0, &&&\nonumber\\
& \ddot{\bfr} +\epsilon\bfr = 0, && &&&\nonumber\\
& \ddot{\bfr} +\dsp{\frac{\lambda}{r^3}}\bfL +\dsp{\frac{\mu}{r^4}} \bfr = 0,
&&\ddot{\bfr} +\dsp{\frac{\lambda}{r^3}}\bfL
+\(\dsp{\frac{\mu}{r^4}} -\epsilon\)\bfr = 0&&&
\label{4.5.2}
\end{align}
there is additional symmetry.  The five-dimensional algebra of (\ref{4.5.2}a)
was found to be $A_2\oplus
so (3) $ in which the $A_2 $ represents invariance under
time translation and under the self-similar transformation
\begin{equation}
t =\gamma\bar{t},\qquad\bfr =\gamma^{- 2/\alpha}\bar{\bfr}.\label{4.5.8}
\end{equation}
The interesting point about (\ref{4.5.2}a) is that, as far as the Lie point
symmetries are concerned, there is no distinction made between the
Kepler problem and any other power law central force apart
from the three-dimensional isotropic harmonic oscillator.

The equations of motion (\ref{4.5.2}b) both possess the algebra
$sl (2,R)\oplus so (3) $ regardless of the
values of the parameters.  They may be regarded as direct extensions
of the results of Moreira \etal
\cite{Moreira 85b} for the algebra of the charge-monopole problem.
Equation (\ref{4.5.2}c) has the twenty-four-dimensional algebra $sl (5,R) $
of the three-dimensional isotropic harmonic oscillator, free particle or repulsor
depending whether $\epsilon >, =, < 0 $ respectively.
Equations (\ref{4.5.2}d) have the algebra $sl (2,R)\oplus so (3) $
and also possess Poincar\'e vectors which can be used, through
 the vector product, to project and rotate them onto equations of
motion equivalent to (\ref{4.5.2}b) respectively in the variable $\bfr\times\bfP $.

The term $\mu\bfr r^{- 4} $ can be interpreted as a centripetal force (Newton--Cotes)
and the term $\epsilon\bfr $ represents a three-dimensional isotropic
harmonic oscillator, free particle or repulsor depending on whether $\epsilon >, =, < 0 $
respectively.  The latter term does not affect the algebra nor the integrability of the equation.
In both equations (\ref{4.5.2}d) $L $ is constant.  The motion remains on the surface of a
cone due to the existence of the Poincar\'e vector.  Another scalar integral
\begin{equation}
I =\half\(\dot{\bfr}\cdot \dot{\bfr} -\frac{\mu}{r^2} -\epsilon r^2\)\label{4.5.9}
\end{equation}
also exists.  Equation (\ref{4.5.2}d2) does not belong to the class of the
problems, \re{4.2.2}, treated by Leach and Gorringe \cite{Leach 88}.  However,
when $\mu = -\lambda^2 $, (\ref{4.5.2}d1) belongs to the class and, in addition
to the three integrals above, there is also the conserved vector
\begin{gather}
\bfJ  = \dot{\bfr}\times\bfL +\frac{\lambda}{r}\bfL
\nonumber\\ \phantom{\bfJ }{}
=\dot{\bfr}\times\bfP\label{4.5.10}
\end{gather}
from which one of the equations specifying the orbit
is easily obtained by taking the scalar product with $\bfr $.

The MICZ problem, \re{3.5.1}, has the four-dimensional algebra, $A_1\oplus
so (3) $, which is surprising as it differs from the five-dimensional algebra,
$A_2\oplus so (3) $, of the Kepler problem and yet the Poincar\'e
vector of the MICZ problem can be used to give the equation of
motion for the Kepler problem in terms of the vector $\bfr\times\bfP $.
We recall that, in the case of the Kepler problem, the self-similar
symmetry is generally associated with Kepler's Third Law.
However, in the MICZ problem, although an analogue of this conservation
law is present, there is no suggestion of this in the Lie point symmetries
of the problem.  A~similar observation can be made in respect of \re{4.2.12}
in the case that $\mu = -\lambda^2 $.  Then the
Poincar\'e vector can be used to express the equation as that of
a three-dimensional isotropic harmonic oscillator in the variable $\bfr\times\bfP $.
Apart from the change in the number of symmetries from six to twenty-four,
 which in itself is sufficient to cause one to pause for wonder,
 the self-similar symmetry of the isotropic harmonic oscillator signals
a Kepler-type Third Law, in this case the oscillator's isochronism.
There is no suggestion of this in \re{4.2.12}.  The obvious geometric
connection between these motions on the surface of a~cone and the
corresponding planar motions is not reflected in the Lie point symmetries.

Finally we note that the equation \re{4.2.2} possesses a Laplace--Runge--Lenz
vector for general functions $h (r) $.  Can it be that there is no connection
between the existence of this extraordinary vector and the underlying
symmetry of the equation of motion apart from its invariance under time translation?
This seems to be most unlikely.

\subsection[The equation $\ddot{\bfr} +f\dot{\bfr} +\mu g\bfr = 0$]{The equation $\boldsymbol{\ddot{\bfr} +f\dot{\bfr} +\mu g\bfr = 0}$}

We turn now to the equation of motion we considered in Chapter~2
for which we found the conserved quantities
\begin{gather}
A  =  L\(gr^3\)^{-\half},\label{4.6.1}\\
\bfK  = \frac{\dot{\bfr}}{L} -\frac{\mu}{A^2}\bht,\label{4.6.2}\\
\bfJ  = \frac{\dot{\bfr}\times\hat{\bfL}}{L} +\frac{\mu}{A^2}\hat{\bfr} \label{4.6.3}
\end{gather}
in the case that
\begin{equation}
f = -\half\(\frac{\dot{g}}{g} + 3\frac{\dot{r}}{r}\),\label{4.6.4}
\end{equation}
where $\dot{g} $ represents the total time derivative of $g $.
We recall that the direction of the angular momentum vector, $\hat{\bfL} $,
is a constant and we can work with plane polar coordinates, $(r,\theta) $.
The equation of the orbit is obtained from the scalar product of \re{4.6.3} and $\bfr $ and is
\begin{equation}
r =\frac{1}{\mu A^{- 2} +J\cos\theta}\label{4.6.44}
\end{equation}
which is a conic section. Gorringe and Leach
\cite{Gorringe 93a} reported just the four Lie point symmetries, \re{4.2.4} and
\re{4.2.5}, in the case that $g $ was a function of $r $, but not
a power law, only (and by consequence $f $ of $r $ and $\dot{r} $ only)
and so we are in a situation which is similar to our considerations
of the previous section. In the case that $g $ is a power law potential
there is also a rescaling symmetry~\cite{Gorringe 93a}.

The Laplace--Runge--Lenz vector is invariant under time translation,
but there is no other Lie point symmetry as in the case of the Kepler problem.
Since we do not have a Lie point symmetry apart from $\upt $ for the
vector $\bfJ $, we look for a symmetry of the form~\cite{Pillay 99}
\begin{equation}
G =\tau\upt +\eta\upr\label{4.6.5}
\end{equation}
with no restrictions made on the coefficient functions.
Note that we are not looking for a general symmetry which
would include a term for $\up{\theta} $.  We are looking
for some generalisation of the similarity symmetry.

We require that \re{4.6.3} be invariant
under the action of the first extension of the symmetry, \ie $G^{\lb 1\rb}\bfJ = 0 $.
The first extension of $G $ is given by
\begin{equation}
G^{\lb 1\rb} =\tau\upt +\eta\upr + 0\up{\theta} +\(\dot{\eta} -\dot{r}\dot{\tau}\)\up{\dot{r}}
-\dot{\theta}\dot{\tau}\up{\dot{\theta}}.\label{4.6.6}
\end{equation}
(As a practical note we include the $\up{\theta} $ term,
even though its coefficient is zero, as a~reminder that there
is a $\up{\dot{\theta}} $ term as well.) Since
\begin{equation}
\hat{\bfr} =\cos\theta\bhi +\sin\theta\bhj\qquad\mbox{\rm and}\qquad \bht
= -\sin\theta\bhi +\cos\theta\bhj,\label{4.6.7}
\end{equation}
we have
\begin{equation}
G^{\lb 1\rb}\hat{\bfr} = 0\qquad\mbox{\rm and}\qquad G^{\lb 1\rb}\bht = 0\label{4.6.8}
\end{equation}
when $\hat{\bfr} $ and $\bht $ appear as geometrical (coordinate frame)
vectors.  When $\hat{\bfr} $ and $\bht $ appear as
a consequence of integration of the equation of motion (as occurs in
\cite{Gorringe 93a}), the calculation is generally more delicate
\cite{Hestenes 86}.  Applying the first extension of $G $ to $\bfJ $
and separating by the coefficients of the independent vectors $\hat{\bfr} $
and $\bht $ we obtain the following equations for the coefficient functions
\begin{gather}
\dot{\tau} +\half\frac{g'}{g}\eta =  0,\label{4.6.9}\\
\dot{\eta}L-\eta r\dot{r}\dot{\theta}  =  0,\label{4.6.10}
\end{gather}
where $g'=\d g/\d r $.

Since $\hat{\bfL} $ is conserved, we can replace $L $
by $r^2\dot{\theta} $ in the solution of these equations.  Hence \re{4.6.10} gives
\begin{equation}
\eta =Cr,\label{4.6.11}
\end{equation}
where $C $ is a constant of integration.  We substitute this into \re{4.6.9}
and integrate the equation to obtain
\begin{equation}
\tau = -\half C\int\frac{g'r}{g}\d t.\label{4.6.12}
\end{equation}
We have, therefore, that
\begin{equation}
G = -\half\lb\int\frac{g'r}{g}\d t\rb\upt +r\upr\label{4.6.13}
\end{equation}
is a symmetry of the (generalised) Laplace--Runge--Lenz vector.  Note that for $g
=r^{\alpha} $ this is a local symmetry and corresponds
to the result of Gorringe and Leach \cite{Gorringe 93a} quoted above.

In the case of elliptical orbits the similarity symmetry
of the Kepler problem is closely connected with Kepler's Third Law.  In this
case we have the period $T $ given by the quadrature
\begin{equation}
T = 2\int_0^{\pi}\frac{\d \theta}{\dot{\theta}} = 2\int_0^{\pi}\(\frac{r}{g}\)^{\half}\frac{\d\theta}{A}.
\label{4.6.14}
\end{equation}
The integral is evaluated by substituting for $r $ from the orbit equation,
\re{4.6.44}.  In the case that $g $ is a power law in $r $ Gorringe
and Leach \cite{Gorringe 93a} were able to evaluate the
integral in \re{4.6.14} in terms of Legendre functions.
One would not expect to be able to evaluate the quadrature
in closed form for general $g $ although this would be the
case were $g $ one of some particular polynomials in $r $.

In terms of cartesian unit vectors in the plane of the orbit
we may rewrite \re{4.6.3}, using \re{4.6.1}, as
\begin{equation}
A^2\bfJ =\bhi\lb\(\frac{\dot{\theta}^2}{g} -\mu\)\cos\theta +
\frac{\dot{r}\dot{\theta}}{gr}\sin\theta\rb
+\bhj\lb\(\frac{\dot{\theta}^2}{g} -\mu\)\sin\theta -\frac{\dot{r} \dot{\theta}}{gr}\cos\theta\rb.
\label{4.6.15}
\end{equation}
(Note that in equation (36) of \cite{Pillay 99} there
is an error which has been corrected in the formula given above.)
The two independent components of the conserved vector, the Ermanno--Bernoulli
constants, can be written in the compact form
\begin{equation}
J_{\pm} =\exp\lb\pm i\theta\rb\(\frac{\dot{\theta}^2}{g} -\mu\pm\frac{\dot{r}\dot{\theta}}{igr}\).\label{4.6.16}
\end{equation}

The components in \re{4.6.16} have a symmetry of the form \re{4.6.5} provided
\begin{equation}
\dot{\eta} -\eta\(\frac{\dot{r}g'}{g} +\frac{\dot{r}}{r}\pm
\frac{i\dot{\theta}rg'}{g}\) = 2\dot{\tau}\(\dot{r}
\pm ir\dot{\theta}\).\label{4.6.17}
\end{equation}
If in \re{4.6.17} we set $\eta = 0 $ and $\eta =r $,
we recover the $\upt $ symmetry and \re{4.6.13} respectively.
When we put $\tau = 0 $, \re{4.6.17} gives
\begin{equation}
\eta =gr\exp\lb\pm i\int\frac{rg'}{g}\d\theta\rb,\label{4.6.18}
\end{equation}
where we recall that $r $ and hence $g (r) $ are functions of $\theta $
and hence must remain in the integral, so that the components have the extra symmetry
\begin{equation}
G_{3\pm} =gr\exp\lb\pm i\int\frac{rg'}{g}\d\theta\rb\upr.\label{4.6.19}
\end{equation}

For a given $g (r) $ the integral in \re{4.6.18} is evaluated using the
expression for $r $ in \re{4.6.44}.  For the Kepler problem $g =r^{- 3} $
and \re{4.6.19} reduces to
\begin{equation}
G_{3\pm} =\frac{1}{r^2}\exp\lb\pm 3i\theta\rb\upr\label{4.6.20}
\end{equation}
which does not seem to have been reported prior to the paper
of Pillay \etal\ \cite{Pillay 99}.  We observe that the 3,2
combination of the similarity symmetry for the Kepler problem recurs in this symmetry.
 We note that this symmetry is a symmetry of the first integral and not
 a symmetry of the differential equation.

In the case of the Kepler problem the Ermanno--Bernoulli constants
have the three Lie point symmetries
\begin{equation}
G_1 =\upt,\qquad G_2 =t\upt +\tth r\upr,\qquad
G_{3\pm} =\frac{1}{r^2}\exp\lb\pm 3i\theta\rb\upr\label{4.6.21}
\end{equation}
with the Lie Brackets
\begin{equation}
\lb G_1,G_2\rb =G_1,\qquad\lb G_1,G_{3\pm}\rb = 0
 \qquad\mbox{\rm and}\qquad\lb G_2,G_{3\pm}\rb = - 2G_{3\pm}\label{4.6.22}
\end{equation}
which is a representation of the algebra $A_{3,5}^{1/2} $.

It would appear that we have now come to the beginning of a resolution of the
quandary which faced us at the end of the previous section.  We simply cannot
expect these more complicated systems to have a Lie point symmetry naturally
associated with a given first integral/invariant.  In the case of simple
systems, to which class most problems arising in a~natural physical context
belong, one has been well-served in the past by point symmetries and, perhaps
consequentially, led to regard point symmetries as the norm.  In cases for which the
analysis of the simple system can be extended to more complicated, albeit
artificial, systems we have the opportunity to extract the essence of the
system and not just the simplicity of the original system.

\subsection{Complete symmetry groups}

In 1994 Krause \cite{Krause 94} introduced a new concept
into the study of the symmetries of ordinary differential equations.
 Krause adopted the terminology
`complete symmetry group', which in the past had been used to describe the
group corresponding to the set of all Lie point symmetries of a differential
equation, defining it by the addition of two properties to the definition of
a Lie symmetry group.  These were that the manifold of solutions is
a~homogeneous space of the group and that the group is specific to the
system, \ie no other system admits it\footnote{There is an exception to this
in that there may be an arbitrary parameter present in the equation which can
be set at any permissible value by means of a transformation, such as
rescaling or translation.}.  To take a simple example the free particle in
one dimension described by the equation of motion
\begin{equation}
\ddot{q} = 0\label{4.9.1}
\end{equation}
has eight Lie point symmetries of which three are
\begin{equation}
G_1 =\upt,\qquad G_2 =\upq,\qquad G_3 =t\upt +q\upq\label{4.9.2}
\end{equation}
with the Lie Brackets
\begin{equation}
\lb G_1,G_2\rb = 0,\qquad \lb G_1,G_3\rb =G_1,\qquad \lb G_2,G_3\rb =G_2\label{4.9.3}
\end{equation}
which is a representation of the algebra $A_{3,3} $, the semidirect product of
dilations and translations,
$D\oplus_sT_2 $.  If we consider the general scalar second order equation
\begin{equation}
\ddot{q} =f (t,q,\dot{q}),\label{4.9.4}
\end{equation}
the action of $G_1 $ is to remove the
dependence of $f $ on $t $ and that of $G_2 $ is to remove the dependence on $q $.
The second extension of $G_3 $ is
\begin{equation}
G^{\lb 2\rb} =t\upt +q\upq + 0\up{\dot{q}} -\ddot{q}\up{\ddot{q}}\label{4.9.5}
\end{equation}
and this requires that the $f (\dot{q}) $ remaining be zero.
Thus the algebra $A_{3,3} $ is a representation of the complete symmetry group of \re{4.9.1}
\cite{Andriopoulos 00}.

The vehicle which Krause used to illustrate the purpose of
the introduction of the complete symmetry group was the Kepler problem.
If we consider the Lie point symmetries of the Kepler problem (in two
dimensions only for the sake of brevity), \viz
\begin{equation}
G_1 =\upt,\qquad G_2 =\up{\theta},\qquad G_3 =t\upt +\tth r\upr,\label{4.9.6}
\end{equation}
and apply them as in the example above,
we find that the two components of the equation of motion are required to have the forms
\begin{gather}
\ddot{r}  = \frac{1}{r^2}f\(r\dot{r}^2,r^3\dot{\theta}^2\),\nonumber\\
\ddot{\theta}  = \frac{1}{r^3}g\(r\dot{r}^2,r^3\dot{\theta}^2\),\label{4.9.7}
\end{gather}
where $f $ and $g $ are arbitrary functions of their arguments.
Evidently the three Lie point symmetries are not adequate to specify
completely the Kepler problem.  To obtain the complete symmetry
group of the Kepler problem Krause introduced a nonlocal symmetry defined by
\begin{equation}
Y =\lb\int\tau (t,x_1,\ldots,x_N)\d t\rb\upt +\eta_i (t,x_i,\ldots,x_N)\up{x_i}.\label{4.9.8}
\end{equation}
{\samepage This type of nonlocal symmetry is a very special type since
only the coefficient function of~$\upt $ contains a nonlocal term.
This type of nonlocal symmetry is potentially very useful to treat autonomous systems for then only the integrand and not the
integral intrudes into the calculation.  In his treatment of the three-dimensional
problem Krause found the standard five Lie point symmetries and
three additional nonlocal symmetries which were needed to specify
completely the Kepler equation, up to the value of the gravitational constant $\mu $.
The three additional symmetries are
\begin{gather}
Y_1  =  2\(\int x_1\d t\)\upt +x_1r\upr,\qquad
Y_2  =  2\(\int x_2\d t\)\upt +x_2r\upr,\nonumber\\
Y_3  =  2\(\int x_3\d t\)\upt +x_3r\upr\label{4.9.9}
\end{gather}
in which $r^2 =x_ix_i $.}

Krause did not give the algebra of these symmetries.  He also remarked
that it was not possible to obtain the symmetries of the complete symmetry
group of the Kepler problem by means of Lie point methods.  In 1996
Nucci~\cite{Nucci 96b} demonstrated that it was indeed possible
to obtain the symmetries by means of Lie point methods.

\subsection{The method of reduction of order}

The technique which Nucci used to show that one could
obtain the symmetries of the complete symmetry group
of the Kepler problem by means of Lie point methods is one
which appears to be generally useful and we summarise it here.
For the purposes of our discussion we
consider a system of $N $ second order ordinary differential equations,
\begin{equation}
\ddot{x}_i = f_i (x,\dot{x}_i),\qquad i = 1,N,\label{11.1}
\end{equation}
in which $t $ is the independent variable and $x_i$, $i = 1,N $ are the $N $
dependent variables.  These equations may be considered
as equations from Newtonian mechanics, which was Krause's
approach and which is appropriate to the context of this paper,
but there is no necessity for that to be the case.  There is no necessity for the
dependent variables to represent cartesian coordinates, nor is there a need for the
system to be of the second order.  There is no requirement that the system
be autonomous.  In the case of a nonautonomous system we can apply the
standard procedure of introducing a new variable $x_{N+ 1} =t $ and an
additional first-order equation of $\dot{x}_{N+ 1} = 1 $ so that the system
becomes formally autonomous.  In our discussion we confine
our attention to autonomous systems.

The first step in the method of reduction of order is to
write the system \re{11.1} as a $2N $-dimensional first
order system by means of the introduction of the variables
\begin{align}
& w_1 =x_1, && w_{N+ 1} =\dot{x}_1, &&&\nonumber\\
& w_2 =x_2, && w_{N+ 2} =\dot{x}_2, \nonumber\\
& \quad \vdots && \quad\vdots &&& \nonumber\\
&w_{N-1} =x_{N-1}, &&w_{2N- 1} =\dot{x}_{N-1},&&&\nonumber\\
& w_{N} =x_N, && w_{2 N} =\dot{x}_N&&&
\label{11.2}
\end{align}
so that the system \re{11.1} becomes
\begin{equation}
\dot{w}_i =g (w),\qquad i = 1,2N,\label{11.3}
\end{equation}
where $g_i =w_{N+i} $ for $i = 1,N $
and $g_i =f_i $ for $i =N+ 1,2N $.  In this first step of the reduction of the
original system, \re{11.1}, we
have simply followed the conventional method used to reduce
a higher order system to a first order system.  Any optimisation is performed in the further
selection of the final variables.  This selection may be motivated
by the existence of a known first integral, such as angular momentum,
or some specific symmetry in the original system, \re{11.1}.

We choose one of the variables $w_i $ to be a new independent variable $y $.
For the purpose of the present development we make the identification $w_N =y $.
 In terms of the new independent variable
we obtain the system of $2N- 1 $ first order equations
\begin{equation}
\dm{w_i}{y} =\frac{g_i}{g_N} =\frac{g_i}{w_{2N}},
\qquad i = 1,\ldots,N- 1,N+ 1,\ldots, 2N.\label{11.4}
\end{equation}
We do not attempt to calculate the Lie point symmetries of the system,
\re{11.4}, because the Lie point symmetries of a first order system are
generalised symmetries and one has to impose some {\it Ansatz} on the form
of the symmetry.  The choice of {\it Ansatz} tends to be more of what appeals
to the mind of the devisor rather than of some intrinsic property of the
system.  Rather we select $n\leq N- 1 $ of the variables to be new dependent
variables and rewrite the system \re{11.4} as a system of $n $ second order
equations plus $2 (N-n) - 1 $ first order equations\footnote{Again there is
no real necessity for the final equations to be a mixture of second and first
order equations.  One could include third or fourth order equations.  However,
from a practical point of view this may not be so productive.}.  The
selection of the new dependent variables, denoted by $u_1$, $u_2$, \ldots
to distinguish them from the intermediate variables $w_1$, $w_2$, \ldots,
is dictated by a number of considerations.  The
first and foremost is that we must be able to eliminate the unwanted
 variables from the system \re{11.4}.  After this condition has been
satisfied, we may look to seek variables which reflect some symmetry system,
for example an ignorable coordinate such as the azimuthal angle in a central force problem.

After the symmetries of the reduced system have been calculated, they can
be translated back to symmetries of the original system.  In the reduction
of order we are hoping that nonlocal symmetries become local,
indeed point, symmetries of the reduced system and that any existing
point symmetries remain as point symmetries of the reduced system.
This means that the nonlocal symmetries are of
Type II \cite{Abraham-Shrauner 93a, Abraham-Shrauner 92,  Abraham-Shrauner 93b,
Abraham-Shrauner 93c}.  As the computation of point symmetries is the easiest of all
computations of symmetries to perform, the existence of Type II
symmetries is desirable.  Unfortunately there is also the possibility
of the occurrence of Type I hidden symmetries, \ie Lie point symmetries
of the original system which become nonlocal on the reduction of order.
If the symmetry associated with the reduction of order is, say, $G_1 $,
and a second symmetry $G_2 $ has the Lie Bracket $\lb G_1,G_2\rb
\neq\lambda G_1 $, where the constant $\lambda $ may be zero, $G_2 $
will become an exponential nonlocal symmetry of the reduced system.
This is the risk which one must accept in  undertaking the search
for symmetry by means of a change of order of the system,
whether that change be a~reduction of order or an increase of order.
Even with simple systems the comings and goings of point
symmetries with an increase or decrease in the order is quite
extraordinary~\cite{Abraham-Shrauner 99}.

\subsection{The symmetries of the reduced Kepler problem}

The two components of the equation of motion for the Kepler problem in two dimensions are
\begin{gather}
\ddot{r} -r\dot{\theta}^2  =  -\frac{\mu}{r^2},\label{3.2}\\
r\ddot{\theta} + 2\dot{r}\dot{\theta}  =  0.\label{3.3}
\end{gather}
We introduce the new variables and their time derivatives
\begin{gather}
w_1 =r,\qquad \dot{w}_1 =w_3,\nonumber\\
w_2 =\theta, \qquad \dot{w}_2 =w_4,\nonumber\\
w_3 =\dot{r}, \qquad \dot{w}_3 =w_1w_4^2-\dsp{\frac{\mu}{w_1^2}},\nonumber\\
w_4 =\dot{\theta},\qquad \dot{w}_4 = -\dsp{\frac{2w_3w_4}{w_1}}.
\label{3.4}
\end{gather}
Since in the original system $\theta $ is an ignorable coordinate,
we select $w_2 $ to be the new independent variable $y $.
The right side of \re{3.4} leads to the reduced system
\begin{gather}
\dm{w_1}{y}  = \frac{w_3}{w_4},\label{3.5}\\
\dm{w_3}{y} = w_1w_4-\frac{\mu}{w_1^2w_4},\label{3.6}\\
\dm{w_4}{y} =  -\frac{2w_3}{w_1}.\label{3.7}
\end{gather}
We proceed now to change the system of three first order
equations into a system of one second order and one first order.  From \re{3.5}
we have $w_3 = w_4w_1'$, where the prime denotes differentiation
with respect to the new independent variable, $y $, and replace \re{3.7} by
\begin{equation}
w_4'= -\frac{2w_4w_1'}{w_1}.\label{3.8}
\end{equation}
In \re{3.6} we make the same replacement to obtain the second order equation
\begin{equation}
w_4w_1''-\frac{w_4w_1'{}^2}{w_1} =w_1w_4-\frac{\mu}{w_1^2w_4}.\label{3.9}
\end{equation}

In the elimination of $w_3 $ we have not precisely decided
that the variables $u_1 $ and $u_2 $ are to be $w_1 $ and $w_4 $.
We observe that the integrating factor $w_1^2 $ makes \re{3.8} exact as
\begin{equation}
\(w_1^2w_4\)'= 0\label{3.10}
\end{equation}
and this prompts the choice of one of the new variables to be $u_2 =w_1^2w_4 $.
We introduce this new variable into \re{3.9} to obtain
\begin{equation}
w_1''= 2\frac{w_1'{}^2}{w_1} +w_1-\frac{\mu w_1^2}{u_2^2}.\label{3.11}
\end{equation}
We rearrange \re{3.11} somewhat as
\begin{equation}
u_2^2\(\frac{w_1''}{w_1^2} -\frac{2w_1'{}^2}{w_1^3}\) =\frac{u_2^2}{w_1} -\mu\label{3.12}
\end{equation}
and immediately recognise a convenient new variable $u_1 =\mu -u_2^2/w_1 $
so that the reduced system  becomes
\begin{gather}
u_1''+u_1  =  0,\nonumber\\
u_2' =  0\label{3.13}
\end{gather}
which is an attractively simple system.  When we calculate the Lie
point symmetries of \re{3.13}, we obtain nine point symmetries.  Some of these contain
sine and cosine terms.  We combine them to give the symmetries
in terms of the exponential of an imaginary variable.  The symmetries are
\begin{gather}
\Gamma_1  = u_2\up{u_2},\nonumber\\
\Gamma_2  = \upy,\nonumber\\
\Gamma_3  = u_1\up{u_1},\nonumber\\
\Gamma_{4\pm}  = \e^{\pm iy}\up{u_1},\nonumber\\
\Gamma_{6\pm}  = \e^{\pm 2iy}\lb\upy\pm iu_1\up{u_1}\rb,\nonumber\\
\Gamma_{8\pm}  = \e^{\pm iy}\lb u_1\upy\pm iu_1^2\up{u_1}\rb.\label{4.10}
\end{gather}
In terms of the original variables we obtain
\begin{gather}
\Gamma_1  =  3t\upt + 2r\upr,\nonumber\\
\Gamma_2  = \up{\theta},\nonumber\\
\Gamma_3  =  2\lb\mu\int r\d t-L^2t\rb\upt +r\(\mu r-L^2\)\upr,\nonumber\\
\Gamma_{4\pm}  =  2\lb\int r\e^{\pm i\theta}\d t\rb\upt +r^2\e^{\pm i\theta}\upr,\nonumber\\
\Gamma_{6\pm}  = 2\lb\int\(\mu r+ 3L^2\)\e^{\pm 2i\theta}\d t\rb\upt +r\(\mu r+ 3L^2\)
\e^{\pm 2i\theta}\upr +L^2\e^{\pm 2i\theta}\up{\theta},\nonumber\\
\Gamma_{8\pm}  =  2\lb\int\left\{2\dot{r}L^3\pm ir\(\mu -r^3\dot{\theta}^2\)\(\mu +r^3\dot{\theta}^2\)
\right \}\e^{\pm i\theta}\d t\rb\upt\nonumber\\
\phantom{\Gamma_{8\pm}  =}{}
+r\lb 2\dot{r}L^3\pm ir\(\mu -r^3\dot{\theta}^2\)\(\mu +r^3\dot{\theta}^2\)
\rb\upr +L^2\(\mu -r^3\dot{\theta}^2\)\e^{\pm i\theta}\up{\theta}
\label{4.11}
\end{gather}
in which a factor of $L^2 $ has generally been included (the exceptions are
$\Gamma_2 $ and $\Gamma_1 $; in the latter a factor
of $L $ has been included) to make the expressions look simpler.

In \re{4.11} we recognise in $\Gamma_1 $ and
$\Gamma_2 $ the two point symmetries which were
not affected by the reduction of order based on the symmetry $\upt $.
The two symmetries $\Gamma_{4\pm} $ are the two nonlocal symmetries
required to be introduced to obtain the complete symmetry group for the
two-dimensional Kepler problem.  The remaining four symmetries are
additional to the number required completely to specify the differential equation.
It is curious that they were not reported by Krause~\cite{Krause 94}
since they are of the same structure as the nonlocal symmetry which he introduced.

If we examine the reduced system \re{3.13}, we find that the symmetries
$\Gamma_1 $, $\Gamma_2 $ and
$\Gamma_{4\pm} $ are sufficient to specify it completely from the general system
\begin{gather}
u_1'' = f (y,u_1,u_2,u_1'),\nonumber\\
u_2' = g (y,u_1,u_2)\label{4.12}
\end{gather}
and so we should not be surprised that in combination with the
symmetry $\upt $ these symmetries completely specify
the two-dimensional Kepler problem.  In the case of the
three-dimensional Kepler problem a third equation, of the first order,
based on the constancy of the $z $-component of the
angular momentum is added to \re{3.13}.  Since there are
at least three representations of the algebra for the complete
symmetry group of a second order linear equation
\cite{Andriopoulos 00}, one would be surprised if this representation be unique.

We have seen in detail for the two-dimensional
Kepler problem that one of the trivial equations of the reduced
system was that of the conservation of angular momentum
and have indicated for the three-dimensional Kepler problem
that there will be two trivial equations,
one for the conservation of the total angular momentum and the other for the
conservation of the $z $-component of the angular momentum.
The nontrivial equation is (\ref{3.13}a), the second order
equation for the simple harmonic oscillator.  In defining
the new variable $u_1 $ we could, although in fact we did
not, have taken the standard transformation of $u = 1/r $ which is used to
convert the radial equation of the Kepler problem to an inhomogeneous
oscillator equation \cite{Whittaker 44} and then introduced the translation to remove
the inhomogeneous term.  The equation (\ref{3.13}a) has two linear first integrals.
In the notation we are using here they are
\begin{equation}
J_{\pm} =\e^{\pm iy}\(u_1\pm iu_1'\)\label{4.13}
\end{equation}
and, if one translates these back into the original coordinates
of the two-dimensional Kepler problem, one finds that these
two invariants for the simple harmonic oscillator are just the
Ermanno--Bernoulli constants, \ie combinations of the
two components of the Laplace--Runge--Lenz vector.
This immediately raises the question whether a similar phenomenon occurs for
any other other problems which possess Laplace--Runge--Lenz vectors.

\subsection[The reduced system for some problems possessing
a conserved vector\\ of Laplace--Runge--Lenz type]{The reduced system for some problems possessing \\
a conserved vector of Laplace--Runge--Lenz type}

\subsubsection{The Kepler problem with drag}

We recall the model proposed by Danby
\cite{Danby 62} for the motion of a satellite at low
altitude subject to a resistive force due to the Earth's
atmosphere described by the equation of motion
\begin{equation}
\ddot{\bfr} +\frac{\alpha\dot{\bfr}}{r^2} +\frac{\mu\bfr}{r^3} = 0,\label{5.1}
\end{equation}
where $\alpha $ and $\mu $ are constants.  Since the
direction of the angular momentum is a constant, we may use
plane polar coordinates, $(r,\theta) $.  The two components of the equation of motion are
\begin{gather}
\ddot{r} -r\dot{\theta}^2+\frac{\alpha\dot{r}}{r^2} +\frac{\mu}{r^2}  =  0,\label{5.2}\\
r\ddot{\theta} + 2\dot{r} \dot{\theta} +\frac{\alpha\dot{\theta}}{r}  =  0.\label{5.3}
\end{gather}
We introduce the same new variables as for the Kepler problem, rewrite the
system \re{5.2} and \re{5.3} as the equivalent system
of four first order equations and again select $w_2 $
to be the new independent variable $y $.  We obtain the three first order equations
\begin{gather}
w_1' = \frac{w_3}{w_4},\label{5.5}\\
w_3' = w_1w_4-\frac{\alpha w_3}{w_1^2w_4} -\frac{\mu}{w_1^2w_4},\label{5.6}\\
w_4' =  -\frac{2w_3}{w_1} -\frac{\alpha w_4}{w_1^2},\label{5.7}
\end{gather}
where the prime denotes differentiation with respect
to the new independent variable, $y $.  The same convention is
used below.

The choice of \re{5.5} to eliminate $w_3 $ is obvious.  We obtain
\begin{gather}
w_1''w_4+w_1'w_4' = w_1w_4-\frac{\alpha w_1'}{w_1^2} -\frac{\mu}{w_1^2w_4},\label{5.8}\\
w_4' =  -\frac{ 2w_1'w_4}{w_1} -\frac{\alpha}{w_1^2}.\label{5.9}
\end{gather}
We can easily manipulate \re{5.9} to obtain
\begin{equation}
\(w_1^2w_4\)'+\alpha = 0\quad\Leftrightarrow\quad w_1^2w_4 = -\(\alpha y+\beta\),\label{5.10}
\end{equation}
which indicates that the angular momentum
is not conserved, and the choice of one of the new variables is then
\begin{equation}
u_2 =w_1^2w_4+\alpha y+\beta.\label{5.11}
\end{equation}
In \re{5.8} we use both forms of \re{5.10} to eliminate $w_4 $ and $w_4'$ to obtain
\begin{equation}
\(\frac{1}{w_1}\)'' +\frac{1}{w_1} =\frac{\mu}{(\alpha y+\beta)^2}\label{5.11a}
\end{equation}
and with the choice of the second new variable
\begin{equation}
u_1 =\frac{1}{w_1} +\mu\int^y\frac{\sin (y-s)\d s}{(\alpha s+\beta)^2}\label{5.12}
\end{equation}
we recover the system \re{3.13}.

The Lie point symmetries of the reduced system are, naturally,
those listed in \re{4.10}.  In terms of the original variables these are
\begin{gather}
\Gamma_1  = \lb\int\frac{\d t}{r^2\dot{\theta}}\rb\upt,\nonumber\\
\Gamma_2  =  2\lb\int rI'\d t\rb\upt +r^2I'\upr +\up{\theta},\nonumber\\
\Gamma_3  =  2\lb t+\int rI\d t\rb\upt +r (1+rI)\upr,\nonumber\\
\Gamma_{4\pm}  =  2\lb\int r\e^{\pm i\theta}\d t\rb\upt +r^2\e^{\pm i\theta}\upr,\nonumber\\
\Gamma_{6\pm} = \lb\int\( 2r (I'\pm iI) -\frac{\alpha}{r^2\dot{\theta}}\)\e^{\pm 2i\theta}\d t\rb\upt
+\lb r\(\pm i+r (I'\pm iI)\)\rb\e^{\pm 2i\theta}\upr -\e^{\pm 2i\theta}\up{\theta},\nonumber\\
\Gamma_{8\pm}  = \lb\int\(\frac{\dot{r}}{r^2\dot{\theta}} +
I'(1+ 2rI)\pm\frac{i}{r} (1+rI)^2-\frac{\alpha}{r^3\dot{\theta}}
(1+rI)\)\e^{\pm i\theta}\d t\rb\upt\nonumber\\
\phantom{\Gamma_{8\pm}  =}{} +\lb (1+rI)\(rI'\pm i (1+rI)'\)
\e^{\pm i\theta}\rb\upr
-\frac{1}{r} (1+rI)\e^{\pm i\theta}\up{\theta},\label{5.13}
\end{gather}
where $I $ stands for the second term in \re{5.12}
and $I'$ is its derivative with respect to $\theta $.

In addition to the symmetries listed in \re{5.13} equation \re{5.1} has the
Lie point symmetry $\upt $ which was the symmetry upon which the reduction
of order was based.  Consequently we can conclude that algebraically the Kepler
problem and the Kepler problem with drag are identical.

\subsubsection[The equation $\ddot{\bfr} -\half\(\frac{\dot{g}}{g} + 3\frac{\dot{r}}{r}\)\dot{\bfr} +\mu g\bfr = 0$]
{The equation $\boldsymbol{\ddot{\bfr} -\half\(\frac{\dot{g}}{g} + 3\frac{\dot{r}}{r}\)\dot{\bfr} +\mu g\bfr = 0}$}

The two components of this equation of motion in plane polar coordinates are
\begin{gather}
\ddot{r}  = r\dot{\theta}^2+\half\(\frac{g'}{g} +\frac{3}{r}\)\dot{r}^2-\mu gr,\nonumber\\
\ddot{\theta}  = \half \dot{r}\dot{\theta}\(\frac{g'}{g} -\frac{1}{r}\),\label{6.1}
\end{gather}
where we have written $\dot{g} (r) =g'(r)\dot{r} $.
We proceed in the same fashion as before with the
same choice of new independent variable and obtain the system of three first-order equations
\begin{gather}
w_1' = \frac{w_3}{w_4}\quad\Leftrightarrow\quad w_3 =w_4w_1',\nonumber\\
w_3' = w_1w_4+\half\(\frac{g'}{g} +\frac{3}{r}\)w_4w'_1{}^2-\mu g\frac{w_1}{w_4},\nonumber\\
w_4' = \frac{w_3}{2w_1}\(\frac{g'}{g}w_1-1\)\label{6.3}
\end{gather}
and again the obvious variable to be eliminated is $w_3 $
from the first of \re{6.3}.  With this substitution the third of \re{6.3} is easily integrated to give
\begin{equation}
A =\(\frac{w_1}{g}\)^{\half}w_4\label{6.4}
\end{equation}
and we use this as the definition of $u_2 $.
The second of \re{6.3} is now (after a certain amount of simplification)
\begin{equation}
u_2w_1''=w_1u_2+ 2\frac{u_2w_1'{}^2}{w_1} -\frac{\mu w_1^2}{u_2}.\label{6.5}
\end{equation}
This equation is just the same equation as one obtains for the Kepler problem
and we know that the correct substitution is $u_1 =\mu -u_2^2/u_1 $.
Again we have recovered the system \re{3.13} and so this problem
is also algebraically equivalent to the Kepler problem.  The symmetries are
\begin{gather*}
\Gamma_1  = \lb\int\frac{rg'}{g}\d t\rb\upt - 2r\upr,\nonumber\\
\Gamma_2  = \up{\theta},\nonumber\\
\Gamma_3  = \half\lb\int r\(\frac{1}{r} -\frac{g'}{g}\)
\(\frac{\mu g}{\dot{\theta}^2} - 1\)\d t\rb\upt
+r\(\frac{\mu g}{\dot{\theta}^2} - 1\)\upr,\nonumber\\
\Gamma_{4\pm}  = \half\lb\int r^2\(\frac{1}{r} -\frac{g'}{g}\)
\e^{\pm i\theta}\d t\rb\upt +r^2\e^{\pm\theta}\upr,\nonumber\\
\Gamma_{6\pm}  = \pm\half i\lb\int\(3+\frac{rg'}{g} +
\frac{\mu (g-g'r)}{\dot{\theta}^2}\)\e^{\pm 2i
\theta}\d t\rb\upt\pm ir\(\frac{\mu g}{\dot{\theta}^2} - 1\)
\e^{\pm 2i\theta}\upr +\e^{\pm 2i\theta}\up{\theta},\nonumber
\end{gather*}
\begin{gather}
\Gamma_{8\pm}  =\half\left\{\int\e^{\pm i\theta}\lb
\frac{gr}{\dot{\theta}^2}\(\mu -\frac{\dot{\theta}^2}{g}\)^2\(
\frac{1}{r} -\frac{g'}{g}\)\pm 2i\(\mu -\frac{\dot{\theta}^2}{g}\) +
\frac{2\dot{r}\dot{\theta}}{rg}\rb\d t
\right\}\upt\nonumber\\
\phantom{\Gamma_{8\pm}  =}{}
+\left\{\e^{\pm i\theta}\frac{gr}{\dot{\theta}^2}\(\mu -\frac{\dot{\theta}^2}{g}\)^2\right\}\upr
+\left\{\e^{\pm i\theta}\(\mu -\frac{\dot{\theta}^2}{g}\)\right\}\up{\theta}.  \label{6.6}
\end{gather}

\subsubsection{An example with an angle-dependent force}

So far we have seen that by the correct choice of variables the
reduced system comes down to be the same one,
\viz \re{3.13}.  One of the variables is related to the Laplace--Runge--Lenz
vector.  We apply the idea to the equation
\begin{equation}
\ddot{\bfr} +g\hat{\bfr} +h\bht = 0,\label{7.2}
\end{equation}
where
\begin{equation}
g =\frac{U''(\theta) +U (\theta)}{r^2} + 2\frac{V'(\theta)}{r^{3/2}}\qquad
\mbox{\rm and}\qquad h =\frac{V (\theta)}{r^{3/2}}.\label{7.3}
\end{equation}

We make the same reduction as in previous cases to arrive at the pair of equations
\begin{gather}
w_1w_1''w_4^2-2w_1'{}^2w_4^2-w_1^2w_4^2 = w_1h-gw_1,\label{7.6}\\
w_1w_4w_4'+ 2w_1'w_4^2  =  -h,\label{7.7}
\end{gather}
where $w_1 =r $ and $w_4 =\dot{\theta} $ as before.

The Laplace--Runge--Lenz vector for equation \re{7.2} is
\begin{equation}
\bfJ =\dot{\bfr}\times\bfL -U\hat{\bfr} -\lb U'+ 2r^{\half}V\rb\bht.\label{7.8}
\end{equation}
If we take the two cartesian components of $\bfJ $, \viz $J_1 $ and $J_2 $,
and combine them we obtain the two Ermanno--Bernoulli constants
\begin{gather}
J_{\pm}  =  -J_1\pm iJ_2,\nonumber\\
 \phantom{J_{\pm}}{}=\lb\(\frac{L^2}{w_1} -U\)\pm i\(\frac{L^2}{w_1} -U\)'\,\rb\e^{\pm iy}.\label{7.9}
\end{gather}
We see that the Ermanno--Bernoulli constants have the same structure
as for the standard Kepler problem.  Immediately we have the
clue to the identification of one of new variables and we set
\begin{equation}
u_1 =\frac{L^2}{w_1} -U\label{7.10}
\end{equation}
so that the Ermanno--Bernoulli constants are given by
\begin{equation}
J_{\pm} = (u_1\pm iu_1')\e^{\pm iy}.\label{7.11}
\end{equation}
The identification of the second variable is more delicate.  Equation \re{7.7}
can be written as
\begin{equation}
LL'= -w_1^{\tha}V (y)\label{7.12}
\end{equation}
and, when \re{7.10} is taken into account, this becomes
\begin{equation}
0 =\frac{L'}{L^2} +\frac{V (y)}{(u_1+U (y))^{\tha}}.\label{7.13}
\end{equation}

From \re{7.11} we have
\begin{equation}
u_1 =\half\(J_+\e^{-iy} +J_-\e^{+iy}\) =J\cos y\label{7.14}
\end{equation}
since $J_- =J_+^*$ and we have written $J =|J_+| =|J_-| $.  We use \re{7.13}
to define the second variable
\begin{equation}
u_2 =\frac{1}{L} -\int\frac{V (y)\d y}{(J\cos y+U (y))^{\tha}}.\label{7.15}
\end{equation}
The reduced system of equations is again \re{3.13}.

In addition to the symmetry $\upt $ which was used to
reduce the original system of equations the reduced system
contributes the following symmetries to the original equation
\begin{gather}
\Gamma_1  =  3\lb\int r^2\dot{\theta}\d t\rb\upt + 2r^3\dot{\theta}\upr,\nonumber\\
\Gamma_2  = \lb\int\(\frac{2U'}{r^2\dot{\theta}^2} +\frac{3Vr^2\dot{\theta}}{U+J\cos\theta}\)\d t\rb\upt
+\lb\frac{U'}{r^2\dot{\theta}^2}
+\frac{2Vr^3\dot{\theta}}{U+J\cos\theta}\rb\upr -\up{\theta},\nonumber\\
\Gamma_3  =  2\lb t-\int\frac{U\d t}{r^3\dot{\theta}^2}\rb\upt +\lb r-\frac{U}{r^2\dot{\theta}^2}\rb
\upr,\nonumber\\
\Gamma_{4\pm}  =  2\lb\int\frac{\e^{\pm i\theta}}{r^3\dot{\theta}^2}\d t\rb\upt +
\frac{\e^{\pm i\theta}}{r^2\dot{\theta}^2}\upr,\nonumber\\
\Gamma_{6\pm}  = \left\{\int\e^{\pm 2i\theta}\lb\frac{3Vr^2\dot{\theta}}{U+
J\cos\theta} +\frac{2 (U'\mp
iU)}{r^3\dot{\theta}^2}\rb\d t\right\}\upt,\nonumber\\
\phantom{\Gamma_{6\pm}  =}{} -\e^{\pm 2i\theta}\lb\frac{2Vr^3\dot{\theta}}
{U+J\cos\theta} \pm ir+\frac{U'\mp iU}{r^2\dot{\theta}^2}\rb\upr
+\e^{\pm 2i\theta}\up{\theta}\nonumber\\
\Gamma_{8\pm}  = \left\{\int\e^{\pm i\theta}\lb 2U'\(1-\frac{U}{r^3\dot{\theta}^2}\) - 3\(r^2\dot{r}
\dot{\theta}\pm ir^3\dot{\theta}^2-2r^{\half}V\mp iU\)\rb \d t\right\}\upt
\label{7.17}\\
\phantom{\Gamma_{8\pm}  =}{}-\e^{\pm i\theta}\!\lb\frac{2Vr^2\dot{\theta}}
{U+J\cos\theta}\(U'\pm i\(r^3\dot{\theta}^2-U\)\)\!\(1-\frac{U}{
r^3\dot{\theta}^2}\)\rb\!\upr +\e^{\pm i\theta}\!\lb
r^3\dot{\theta}^2-U\rb\!\up{\theta}.\nonumber
\end{gather}

Again we have a system possessing a Laplace--Runge--Lenz
vector having the same algebraic properties as the standard Kepler problem.
The standard Kepler problem has a Hamiltonian structure whereas \re{7.2}
was shown by Sen \cite{Sen 87} to be Hamiltonian only under
considerable restrictions on the functions $U (\theta) $ and $V (\theta) $.
It is presently an open question whether one can construct
a Hamiltonian through these transformations.

\subsubsection{A truly three-dimensional motion}

We mentioned above that the three-dimensional treatment
of the standard Kepler problem would require simply the addition
of a second first order equation to the system \re{3.13} and that
the suitable variable for this equation would be the $z $-component
of the angular momentum.  In the case of \re{3.1.15} we did not have a conserved
$z $-component of the angular momentum.  The magnitude of the
angular momentum vector was conserved, but the direction of
angular momentum was not constant and the motion was truly three-dimensional.
This equation gives us a good chance to test the idea put forward above
that one of the useful variables in the reduction is somehow related
to the Laplace--Runge--Lenz vector in the case in which we cannot reduce
the problem to one of two dimensions.  We recall that the equation of motion
\begin{equation}
\ddot{\bfr} +\frac{h'}{r}\bfL +\(hh'+\frac{k}{r^2}\)\hat{\bfr} = 0\label{8.1}
\end{equation}
has the Laplace--Runge--Lenz vector
\begin{equation}
\bfJ =\dot{\bfr}\times\bfL -h (r)\bfL -k{\bfr}.\label{8.2}
\end{equation}
In addition the square of the magnitude of the angular momentum, \viz
\begin{equation}
L^2 =r^4\(\dot{\theta}^2+\dot{\phi}^2\sin^2\theta\),\label{8.3}
\end{equation}
is a conserved scalar.  In spherical polar coordinates
the three components of the equation of motion, \re{8.1},
are
\begin{gather}
\ddot{r}  = r\(\dot{\theta}^2+\dot{\phi}^2\sin^2\theta\)
-hh'-\frac{k}{r^2},\label{8.4}\\
\ddot{\theta}  =  - 2\frac{\dot{r}\dot{\theta}}{r}
+\dot{\phi}^2\sin\theta\cos\theta +h'\sin\theta\dot{\phi},\label{8.5}\\
\ddot{\phi}\sin\theta  =
 - 2\frac{\dot{r}\dot{\phi}}{r}\sin\theta - 2\dot{\theta}\dot{\phi}\cos\theta -h'\dot{\theta}.\label{8.6}
\end{gather}

The three components of the Laplace--Runge--Lenz vector are
\begin{gather}
J_x  = \(\frac{L^2}{r} -k\)\sin\theta\cos\phi -r^2\(\dot{r}\dot{\theta}
-h\dot{\phi}\sin\theta\)\cos\theta\cos\phi +r^2\(\dot{r}\dot{\phi}\sin\theta
+h\dot{\theta}\)\sin\phi,\nonumber\\
J_y  = \(\frac{L^2}{r} -k\)\sin\theta\sin\phi -r^2\(\dot{r}\dot{\theta}
-h\dot{\phi}\sin\theta\)\cos\theta\sin\phi -r^2\(\dot{r}\dot{\phi}\sin\theta
+h\dot{\theta}\)\cos\phi,\nonumber\\
J_z  = \(\frac{L^2}{r} -k\)\cos\theta +r^2\(\dot{r}\dot{\theta} -h\dot{\phi}\sin\theta\)\sin\theta.\label{8.9}
\end{gather}
To obtain the Ermanno--Bernoulli constants we take the combinations
\begin{gather}
J_{\pm}  = J_x\pm iJ_y\label{8.10}\\
\phantom{J_{\pm} }{} =\e^{\pm i\phi}\lb \(\(\frac{L^2}{r} -k\)\sin\theta
-r^2\(\dot{r}\dot{\theta} -h\dot{\phi}\sin\theta\)\cos\theta\)
\pm i\(-r^2\(\dot{r}\dot{\phi}\sin\theta +h\dot{\theta}\)\)\rb\!.\!
\nonumber
\end{gather}
Since $\phi $ does not appear in the components of the equation of motion,
we choose it to be the new independent variable.  It is a simple matter
to verify that the derivative of the real part within crochets in \re{8.10}
with respect to $\phi $ gives the imaginary part and so we define the new variable
\begin{equation}
u_1 = \(\frac{L^2}{r}-k\)\sin\theta -
r^2\(\dot{r}\dot{\theta}-h\dot{\phi}\sin\theta\)\cos\theta
..\label{8.11}
\end{equation}
Thus we have the Ermanno--Bernoulli constants in the standard form
\begin{equation}
J_{\pm} =\e^{\pm i\phi}\(u_1\pm iu_1'\),\label{8.12}
\end{equation}
where the new variable $u_1 $ satisfies the second order equation
\begin{equation}
u_1''+u_1 = 0.\label{8.13}
\end{equation}

The other two variables for the reduced system can be taken
to be the remaining component of the Laplace--Runge--Lenz vector and the
square of the magnitude of the angular momentum.  We write
\begin{align}
& u_2 =r^4\(\dot{\theta}^2+\dot{\phi}^2\sin^2\theta\),&& u_2'= 0,&&&\nonumber\\
& u_3 =\({\frac{L^2}{r}} -k\)\cos\theta +r^2\(\dot{r}\dot{\theta} -
h\dot{\phi}\sin\theta\)\sin\theta, && u_3'= 0,&&&\label{8.15}
\end{align}
and so we have the complete reduction of \re{8.1} to a system of two
first-order equations and the simple harmonic oscillator.  The algebra of the Lie
point symmetries is $2 A_1\oplus sl (3,R) $ if one takes the two symmetries
\begin{equation}
G_9 =A (u_2,u_3)\up{u_2}\qquad\mbox{\rm and}\qquad G_{10} =B (u_2,u_3)\up{u_3},\label{8.16}
\end{equation}
where $A $ and $B $ are arbitrary functions of the two first integrals,
to be equivalent to just $\up{u_2} $ and $\up{u_3} $.
The involved expressions for the new variables and
the additional complication of an extra dimension make the computation
of the symmetries in the original variables a task of somewhat greater
length than the result justifies and we omit them.

Finally we note that, of the Lie point symmetries of the system of
differential equations, those associated with $J_{\pm} $ are
\begin{gather}
G_{1\pm} =\pm i\upy +u_1\up{u_1},\nonumber\\
 G_{2\pm}
=\e^{\pm 2iy}\(\upy\pm iu_1\up{u_1}\)\qquad\mbox{\rm and}\qquad
G_{3\pm} =\e^{\pm iy}\up{u_1}\label{8.17}
\end{gather}
in addition to the trivial $G_9 $ and $G_{10} $.  To these three
symmetries in the original representation there is added $\upt $.

\subsection*{Acknowledgments}
PGLL thanks the Director of GEODYSYC, Dr S Cotsakis, and the
Department of Mathematics, University of the Aegean, for their
hospitality and provision of facilities while this work was
undertaken and acknowledges the continued support of the National
Research Foundation of the Republic of South Africa and the
University of Natal.

\label{leach-lastpage}

\end{document}